\documentclass[a4paper,iop,twocolumn,numberedappendix]{emulateapj} 

\usepackage[colorlinks=true,linkcolor=blue,anchorcolor=blue,citecolor=blue,urlcolor=blue,draft=false]{hyperref}

\usepackage{color}
\usepackage{graphicx}
\usepackage{amsmath} 
\usepackage{amssymb}

\usepackage{placeins} 
\usepackage{times}

\usepackage{xcolor}

\usepackage{natbib}
 
\newcommand{\simgt}{\lower.5ex\hbox{$\; \buildrel > \over \sim \;$}}
\newcommand{\simlt}{\lower.5ex\hbox{$\; \buildrel < \over \sim \;$}}
\newcommand{\Om}{\Omega_m}
\newcommand{\OL}{\Omega_\Lambda}

\newcommand{\Mhalo}{M_{200\mathrm{c}}}
\newcommand{\cvir}{c_\mathrm{vir}}
\newcommand{\chalo}{c_{200\mathrm{c}}}
\newcommand{\rvir}{r_\mathrm{vir}}
\newcommand{\rhalo}{r_{200\mathrm{c}}}
\newcommand{\Nbin}{N_\mathrm{bin}}
\newcommand{\Npix}{N_\mathrm{pix}}
\newcommand{\fgas}{f_\mathrm{gas}}
\newcommand{\kpch}{kpc\,$h^{-1}$}
\newcommand{\Mpch}{Mpc\,$h^{-1}$}
\newcommand{\Munit}{10^{15}M_\odot h^{-1}}

\def\bm{\mbox{\boldmath $m$}}
\def\bc{\mbox{\boldmath $c$}}
\def\btheta{\mbox{\boldmath $\theta$}} 
\def\bnabla{\mbox{\boldmath $\nabla$}}

\def\bSigma{\mbox{\boldmath $\Sigma$}}

\def\bd{\mbox{\boldmath $d$}}

\def\bo{\mbox{\boldmath $o$}}
\def\bs{\mbox{\boldmath $s$}}
\def\bp{\mbox{\boldmath $p$}}
\def\br{\mbox{\boldmath $r$}}

\lefthead{Umetsu et al.}
\righthead{Three-Dimensional Multi-probe Analysis of A1689}  

\usepackage{datetime}
\usepackage{version}	

\begin{document}

\title{Three-dimensional Multi-probe Analysis of the Galaxy Cluster A1689\altaffilmark{*}}
\shorttitle{Three-dimensional Multi-Probe Analysis of A1689}
\hypersetup{pdfauthor={Keiichi Umetsu}, pdftitle={Three-dimensional
Multi-probe Analysis of A1689}}

\author{Keiichi Umetsu\altaffilmark{1}}      
\author{Mauro Sereno\altaffilmark{2,3}}      

\author{Elinor Medezinski\altaffilmark{4,5}} 
\author{Mario Nonino\altaffilmark{6}}        
\author{Tony Mroczkowski\altaffilmark{7}}    
\author{Jose M. Diego\altaffilmark{8}}       
\author{Stefano Ettori\altaffilmark{3,9}}    
\author{Nobuhiro Okabe\altaffilmark{10,11}}  
\author{Tom Broadhurst\altaffilmark{12}}     
\author{Doron Lemze\altaffilmark{5}}         

\altaffiltext{*}
 {Based in part on data collected at the Subaru Telescope,
  which is operated by the National Astronomical Society of Japan.}
\altaffiltext{1}
 {Institute of Astronomy and Astrophysics, Academia Sinica,
  P.~O. Box 23-141, Taipei 10617, Taiwan; \email{keiichi@asiaa.sinica.edu.tw}}
\altaffiltext{2}
 {Dipartimento di Fisica e Astronomia, 
 Alma Mater Studiorum -- Universit\`a di Bologna
 Viale Berti Pichat 6/2, 40127 Bologna, Italia}
\altaffiltext{3}{INAF-Osservatorio Astronomico di Bologna, via Ranzani
 1, I-40127 Bologna, Italy} 
\altaffiltext{4}
 {The Racah Institute of Physics,The Hebrew University of Jerusalem,
Jerusalem 91904, Israel}
\altaffiltext{5}
 {Department of Physics and Astronomy, The Johns Hopkins
University, 3400 North Charles Street, Baltimore, MD 21218, USA}
\altaffiltext{6}
 {INAF-Osservatorio Astronomico di Trieste, via G.B. Tiepolo 11, I-34143
 Trieste, Italy} 
\altaffiltext{7}
 {National Research Council Fellow at the U.S. Naval Research Laboratory,
 4555 Overlook Ave SW, Washington, DC 20375, USA} 
\altaffiltext{8}{IFCA, Instituto de F\'isica de Cantabria (UC-CSIC),
 Av. de Los Castros s/n, 39005 Santander, Spain} 
\altaffiltext{9}{INFN, Sezione di Bologna, Viale Berti Pichat 6/2,
 I-40127 Bologna, Italy}
\altaffiltext{10}{Department of Physical Science, Hiroshima University,
 1-3-1 Kagamiyama, Higashi-Hiroshima, Hiroshima 739-8526, Japan}
\altaffiltext{11}{Kavli Institute for the Physics and Mathematics of the
 Universe (WPI), Todai Institutes for Advanced Study,University of
 Tokyo, 5-1-5 Kashiwanoha, Kashiwa, Chiba 277-8583, Japan} 
\altaffiltext{12}{Ikerbasque, Basque Foundation for Science, Alameda
 Urquijo, 36-5 Plaza Bizkaia, E-48011 Bilbao, Spain} 

\begin{abstract}
We perform a three-dimensional multi-probe analysis of the rich galaxy
 cluster A1689, 
one of the most powerful known lenses on the sky,  
by combining improved weak-lensing data from new wide-field  
$BVR_\mathrm{C}i'z'$ Subaru/Suprime-Cam observations with 
strong-lensing, X-ray, and Sunyaev-Zel'dovich effect (SZE) data sets.
We reconstruct the projected matter distribution from a
 joint weak-lensing analysis of two-dimensional shear and
 azimuthally integrated magnification constraints, the combination of
 which allows us to break the mass-sheet degeneracy.
The resulting mass distribution reveals elongation with an axis ratio of
 $\sim 0.7$ in projection, aligned well with the 
 distributions of cluster  galaxies and intracluster gas.
When assuming a spherical halo, our full weak-lensing analysis 
yields a projected halo concentration of $\chalo^\mathrm{2D}=8.9\pm 1.1$ 
($\cvir^\mathrm{2D} \sim 11$),
consistent with and improved from earlier weak-lensing work. 
We find excellent consistency between independent weak and strong
 lensing in the region of overlap. 
In a parametric triaxial framework, we constrain the intrinsic
 structure and geometry of the matter and gas  distributions,
 by combining weak/strong lensing and X-ray/SZE data with minimal
 geometric assumptions. 
We show that the data favor a triaxial geometry with minor--major axis
 ratio $0.39\pm 0.15$ and major axis closely aligned with the line of
 sight ($22^\circ\pm 10^\circ$).
We obtain
a halo mass  
$\Mhalo=(1.2\pm 0.2)\times \Munit$
and
a halo concentration 
$\chalo=8.4\pm 1.3$,
which overlaps with the $\simgt 1\sigma$ tail of the predicted
 distribution. 
The shape of the gas is rounder than the underlying
 matter but quite elongated with minor--major axis ratio 
$0.60\pm 0.14$. 
The gas mass fraction within 0.9\,Mpc is $10^{+3}_{-2}\%$,
a typical value for high-mass clusters.
The thermal gas pressure contributes to $\sim 60\%$ of the equilibrium
 pressure, indicating a significant level of non-thermal pressure support. 
When compared to {\em Planck}'s hydrostatic mass
 estimate, our lensing measurements yield a 
spherical mass ratio of $M_\mathrm{\it Planck}/M_\mathrm{GL}=0.70\pm 0.15$ and
 $0.58\pm 0.10$ with and without corrections for lensing projection
 effects, respectively. 
\end{abstract}


\section{Introduction} 
\label{sec:intro}

The evolution of the abundance of galaxy clusters with cosmic epoch is
sensitive to the amplitude and growth rate of primordial density
fluctuations as well as to the cosmic volume-redshift relation
because massive clusters lie in the high-mass exponential tail of the
halo mas function \citep{2001ApJ...553..545H,Watson+2014}.
Therefore, large cluster samples defined from cosmological
surveys can provide an independent means of examining any viable
cosmological model, including the current concordance $\Lambda$ cold
dark matter ($\Lambda$CDM) model defined in the framework of general
relativity, complementing cosmic microwave background (CMB),
large-scale galaxy clustering, and supernova observations.  

Clusters provide various probes of the role and nature of ``dark
matter'' (DM) that dominates the material universe
\citep{Clowe2006Bullet},  
or modified gravity theories as an alternative to DM
\citep{Rapetti+2011}, physics governing the final state of
self-gravitating collisionless systems in an expanding universe  
\citep{1996ApJ...462..563N,1997ApJ...490..493N,Taylor+Navarro2001,Hjorth+2010DARKexp},  
and screening mechanisms in long-range modified models of gravity
whereby general relativity is restored \citep{Narikawa+2013}.
 
Substantial progress has been made in recent years in constructing
statistical samples of clusters thanks to dedicated surveys
\citep[e.g.,][]{Planck2014XX,Planck2015XXIV,SPT2015sze}. 
Cluster samples are often defined by X-ray or
Sunyaev-Zel'dovich effect (SZE) observables,
so that the masses are indirectly inferred from scaling relations, which
are often based on the assumption of hydrostatic 
equilibrium (HSE) and then statistically calibrated using weak lensing
or internal dynamics using a subset of massive clusters at lower
redshifts \citep{Rines2013,2014MNRAS.442.1507G}. 
Since the level of mass bias from indirect observations assuming
HSE is likely mass dependent \citep{CoMaLit2} and sensitive to
calibration systematics of the instruments
\citep{Donahue2014clash,Israel+2015}, a systematic effort is needed to  
enable a self-consistent calibration of mass--observable relations
using robust, direct cluster mass measurements
\citep{vonderLinden2014calib,Umetsu2014clash,Merten2014clash,Ford2014cfhtlens,Jimeno2015,Hoekstra2015CCCP,Simet2015}
and well-defined selection functions \citep[e.g.,][]{JPAS2014}.

The great attraction of gravitational lensing in the cluster regime is
its ability to map the mass distribution on an individual cluster basis,
independent of and free from assumptions about the physical and dynamical
state of the cluster system \citep{2007ApJ...669..714M,Okabe+Umetsu2008,Hamana2009}.
Clusters act as efficient gravitational lenses, producing various
observable effects, including deflection, distortion, and magnification
of the images of background sources \citep{2001PhR...340..291B}.
In the weak regime, the lensing signals are approximately linearly
related to the gravitational potential, so that one can determine the
distribution of lensing matter at large scales in a model-independent
manner \citep[e.g.,][]{1999PThPS.133...53U,Umetsu+2011}.
In the strong regime, several sets of multiply-lensed images with known
redshifts can be used to constrain the mass distribution in the cluster
cores \citep[e.g.,][]{Jauzac2014,Zitrin2015clash}. 


A practical difficulty of obtaining precise mass measurements from
cluster lensing, however, is significant scatter present in the
projected lensing 
signals due to inherent variations (at a fixed halo mass) in halo
concentration, asphericity, orientation, and the presence of correlated
large scale structure \citep{Rasia+2012}.
The projection effects due to such intrinsic profile variations alone
can produce a $\simlt 20\%$ uncertainty in lensing mass estimates  
for $\sim 10^{15}M_\odot$ clusters \citep{Becker+Kravtsov2011,Gruen2015}.  

A possible way to overcome this problem
 is to simultaneously determine the mass, concentration, shape, and
orientation of a given cluster by combining lensing data with
independent probes or information about its line-of-sight elongation
 \citep{2007MNRAS.380.1207S,Corless2009triaxial,Limousin2013}.
Gravitational lensing probes the structure and morphology of the matter
 distribution in projection. 
X-ray observations constrain the characteristic size and
orientation of the intracluster medium (ICM) in the sky plane.
The elongation of the ICM along the line of sight can be constrained
from the combination of X-ray and thermal SZE observations
 \citep{DeFilippis2005,Sereno+Ettori+Baldi2012}.
Recently,
\citet{Sereno2013glszx} developed a parametric triaxial framework
to combine and couple independent morphological constraints from lensing
and X-ray/SZE data, using minimal geometric assumptions about the matter
and gas distributions but without assuming HSE.


The first critical step in a three-dimensional (3D) cluster analysis 
is an unbiased, direct recovery of the projected cluster mass
distribution from weak lensing.
A fundamental limitation of measuring
shear only is the {\em mass-sheet} degeneracy
\citep{Schneider+Seitz1995}. 
This degeneracy can be 
broken by using the complementary combination of shear and magnification 
\citep{Schneider+2000,Umetsu+2011,Umetsu2013}.
\citet{Umetsu+2011} have shown that the magnification effect can
significantly enhance the accuracy and precision of lensing-derived
cluster mass profiles when added to weak-lensing shear measurements. 

Our aim in this paper is to develop and apply a comprehensive set of
techniques and methods for 3D analysis of galaxy clusters
based on the multi-probe framework of \citet{Sereno2013glszx}.
For this aim, we first generalize the one-dimensional (1D)
weak-lensing inversion method of \citet{Umetsu+2011}  
to a two-dimensional (2D) description of the mass distribution
without assuming particular functional forms,
i.e., in a {\em free-form} fashion.
In this approach, we combine the spatial shear pattern with azimuthally
averaged magnification information, imposing integrated constraints on
the mass distribution. 

Taking advantage of new $BVR_\mathrm{C}i'z'$ imaging obtained with
Suprime-Cam on the 8.3\,m Subaru Telescope, we perform a new
weak-lensing analysis of the rich cluster A1689 at $z=0.183$
and then apply our methods to weak-lensing, strong-lensing, X-ray, and
SZE data sets we have obtained for the cluster. 
The cluster is among the best studied clusters
\citep{Tyson1995,1998ApJ...501..539T,Andersson+Madjeski2004,2005ApJ...621...53B,2006MNRAS.372.1425H,2007ApJ...668..643L,UB2008,Peng+2009,Kawaharada+2010,Coe+2010,Sereno+Ettori+Baldi2012,Nieuwenhuizen+Morandi2013,Sereno2013glszx} 
and 
one of the most powerful known lenses on the sky,
characterized by a large Einstein radius of
$\theta_\mathrm{Ein}=47.0\arcsec\pm 1.2\arcsec$ for a fiducial source at
$z_\mathrm{s}=2$ \citep[see Table \ref{tab:cluster};][]{Coe+2010};
this indicates a high degree of mass concentration in projection
\citep{Broadhurst+Barkana2008}. 
To date, 61 candidate systems of 165 multiply-lensed images have been identified 
\citep{2005ApJ...621...53B,Coe+2010,Diego2015a1689}
from Advanced Camera for Surveys (ACS) observations with the
{\em Hubble Space Telescope} ({\em HST}).
Despite significant efforts, the degree of concentration inferred from
different lensing analyses is somewhat controversial
\citep[see][]{Coe+2010,Sereno2013glszx}, and it is still unclear if and
to what degree this cluster is over-concentrated.


\begin{deluxetable}{lc}
\tablecolumns{2}
\tablecaption{
 \label{tab:cluster}
Properties of the galaxy cluster A1689
} 
\tablewidth{0pt}  
\tablehead{ 
 \multicolumn{1}{c}{Parameter} &
 \multicolumn{1}{c}{Value} 
} 
\startdata
ID .............................................. &  A1689 \\
Optical center position (J2000.0) & \\
\ \ \ \ R.A. ...................................... &  13:11:29.52\\
\ \ \ \ Decl. ..................................... & -01:20:27.59\\
X-ray center position (J2000.0) & \\
\ \ \ \ R.A. ...................................... &  13:11.29.50\\
\ \ \ \ Decl. ..................................... & -01:20:29.92\\
SZE center position (J2000.0) & \\
\ \ \ \ R.A. ...................................... &  13:11.29.57\\
\ \ \ \ Decl. ..................................... & -01:20:29.87\\
Redshift .................................... & $0.183$\\
X-ray temperature  (keV) ..........& $10.4$\\
Einstein radius ($\arcsec$) ....................& $47.0\pm 1.2$ at $z_s=2$
\enddata
\tablecomments{
The optical cluster center is defined as the center of the BCG 
from Ref.~[2].
Units of right ascension are hours, minutes, and
seconds, and units of declination are degrees, arcminutes, and
arcseconds. 
The X-ray properties were taken from Ref.~[3].
The X-ray center is defined as the X-ray emission centroid derived from 
{\it XMM-Newton} observations.
See also Ref.~[1].
The SZE center is determined from the joint analysis of interferometric
 BIMA/OVRO/SZA observations described in Section \ref{subsec:szx}.
The BCG is located within $2.3\arcsec$ 
($\simeq 5$\,kpc\,$h^{-1}$) of the X-ray center.
The X-ray and SZE centroid positions agree to within $1\arcsec$.
The Einstein radius is constrained by detailed strong
 lens modeling by Ref.~[4].
}
\tablerefs{ 
 [1] \cite{Andersson+Madjeski2004};
 [2] \cite{2007ApJ...668..643L};
 [3] \cite{Kawaharada+2010};
 [4] \cite{Coe+2010}.
 }
\end{deluxetable}

The paper is organized as follows. 
After summarizing the basic theory of cluster weak lensing, we
present in Section \ref{sec:method} the formalism that we use for
our weak-lensing analysis.
In Section \ref{sec:subaru}, we describe our Subaru observations and
data processing.
In Section \ref{sec:wlana}, we present our Subaru weak-lensing analysis.
Section \ref{sec:slana} presents our {\em HST} strong-lensing analysis.
In Section \ref{sec:3dmodel} we outline the triaxial modeling and
describe the statistical framework for the 3D cluster analysis.
In Section \ref{sec:results} we present the multi-probe analysis 
of lensing and X-ray/SZE data.
In Section \ref{sec:discussion} we discuss the results and their
implications for the intrinsic properties of A1689. 
Finally, a summary of our work is given in Section \ref{sec:summary}.
 
Throughout this paper, we use the AB magnitude system and adopt a
concordance $\Lambda$CDM cosmology with $\Om=0.3$, $\OL=0.7$, and 
$h\equiv 0.7 h_{70}=0.7$ 
where $H_0=h\times 100$\,km\,s$^{-1}$\,Mpc$^{-1}$. 
In this cosmology, $1\arcmin$ corresponds to 
129\,kpc\,$h^{-1}\simeq 185$\,kpc\,$h_{70}^{-1}$ for this cluster. 
The reference sky position is the center of  the brightest cluster
galaxy (BCG): 
$\mathrm{R.A.(J2000.0)}=13:11:29.52$,
$\mathrm{Decl.(J2000.0)}=-01:20:27.59$ (Table \ref{tab:cluster}).
We use the standard notation $r_\Delta$ to denote the spherical
overdensity radius
within which the mean interior density is $\Delta$
times the critical density $\rho_\mathrm{c}$
of the universe at the cluster redshift. For its ellipsoidal
counterpart $R_\Delta$, see Section \ref{subsec:matter}.
All quoted errors are 68.3\% ($1\sigma$) confidence
limits (CL) unless otherwise stated.

\section{Weak-lensing Methodology}
\label{sec:method}


\subsection{Weak Lensing Basics}
\label{subsec:basics}

In the cluster regime, the lensing convergence,
$\kappa=\Sigma/\Sigma_\mathrm{c}$,
is the projected mass density $\Sigma(\btheta)$
in units of the critical surface density for lensing,
$\Sigma_\mathrm{c}=(c^2 D_\mathrm{s})/(4\pi GD_\mathrm{l}D_\mathrm{ls})\equiv c^2/(4\pi GD_\mathrm{l}\beta)$
with $D_\mathrm{l}$, $D_\mathrm{s}$, and $D_\mathrm{ls}$ 
the lens, source, and lens-source angular diameter distances,
respectively;
$\beta(z) = D_\mathrm{ls}(z)/D_\mathrm{s}(z)$ represents the geometric lensing
strength for a source at redshift $z$, 
where $\beta(z)=0$ for $z\le z_\mathrm{l}$.

The gravitational shear $\gamma=\gamma_1+i\gamma_2$ 
can be directly observed from ellipticities of background galaxies in the weak
regime, $\kappa\ll 1$.
The shear and convergence are related by
\begin{equation}
\label{eq:kappa2gamma}
\gamma(\btheta) =
\int\!d^2\theta'\,D(\btheta-\btheta')\kappa(\btheta')
\end{equation}
with 
$D(\btheta)=(\theta_2^2-\theta_1^2-2i\theta_1\theta_2)/(\pi|\btheta|^4)$
\citep{1993ApJ...404..441K}.
The observable quantity for quadrupole weak lensing
in general is not $\gamma$ but the complex reduced shear,
\begin{equation}
\label{eq:redshear}
g(\btheta)=\frac{\gamma(\btheta)}{1-\kappa(\btheta)}.
\end{equation}
The $g$ field is invariant under 
$\kappa(\btheta)\to \lambda \kappa(\btheta) + 1-\lambda$
and 
$\gamma(\btheta)\to \lambda \gamma(\btheta)$
with an arbitrary constant $\lambda\ne 0$, known as the mass-sheet 
degeneracy  \citep{Schneider+Seitz1995}.
This degeneracy can be broken, for example,
by measuring the magnification $\mu(\btheta)$ in the subcritical regime,
\begin{equation}
\label{eq:mu}
\mu(\btheta) = \frac{1}{[1-\kappa(\btheta)]^2-|\gamma(\btheta)|^2} \equiv \frac{1}{\Delta(\btheta)},
\end{equation}
which transforms as $\mu(\btheta)\to \lambda^2\mu(\btheta)$.

Let us consider a population of
source galaxies described by their redshift distribution function,  
$\overline{N}(z)$. 
In general, we apply different size, magnitude, and color cuts in background selection for
measuring shear and magnification,
which results in different $\overline{N}(z)$.
In contrast to the former effect, the
latter does not require source galaxies to be spatially resolved, but
it requires a stringent flux limit against incompleteness effects. 
The mean lensing depth for a given population ($X=g,\mu$) is
\begin{equation}
\label{eq:depth}
\langle\beta\rangle_X =\left[
\int_0^\infty\!dz\, w(z)\overline{N}_X(z) \beta(z)\right]
\left[
\int_0^\infty\!dz\, w(z)\overline{N}_X(z)
\right]^{-1},
\end{equation} 
where $w(z)$ is a weight factor (see Section \ref{subsec:back}).

We introduce the relative lensing strength of a given source population
relative to a fiducial source in the far background as  
$\langle W\rangle_X   = \langle\beta\rangle_X  / \beta_\infty$ 
\citep{2001PhR...340..291B}
with $\beta_\infty\equiv \beta(z\to \infty; z_\mathrm{l})$.
The associated critical density is
$\Sigma_{\mathrm{c},\infty}(z_\mathrm{l})=c^2/(4\pi G D_\mathrm{l}\beta_{\infty})$.
Hereafter, we use the far-background fields
$\kappa_\infty(\btheta)$ and 
$\gamma_\infty(\btheta)$
to describe the projected cluster mass distribution.

\subsection{Discretized Mass Distribution}
\label{subsec:massmodel}

We discretize the convergence field 
$\kappa_\infty(\btheta)=\Sigma_{\mathrm{c},\infty}^{-1}\Sigma(\btheta)$ into
a regular grid of pixels and
approximate $\kappa_\infty(\btheta)$ by a linear combination of basis
functions $B(\btheta-\btheta')$
as
\begin{equation}
\label{eq:basis}
\kappa_\infty(\btheta) 
=\Sigma_{\mathrm{c},\infty}^{-1}
\sum_{m=1}^{N_\mathrm{pix}}
 B(\btheta-\btheta_m)\, \Sigma_m,
\end{equation}
where our model (signal) 
$\bs = \left\{\Sigma_m\right\}_{m=1}^{N_\mathrm{pix}}$ is a vector of 
parameters containing mass coefficients.
To avoid the loss of information due to oversmoothing,
we take the basis function to be the Dirac delta function
$B(\btheta-\btheta_m)=(\Delta\theta)^2\delta^2_\mathrm{D}(\btheta-\btheta_m)$
with $\Delta\theta$ a constant spacing,
so that $\bs$ represents the cell-averaged projected mass density.
The $\gamma_\infty(\btheta)$ field can be expressed as 
\begin{equation}
\label{eq:shear2m}
\gamma_\infty(\btheta)= \Sigma_{\mathrm{c},\infty}^{-1}\sum_{m=1}^{\Npix} 
{\cal D}(\btheta-\btheta_m)\Sigma_m
\end{equation}
with ${\cal D}\equiv D\otimes B$ an effective kernel
(Equation (\ref{eq:kappa2gamma})).
Hence, both $\kappa_\infty$ and $\gamma_\infty$ can be written as linear
combinations of  $\bs$. 

Because of the choice of the basis function,
an unbiased extraction of mass coefficients
$\{\Sigma_m\}_{m=1}^{\Npix}$ (or certain linear combinations of $\Sigma_m$)
can be done by performing a spatial integral of Equation
(\ref{eq:basis}) over a certain area. In practical applications, such
operations include 
smoothing (Figure \ref{fig:subaru}),
azimuthal averaging for a mass profile reconstruction (Section
\ref{subsec:wlslcomp}), 
and profile fitting with smooth functions (Section \ref{sec:results}).

\subsection{Weak-lensing Observables}

\subsubsection{Reduced Shear}
\label{subsubsec:shear}

The quadrupole image distortion due to
 lensing is described by the reduced shear, $g=g_1+i g_2$.
We calculate the weighted average $g_m\equiv g(\btheta_m)$
of individual shear estimates on a regular cartesian grid
 ($m=1,2,..., \Npix$) as
\begin{equation}
\label{eq:bin_shear} 
g_m
=
\left[
\displaystyle\sum_k
S(\btheta_{(k)},\btheta_m)
w_{(k)}g_{(k)}
\right]
\left[
\displaystyle\sum_{k} 
S(\btheta_{(k)},\btheta_m)w_{(k)}
\right]^{-1} 
\end{equation}
where $S(\btheta_{(k)},\btheta_m)$ is a spatial window function,
$g_{(k)}$ is an estimate of $g(\btheta)$ for the $k$th
object at $\btheta_{(k)}$, 
and 
$w_{(k)}$ is its statistical weight 
given by 
$w_{(k)} = 1/(\sigma^2_{g(k)}+\alpha^2_g)$
with $\sigma^2_{g(k)}$ the error variance of $g_{(k)}$
 and $\alpha_g^2$ the softening constant variance.
We choose $\alpha_g=0.4$, a typical value of
the mean rms $\sqrt{\overline{\sigma_g^2}}$ found in Subaru observations
\citep[e.g.,][]{Umetsu+2009}.

The source-averaged theoretical expectation for the estimator
(\ref{eq:bin_shear}) is approximated by (see Appendix \ref{appendix:nonlin_g})
\begin{equation}
\label{eq:g_ave}
\hat{g}(\btheta_m) = \frac{\langle W\rangle_g
 \gamma_\infty(\btheta_m)}{1-f_{W,g} \langle W \rangle_g \kappa_\infty(\btheta_m)},
\end{equation}
where $\langle W\rangle_g$ is the source-averaged relative lensing
strength (Section \ref{subsec:basics}), and
$f_{W,g}=\langle W^2\rangle_g/\langle W\rangle_g^2$ is a dimensionless
quantity of the order unity.
The variance  $\sigma_{g,m}^2\equiv \sigma_{g}^2(\btheta_m)$ 
for $g_m=g_{1,m}+ig_{2,m}$ is expressed as
\begin{equation}
\label{eq:bin_shearvar}
\sigma^2_{g,m}=
\left[
\sum_k S^2(\btheta_{(k)},\btheta_m)w_{(k)}^2 \sigma^2_{g(k)}
\right] 
\left[
 {\sum_k S^2(\btheta_{(k)},\btheta_m)w_{(k)}}
\right]^{-2}.
\end{equation}
In this work, we adopt the top-hat window
of radius $\theta_\mathrm{f}$ \citep{Merten+2009},
$S(\btheta,\btheta')=H(\theta_\mathrm{f}-|\btheta-\btheta'|)$, with
$H(x)$ the Heaviside function defined such that
$H(x)=1$ if $x\ge 0$ and $H(x)=0$ otherwise.
The covariance matrix for $g_m$ is 
\begin{equation}
  \mathrm{Cov}(g_{\alpha,m},g_{\beta,n}) 
 \equiv 
  \delta_{\alpha\beta}\left(C_g\right)_{mn}
 = 
  \frac{\delta_{\alpha\beta}}{2} \sigma_{g,m} \sigma_{g,n}
  \xi_{H}(|\btheta_m-\btheta_n|),
\end{equation}
where $\xi_H(x; \theta_\mathrm{f})$ is the autocorrelation of a pillbox of 
radius 
$\theta_\mathrm{f}$ \citep{1999ApJ...514...12W,Park+2003}, given by
\begin{equation}
 \xi_{H}(x)=\frac{2}{\pi}\left[\cos^{-1}\left(\frac{x}{2\theta_{\rm
  f}}\right)-\left(\frac{x}{2\theta_{\rm
  f}}\right)\sqrt{1-\left(\frac{x}{2\theta_\mathrm{f}}\right)^2}\right]
\end{equation}
for $|x|\le 2\theta_\mathrm{f}$
and $\xi_{H}(x)=0$ for $|x|>2\theta_\mathrm{f}$.


\subsubsection{Magnification Bias}
\label{subsubsec:magbias}

Deep multi-band photometry allows us to explore the faint end of the
luminosity function of red quiescent galaxies at $z\sim 1$ 
\citep{Ilbert2010},
for which the effect of magnification bias is dominated by the geometric
area distortion and thus not sensitive to the
exact form of the source luminosity function.
In this work, we perform magnification measurements 
using a flux-limited sample of red background galaxies.

If the magnitude shift $\delta m=2.5\log_{10}\mu$ due to 
magnification is small compared to that on which 
the logarithmic slope of the luminosity function
varies, 
their number counts can be locally approximated by a power law at 
the limiting flux \citep{1995ApJ...438...49B}.
The expectation value for the source counts $N_\mu(\btheta_m)$ on a grid of
equal-area cells ($m=1,2,...$) is modified by lensing magnification
as (see Appendix \ref{appendix:nonlin_mu})
\begin{equation}
 \label{eq:magbias}
 \begin{aligned}
 & E[N_\mu(\btheta_m)] = 
  \overline{N}_\mu \Delta^{1-\alpha}(\btheta_m),\\
 &\Delta(\btheta) =  \left[1-\langle W\rangle_\mu\kappa_\infty(\btheta)\right]^2-\langle
  W\rangle_\mu^2|\gamma_\infty(\btheta)|^2,
  \end{aligned}
\end{equation}
where $\overline{N}_\mu$ is the unlensed 
mean source counts per cell, $\alpha$ is the
unlensed count slope evaluated at the flux limit $F$,
$\alpha=-d\log\overline{N}_\mu(>F)/d\log F$,\footnote{In the
weak-lensing literature, 
$s\equiv d\log_{10}N(<m)/dm=0.4\alpha$ in terms of
the limiting magnitude $m $ is often
used instead of $\alpha$ \citep[e.g.,][]{Umetsu+2011,Umetsu2014clash,Medezinski+2013}.} 
and $\langle W\rangle_\mu$ is the source-averaged 
relative lensing strength (Section \ref{subsec:basics}).
%

The net magnification effect on the source counts vanishes when $\alpha=1$.
In the regime where $\alpha\ll 1$, 
the bias is dominated by the expansion of
the sky area, producing a net count depletion.
For a population with $\alpha>1$,
the bias is positive, and
a net density enhancement results \citep[e.g.,][]{Hildebrandt+2011,Ford+2012,Ford2014cfhtlens}.
The faint blue population
lying at $z\sim 2$ \citep[e.g.,][]{Lilly+2007,Medezinski+2010,Medezinski+2013}
tends to have a steep intrinsic slope close to the lensing-invariant
one, $\alpha=1$.


The covariance matrix of $N_\mu(\btheta)$ includes both sample
covariance and Poisson variance \citep{Hu+Kravtsov2003}:
\begin{equation}
 \mathrm{Cov}[N_\mu(\btheta_m),N_\mu(\btheta_n)] 
\equiv
\left(C_N\right)_{mn}
=
(\overline{N}_\mu)^2
\omega_{mn} +
\delta_{mn} 
N_\mu(\btheta_m),
\end{equation}
where $\omega_{mn}$ is the cell-averaged angular 
correlation function
\begin{equation}
\omega_{mn}=
 \frac{1}{\Omega_\mathrm{cell}^2}
 \int\!
 \!d^2\theta\,d^2\theta'\,
 S_m(\btheta)
 S_n(\btheta')
 \omega(\btheta-\btheta')
\end{equation}
with $\omega(\btheta)$ the angular two-point correlation function of the
source galaxies,
$S_m(\btheta)$ the boxcar window function of the $m$th cell,
and 
$\Omega_\mathrm{cell}=\int\!d^2\theta\,S_m(\btheta)$.
For deep lensing observations,
the angular correlation length of background galaxies
can be small \citep[e.g.,][]{Connolly+1998}
compared to the typical resolution $\sim 1\arcmin$
of reconstructed mass maps.
Therefore, the correlation between different cells can be
generally ignored, 
whereas the unresolved correlation on small angular scales accounts for increase of the
variance of $N_\mu(\btheta)$ \citep{vanWaerbeke2000}. 
We thus approximate $C_N$ by
\begin{equation}
\label{eq:covN}
\left(C_N\right)_{mn} \approx
\left[
\langle \delta N_\mu^2(\btheta_m)\rangle +  
N_\mu(\btheta_m)
\right]\delta_{mn},
\end{equation}
with $\langle \delta N_\mu^2(\btheta_m)\rangle$ the variance of the
$m$th counts.

To enhance the signal-to-noise ratio,
we azimuthally average $N_\mu(\btheta)$ in contiguous, concentric annuli and 
calculate the surface number density 
$\{n_{\mu,i}\}_{i=1}^{\Nbin}$
of background galaxies 
as a function of clustercentric radius:
\begin{equation}
\label{eq:nb}
  n_{\mu,i} =
  \frac{\eta_i}{\Omega_\mathrm{cell}}
  \sum_{m} {\cal P}_{im}N_\mu(\btheta_m)
\end{equation}
with ${\cal P}_{im}=(\sum_m A_{mi})^{-1}A_{mi}$ the radial projection matrix normalized as
$\sum_m {\cal P}_{im}=1$.
Here $A_{mi}$ represents the fraction of the area of
the $m$th cell lying within the $i$th annular bin  ($0\le A_{mi} \le 1$), 
and $\eta_i (\ge 1)$ is the mask correction factor for the $i$th annular bin,
$\eta_i = \left[ \sum_{m} (1-f_m)A_{mi}\right]^{-1} \sum_{m} A_{mi}$,
with $f_m$ the fraction of the mask area
in the $m$th cell, 
due to bad pixels, saturated objects, foreground and cluster member galaxies
\citep[see Section 3.2 of][]{Umetsu2014clash}.  

The theoretical expectation for the estimator (\ref{eq:nb}) is 
\begin{equation}
 \label{eq:nb_th}
\hat{n}_{\mu,i} =
\overline{n}_\mu
\sum_m {\cal P}_{im} \Delta^{1-\alpha}(\btheta_m)
\end{equation}
with $\overline{n}_\mu =\overline{N}_\mu/\Omega_\mathrm{cell}$.
The bin-to-bin covariance matrix for the estimator (\ref{eq:nb}) is
obtained as
\begin{equation}
\label{eq:Covn}
\mathrm{Cov}(n_{\mu,i},n_{\mu,j})
\equiv \left(C_{\mu}\right)_{ij}
=
\frac{\eta_i\eta_{j}}{\Omega_\mathrm{cell}^2}
\sum_{m,n}
{\cal P}_{im}{\cal P}_{jn} 
\left(C_N\right)_{mn}.
\end{equation}
Note that since $C_N$ is diagonal, $C_\mu$ is also diagonal:
\begin{equation}
\label{eq:Covmu}
(C_\mu)_{ij} \equiv \sigma_{\mu,i}^2\delta_{ij}.
\end{equation}

\subsection{Mass Reconstruction}
\label{subsec:MLE}

Given a model $\bm$ and observed (fixed) data $\bd$, the
posterior probability  $P(\bm|\bd)$ is proportional to the product of
the likelihood ${\cal L}(\bm)\equiv P(\bd|\bm)$ and the prior probability
$P(\bm)$.
In our 2D inversion problem, $\bm$ is a vector containing the
signal parameters $\bs$ (Section \ref{subsec:massmodel})
and calibration parameters $\bc$ (Section \ref{subsubsec:calib}), 
$\bm\equiv (\bs,\bc)$.

The total likelihood function ${\cal L}$ for combined weak-lensing data
$\bd$ is given as a product of the two separate likelihoods, 
${\cal L}={\cal L}_g {\cal L}_\mu$, where ${\cal L}_g$ and 
${\cal L}_\mu$ are the likelihood functions for shear and magnification,
respectively.  
We assume that the errors on the data follow a Gaussian distribution, 
so that ${\cal L}\propto \exp(-\chi^2/2)$, with $\chi^2$ the standard
misfit statistic.

\subsubsection{Shear Log-likelihood Function}
\label{subsubsec:lg}

The log-likelihood function 
$l_g\equiv -\ln{\cal L}_g$ for 2D shear data
can be written in the general form (ignoring constant terms) as 
\citep{Oguri2010LoCuSS,Umetsu+2012} 
\begin{equation}
l_g = \frac{1}{2}
 \sum_{m,n=1}^{\Npix}
 \sum_{\alpha=1}^{2}
[g_{\alpha,m}-\hat{g}_{\alpha,m}(\bm)]
\left({\cal W}_g\right)_{mn}%
 [g_{\alpha,n}-\hat{g}_{\alpha,n}(\bm)]
\end{equation}
where  $\hat{g}_{\alpha,m}(\bm)$
is the theoretical expectation for $g_{\alpha,m}=g_\alpha(\btheta_m)$,
and $({\cal W}_g)_{mn}$ is the shear weight 
matrix,
\begin{equation}
\left({\cal W}_g\right)_{mn} =
M_m M_n \left(C_g^{-1}\right)_{mn},
\end{equation}
with $(C_g^{-1})_{mn}$ the inverse covariance matrix for the 2D
shear data and $M_m$ a mask weight, defined such that $M_m=0$ if the 
$m$th cell is masked out and $M_m=1$ otherwise.

\subsubsection{Magnification Log-likelihood Function}

Similarly, the log-likelihood function for magnification-bias data
$l_{\mu}\equiv -\ln{\cal L}_\mu$ can be written as
\begin{equation}
l_\mu= \frac{1}{2}\sum_{i=1}^{\Nbin}
[n_{\mu,i}-\hat{n}_{\mu,i}(\bm)]
\left({\cal W}_\mu\right)_{ij}
[n_{\mu,j}-\hat{n}_{\mu,j}(\bm)],
\end{equation}
where $\hat{n}_{\mu,i}(\bm)$ is the theoretical prediction for the observed counts
$n_{\mu,i}$ (see Equations (\ref{eq:nb}) and (\ref{eq:nb_th})), and 
$({\cal W}_\mu)_{ij}$ is the magnification weight matrix,
\begin{equation}
\left({\cal W}_\mu\right)_{ij} = \left(C_\mu^{-1}\right)_{ij} =
\frac{\delta_{ij}}{\sigma_{\mu,i}^2}
\end{equation}
(Equations (\ref{eq:Covn}) and (\ref{eq:Covmu})).
We use Monte Carlo integration to calculate the radial projection matrix
${\cal P}_{im}$ (Equation \ref{eq:nb})
of size $\Nbin \times \Npix$, which is needed to predict
$\{\hat{n}_{\mu,i}(\bm)\}_{i=1}^{\Nbin}$
for a given $\bm=(\bs,\bc)$.

The $l_\mu$ function imposes a set of azimuthally integrated constraints
on the underlying projected mass distribution. Since
magnification is locally related to $\kappa$, 
this will essentially provide the otherwise unconstrained normalization of
$\Sigma(\btheta)$ over a set of concentric rings where count
measurements are available.
We note that no assumption is made of azimuthal symmetry or isotropy of the
2D mass distribution $\Sigma(\btheta)$.

\subsubsection{Calibration Parameters}
\label{subsubsec:calib}

We account for the calibration uncertainty in the
observational nuisance parameters,
\begin{equation}
\label{eq:calib}
\bc=(\langle W\rangle_g, f_{W,g}, \langle W\rangle_\mu,
\overline{n}_\mu,\alpha).
\end{equation}
To do this, we include in our analysis 
Gaussian priors on $\bc$
given by means of quadratic penalty terms
with mean values and errors directly estimated from data.

\subsubsection{Best-fit Solution and Covariance Matrix}
\label{subsubsec:cmat}

The log posterior $F(\bm) = -\ln{P(\bm|\bd)}$ is 
expressed as a linear sum of the log-likelihood and prior terms.
The maximum-likelihood (ML) solution, $\hat{\bm}$, is obtained by minimizing
$F(\bm)$ with respect to $\bm$.
In our implementation we use the conjugate-gradient method 
\citep{1992nrfa.book.....P} to find the solution.
Here we employ an 
analytic expression for the gradient function $\bnabla F(\bm)$ obtained
in the nonlinear, subcritical regime.
To be able to quantify the errors on the reconstruction, we
evaluate the Fisher matrix at $\bm=\hat{\bm}$, as
\begin{equation}
{\cal F}_{pp'} = 
\left\langle \frac{\partial^2 F(\bm)}{\partial m_p \partial m_{p'}}
\right\rangle\Big|_{\bm=\hat{\bm}}
\end{equation}
where the angular brackets represent an ensemble average, and the
indices $(p,p')$ run over all model parameters $\bm=(\bs,\bc)$.
We estimate the error covariance matrix as
\begin{equation}
\mathrm{Cov}(m_p,m_{p'})\equiv C_{pp'} = \left({\cal F}^{-1}\right)_{pp'}.
\end{equation}

\section{Subaru Observations}
\label{sec:subaru}

Here we present a description of our data analysis
of A1689 based on deep Subaru $BVR_\mathrm{C}i'z'$ images.
In this study, we analyze the data using
the same methods and procedures as in \citet{Umetsu2014clash}, who
performed a weak-lensing analysis of 20 high-mass clusters
selected from the CLASH survey \citep{Postman+2012CLASH}. For details of
our reduction and analysis pipelines,
we refer to Section 4 of \citet{Umetsu2014clash}.

\subsection{Data and Photometry}
\label{subsec:data}


\begin{deluxetable}{cccc}
\tablecolumns{4}
\tablecaption{
 \label{tab:subaru}
Subaru/Suprime-Cam data
} 
\tablewidth{0pt} 
\tablehead{ 
 \multicolumn{1}{c}{Filter} &
 \multicolumn{1}{c}{Exposure time\tablenotemark{a}} &
 \multicolumn{1}{c}{Seeing\tablenotemark{b}} &
 \multicolumn{1}{c}{$m_\mathrm{lim}$\tablenotemark{c}}  
\\
 \colhead{} &
 \multicolumn{1}{c}{(ks)} &
 \multicolumn{1}{c}{(arcsec)} &
 \multicolumn{1}{c}{(AB mag)} 
} 
\startdata
 $B$   & 2.40 & 0.91   & 27.1\\
 $V$   & 4.08 & 0.84  & 27.0\\
 $R_\mathrm{C}$ & 6.42 & 0.70 (0.60) & 27.0 \\ 
 $i'$  & 4.08 & 0.84 & 26.4 \\
 $z'$  & 8.02 & 0.81  & 26.2 
\enddata
\tablenotetext{a}{Total exposure time.} 
\tablenotetext{b}{Seeing FWHM in the full stack of images.}
\tablenotetext{c}{Limiting magnitude for a $3\sigma$ detection within a
 $2\arcsec$ aperture.}
\tablecomments{The $R_\mathrm{C}$ band is used as the filter to measure
 object shapes for the weak-lensing analysis,
where we separately stack data from different epochs.
The $R_\mathrm{C}$-band seeing in parentheses is the average of values
 derived from separate stacks.}  
\end{deluxetable}

\begin{figure*}[!htb] 
 \begin{center}
  \includegraphics[width=\textwidth,angle=0,clip]{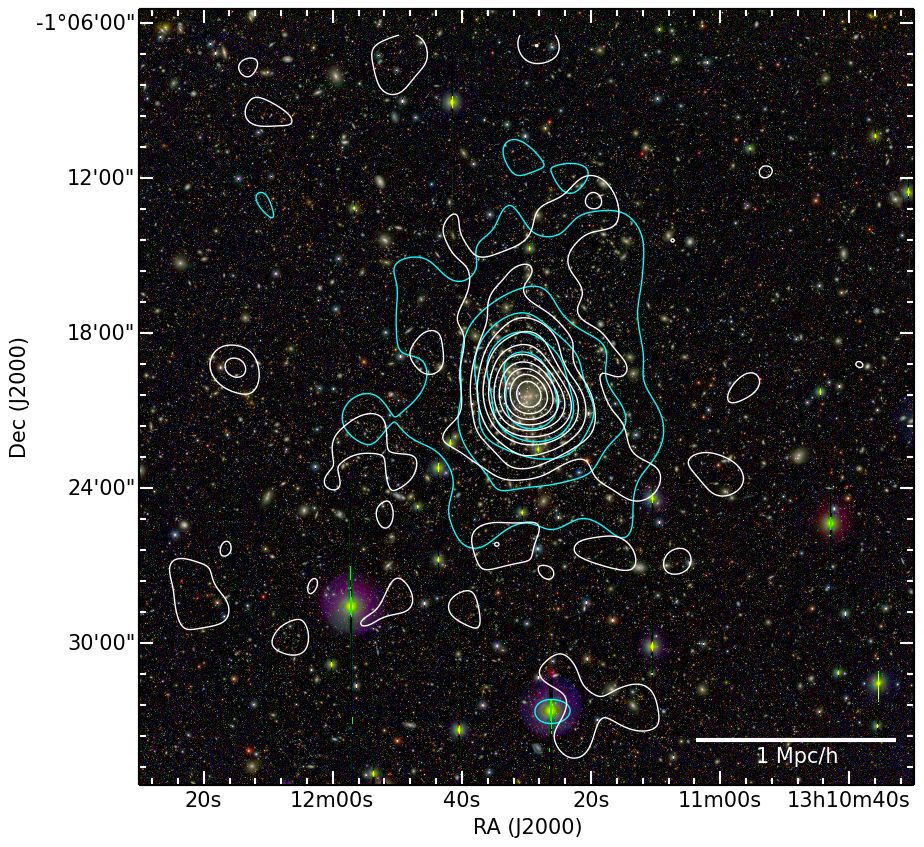}
 \end{center}
\caption{
\label{fig:subaru}
Subaru $BVR_\mathrm{C}i'z'$ composite color image centered on the galaxy
 cluster A1689 ($z=0.183$), overlaid with mass contours from our joint
 shear-and-magnification weak-lensing analysis of  Subaru data.
The image is $30\arcmin\times 30\arcmin$ in size.
The mass map is smoothed with a Gaussian of $\mathrm{FWHM}=1.5\arcmin$.
The horizontal bar represents 1\,\Mpch at the cluster redshift.
The lowest contour level and the contour interval are
 $\Delta\kappa=0.06$. 
The cyan contours show the smoothed projected distribution of cluster
 red-sequence galaxies.
} 
North is up and east is to the left.
\end{figure*} 

We analyze deep $BVR_\mathrm{C}i'z'$ images of A1689 observed
with the wide-field camera Suprime-Cam 
\citep[$34\arcmin\times 27\arcmin$;][]{2002PASJ...54..833M}
at the prime focus of the 8.3\,m Subaru Telescope.
We combine both existing archival data taken from
SMOKA\footnote{\href{http://smoka.nao.ac.jp}{http://smoka.nao.ac.jp}}
with observations acquired by the team on the nights of 
2010 March 17--18 (S10A-019).
The observation details of A1689 are summarized in Table \ref{tab:subaru}. 
Figure \ref{fig:subaru} shows a $BVR_\mathrm{C}i'z'$
composite color image of the cluster field, produced using
the publicly available {\sc Trilogy} software \citep{Coe+2012A2261}.
The image is overlaid by mass contours determined from our weak-lensing
analysis (see Section \ref{subsec:wl2d}).

Our imaging reduction pipeline derives from
\cite{Nonino+2009} and has been optimized separately for accurate
photometry and shape measurements.
For multi-band photometry,
standard reduction steps include bias
subtraction, super-flat-field correction, and point-spread-function
(PSF, hereafter) matching between exposures in the same band.   
An accurate astrometric solution is derived with the {\sc SCAMP} software 
\citep{SCAMP},
using the 
the Sloan Digital Sky Survey 
 \citep[SDSS,][]{SDSSDR6photometry} as an external reference catalog.\footnote{This
 research has made use of the VizieR catalog access tool, CDS,
 Strasbourg, France.} 
The {\sc Swarp} software \citep{Bertin+2002Swarp} is used to stack
 individual exposures on a common World Coordinate System (WCS) grid
 with pixel scale of $0.2\arcsec$.   


The photometric zero-points for the co-added images were derived using
{\em HST}/ACS magnitudes of cluster elliptical-type galaxies.
These zero points were further refined by fitting SED (spectral energy
distribution) templates with the BPZ code
\citep[Bayesian photometric redshift
estimation;][]{Benitez2000,Benitez+2004} 
to 1445 galaxies having spectroscopic redshifts.\footnote{The data used here
are part of an extensive multi-object spectroscopy survey carried
out with the VIMOS spectrograph on the VLT \citep{Czoske2004}. For
details, see \citet{Lemze+2009}.}
This leads to a final photometric accuracy of $\sim 0.01$\,mag in all  
passbands.
The magnitudes were corrected for Galactic extinction according to
\citet{1998ApJ...500..525S}.
The multi-band photometry was measured
using SExtractor \citep{SExtractor} in dual-image mode on
PSF-matched images created by ColorPro
\citep{colorpro}. 

\subsection{Shape Measurement}
\label{subsec:shape}

We use our shear analysis pipeline based on the {\sc IMCAT}
package \citep[][KSB]{1995ApJ...449..460K}   
incorporating improvements developed by \citet{Umetsu+2010CL0024}.
On the basis of simulated Subaru/Suprime-Cam images 
\citep{Oguri+2012SGAS,2007MNRAS.376...13M},
\citet{Umetsu+2010CL0024} showed 
that the lensing signal can be recovered with $|m|\sim 5\%$ of the
multiplicative shear calibration bias 
\citep[as defined by][]{2006MNRAS.368.1323H,2007MNRAS.376...13M},
and $c\sim 10^{-3}$ of
the residual shear offset, which is about one order of magnitude
smaller than the typical shear signal in cluster outskirts. 
Accordingly, we include for each galaxy a shear
calibration factor of $1/0.95$ ($g\to g/0.95$)
to account for residual calibration.

In this work, we perform weak-lensing shape analysis 
using the same procedures adopted in the CLASH weak-lensing analysis of 
\citet{Umetsu2014clash}. 
Here, we only highlight key aspects of our analysis pipeline:
\begin{itemize}
\item {\em Object detection}. Objects are detected using the {\sc IMCAT}
      peak finder, {\em hfindpeaks}, using a set of Gaussian kernels of
      varying sizes. This algorithm produces object
      parameters such as the peak position, the {\em best-matched}
      Gaussian scale length, $r_g$, 
      and an estimate of the significance of the peak detection, $\nu$.

\item {\em Crowding effects}. 
      Objects having any detectable
      neighbors within $3r_g$ are identified. All such close pairs of
      objects are rejected to avoid possible shape measurement errors
      due to crowding. The detection threshold is set to
      $\nu=7$ for close-pair identification. After this close-pair
      rejection, objects with low detection significance $\nu<10$
      are excluded from our analysis. 

\item {\em Shear calibration}. We calibrate KSB's isotropic correction
      factor $P_g$ as a function of object size ($r_g$) and magnitude,  
      using galaxies detected with high significance $\nu>30$ \citep{Umetsu+2010CL0024}.
      This is to minimize the inherent shear calibration bias in the
      presence of noise. We correct for the isotropic smearing effect
      caused by seeing as well as by the window function used in the
      shape estimate as $g_\alpha=e_\alpha/P_g$ with $e_\alpha$ the
      anisotropy-corrected object ellipticity.
\end{itemize}

To measure the shapes of background galaxies,
we use the $R_\mathrm{C}$ band data, which have the best image quality
in our data sets (Table \ref{tab:subaru}).
Two separate co-added $R_\mathrm{C}$-band images are created, one from 2009
(observed by Matsuda et al.)
and another from 2010 (observed by Umetsu et al.).
We separately stack data obtained at different epochs.
We do not smear individual exposures before stacking, so as not to
degrade the weak-lensing signal.
After PSF anisotropy correction, the mean residual stellar ellipticity is
consistent with zero, and the rms residual stellar
ellipticity in each stack is 
$\sigma(\delta e^*_\alpha)\sim 2.5\times 10^{-3}$ per component. 
A shape catalog is created for each epoch separately.
These subcatalogs are then combined by properly
weighting and stacking the calibrated shear estimates
for galaxies in the overlapping region \citep[see Section 4.3 of][]{Umetsu2014clash}.

\subsection{Background Galaxy Selection}
\label{subsec:back}


\begin{deluxetable}{cccccc}
\tabletypesize{\footnotesize}
\tablecolumns{6} 
\tablecaption{
 \label{tab:gsample}
Background Galaxy Samples for Weak-lensing Shape Measurements
}  
\tablewidth{0pt}  
\tablehead{ 
 \multicolumn{1}{c}{Sample} &
 \multicolumn{1}{c}{$N_g$} &
 \multicolumn{1}{c}{$\overline{n}_g$\tablenotemark{a}} & 
 \multicolumn{1}{c}{$\overline{z}_\mathrm{eff}$\tablenotemark{b}} &
 \multicolumn{1}{c}{$\langle D_\mathrm{ls}/D_\mathrm{s}\rangle$} &
 \multicolumn{1}{c}{$f_W$} 
\\
 \colhead{} & 
 \colhead{} &
 \multicolumn{1}{c}{(arcmin$^{-2}$)} &
 \colhead{} &
 \colhead{} &
 \colhead{} 
} 
\startdata
 Red      & 12674  & 12.0 & 1.10 & $0.79\pm 0.04$ & 1.00 \\
 Blue     & 9238   & 8.7  & 1.62 & $0.84\pm 0.04$ & 1.01\\
 Blue+red & 21912  & 20.7 & 1.22 & $0.80\pm 0.04$ & 1.01
\enddata 
\tablenotetext{a}{Mean surface number density of source background galaxies.} 
\tablenotetext{b}{Effective source redshift corresponding to the mean
 lensing depth 
$\langle \beta\rangle =\langle D_\mathrm{ls}/D_\mathrm{s}\rangle$, defined as 
$\beta(\overline{z}_\mathrm{eff})=\langle \beta\rangle$.}
\end{deluxetable}


\begin{deluxetable*}{cccccccc}
\tabletypesize{\footnotesize}
\tablecolumns{8} 
\tablecaption{
 \label{tab:musample}
Background Galaxy Samples for Magnification-bias Measurements
}  
\tablewidth{0pt}  
\tablehead{ 
 \multicolumn{1}{c}{Sample} &
 \multicolumn{1}{c}{$z'_\mathrm{cut}$\tablenotemark{a}} &
 \multicolumn{1}{c}{$N_\mu$} &
 \multicolumn{1}{c}{$\overline{n}_\mu$\tablenotemark{b}} & 
 \multicolumn{1}{c}{$\alpha$\tablenotemark{c}} & 
 \multicolumn{1}{c}{$\langle z\rangle$\tablenotemark{d}} &
 \multicolumn{1}{c}{$\overline{z}_\mathrm{eff}$\tablenotemark{e}} &
 \multicolumn{1}{c}{$\langle D_\mathrm{ls}/D_\mathrm{s}\rangle$} 
 \\
 \colhead{} & 
 \multicolumn{1}{c}{(AB mag)} &
 \colhead{} &
 \multicolumn{1}{c}{(arcmin$^{-2}$)} &
 \colhead{} &
 \colhead{} 
} 
\startdata  
 Red      & $25.6$ & 26136  & $19.0\pm 0.5$ & $0.39\pm 0.08$ & 1.13& 1.05 & $0.73\pm 0.04$\\
 Blue     & $25.6$ & 12143  & $8.8\pm 0.3$  & $0.82\pm 0.12$ & 1.81& 1.39 & $0.82\pm 0.04$
\enddata 
\tablenotetext{a}{Fainter magnitude cut of the background
 sample. Apparent magnitude cuts are applied in the reddest CC-selection
 band available ($z'$) to avoid incompleteness near the detection limit.}
\tablenotetext{b}{Coverage- and mask-corrected normalization of unlensed
 background source counts.} 
\tablenotetext{c}{Logarithmic slope of the unlensed source
 counts $\alpha=2.5 \left[
 d\log_{10}\overline{N}_\mu(<z')/dz'\right]_{z'=z'_\mathrm{cut}}$.}
\tablenotetext{d}{Mean photometric redshift of the sample obtained with the BPZ code, defined similarly to Equation (\ref{eq:depth}).}
\tablenotetext{e}{Effective source redshift corresponding to the mean
 lensing depth 
$\langle \beta\rangle =\langle D_\mathrm{ls}/D_\mathrm{s}\rangle$, defined as 
$\beta(\overline{z}_\mathrm{eff})=\langle \beta\rangle$.}
\end{deluxetable*}

\begin{figure}[!htb] 
 \begin{center}
  \includegraphics[width=0.45\textwidth,angle=0,clip]{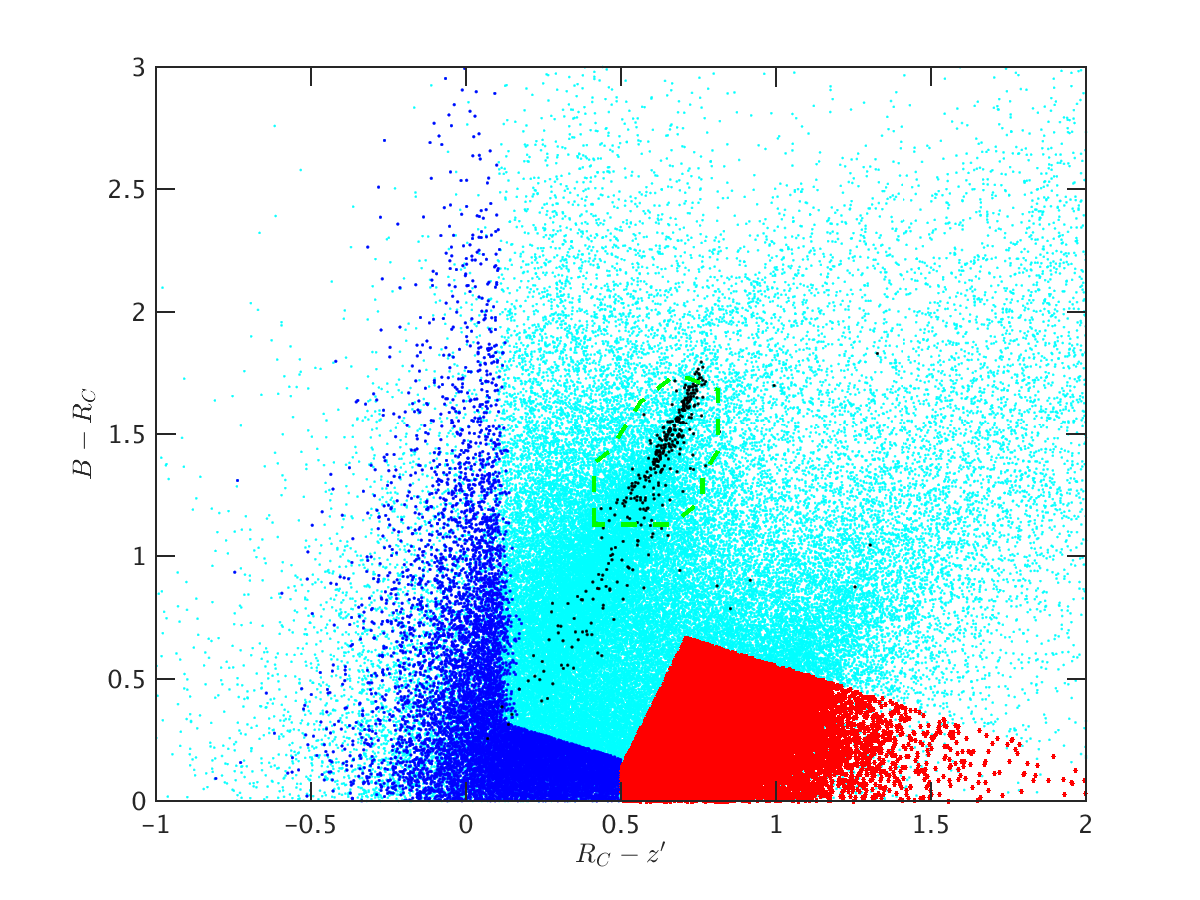}
 \end{center}
\caption{\label{fig:CCsel}
``Blue'' and ``red'' background galaxy samples selected for the weak-lensing analysis
(lower-left blue and lower-right red regions, respectively) on the
 basis of Subaru $BR_\mathrm{C}z'$ color-color-magnitude selection. All galaxies (cyan) are
shown in the diagram. At small clustercentric radius ($<4\arcmin$), an
 overdensity of cluster galaxies is identified as our ``green'' sample
 (green), comprising mostly the red sequence of cluster ellipticals and
 some blue trail of  later-type cluster members. 
The background samples are well isolated from the green region and satisfy
other criteria as discussed in Section \ref{subsec:back}.
The black dots represent a dynamically-selected spectroscopic
sample of 377 cluster galaxies found within a projected distance of
 $12\arcmin$ ($\sim r_\mathrm{200c}$) from the cluster center.
Our background selection successfully excludes all except 2
 spectroscopically confirmed cluster members (see Section \ref{subsec:back}).
} 
\end{figure} 

\begin{figure}[!htb] 
 \begin{center}
  \includegraphics[width=0.45\textwidth,angle=0,clip]{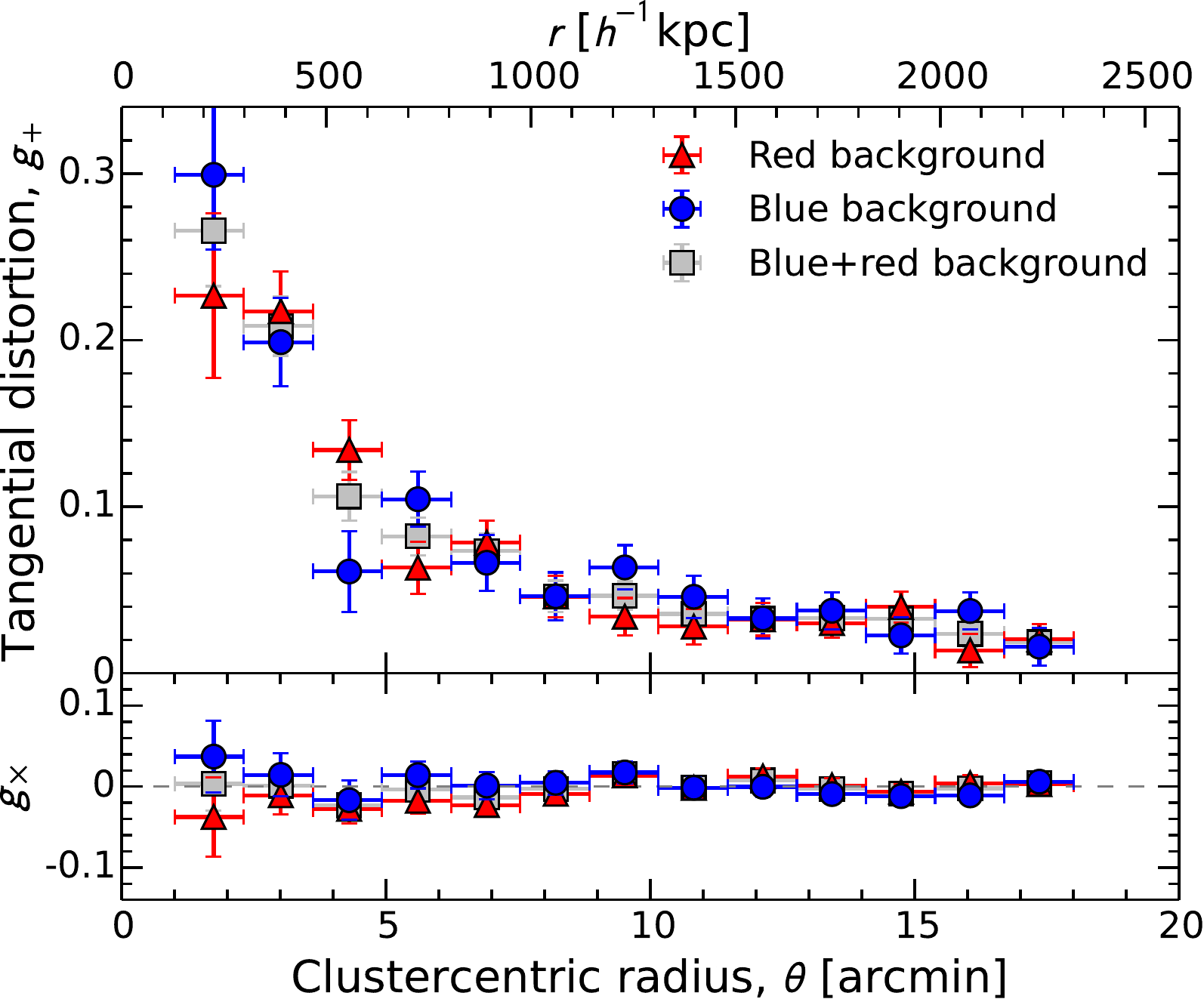}
 \end{center}
\caption{
\label{fig:RGB}
Azimuthally averaged radial profiles of the tangential lens distortion
$g_+$ (upper panel) and the $45^\circ$ rotated ($\times$) component
 $g_\times$ (lower panel) for our 
red (triangles), blue (circles), and blue+red (squares) galaxy
samples derived from Subaru multi-color photometry (Table \ref{tab:gsample}).
} 
\end{figure} 


\begin{figure}[!htb] 
 \begin{center}
  \includegraphics[width=0.45\textwidth,angle=0,clip]{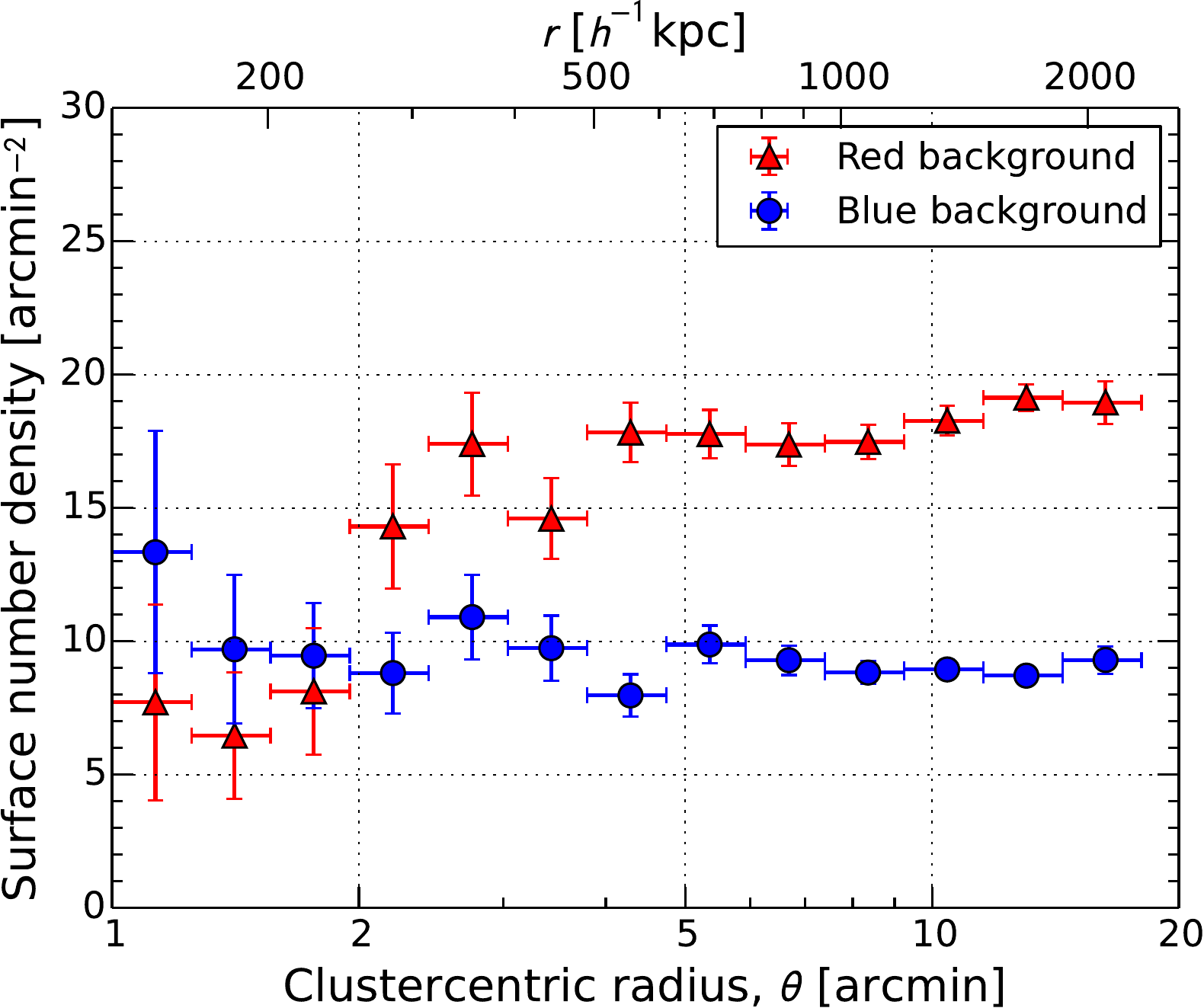}
 \end{center}
\caption{
\label{fig:nplot}
Coverage- and mask-corrected surface number density profiles of Subaru 
$BR_\mathrm{C}z'$-selected galaxy samples (Table \ref{tab:musample}).
The results are shown for our red (triangles) and blue (circles)
 background samples.
The error bars
 include contributions from Poisson counting uncertainties and
 contamination due to intrinsic clustering of each source population.
For the red sample, a systematic radial depletion of the source counts is
 seen toward the cluster center owing to magnification of the
 sky area, while the faint blue counts are nearly constant with the distance
 from the cluster center. 
See also Figure \ref{fig:wlplot}.
}
\end{figure} 

A careful background selection is critical for a cluster weak-lensing
analysis, so that unlensed objects do not dilute the true lensing signal
of the background \citep{Medezinski+07,UB2008,Okabe+2013,Hwang2014}. 
In particular, dilution due to contamination by cluster members can lead
to a substantial underestimation of the true signal at small cluster radii,
$r \simlt r_\mathrm{2500c}$ \citep{Medezinski+2010,Okabe+2010WL}.
The relative
importance of the dilution effect indicates that, the impact of 
background purity and depth is more important than that of
shot noise ($\propto \overline{n}_g^{-1/2}$).      
 
We use the color-color (CC) selection method of \citet{Medezinski+2010} to
define uncontaminated samples of background galaxies from which to measure
the shear and magnification effects.
Here we refer the reader to \citet{Medezinski+2010} for further details.
Our multi-color approach and its variants have been successfully applied to a
large number of clusters 
\citep{Medezinski+2010,Medezinski+2011,Medezinski+2013,Umetsu+2010CL0024,Umetsu+2011,Umetsu+2012,Umetsu2014clash,Coe+2012A2261,Oguri+2012SGAS,Covone2014,Sereno2014s8}.

We use the Subaru $BR_\mathrm{C}z'$ photometry, which spans the full
optical wavelength range, to perform CC selection of background samples. 
In Figure \ref{fig:CCsel}, we show the $B-R_\mathrm{C}$ versus 
$R_\mathrm{C}-z'$ distribution of all galaxies to our limiting magnitudes
(cyan). 
We select two distinct populations
that encompass the red and blue branches of background galaxies in CC
space, each with typical redshift distributions peaked around $z \sim 1$
and $\sim 2$, respectively 
\citep[see Figures 5 and 6 of][]{Medezinski+2011,Lilly+2007}. 
The color boundaries of our ``blue'' and ``red'' background 
samples are shown in Figure \ref{fig:CCsel}.

As a cross-check we calculate
the tangential ($g_+$) and cross ($g_\times$) reduced-shear components
in clustercentric radial bins, which we show in Figure \ref{fig:RGB}.
In the absence of higher-order effects, weak lensing produces only
curl-free tangential distortions, $g_+$. 
The presence of $\times$ modes can thus be used to check for systematic
errors. 
Using the weak-lensing-matched blue and red samples, we find a consistent,
rising distortion signal all the way to the cluster center.
For all cases, the $\times$-component is
 consistent with a null signal detection well within $2\sigma$ at all radii.

For the number counts to measure magnification, 
we define flux-limited photometry samples of background galaxies.
Here we limit the data to $z'=25.6$\,mag in the reddest band (Table
\ref{tab:musample}), corresponding to the $5\sigma$ limiting magnitude
within $2\arcsec$ diameter aperture. 
We plot in Figure \ref{fig:nplot} the coverage- and mask-corrected 
surface number density as a function of clustercentric radius,
for the blue and red samples. No clustering is observed toward the center,
demonstrating that there is no detectable contamination
by cluster members in the background samples. The red sample reveals a
systematic decrease in their counts toward
the cluster center, caused by magnfication of the sky area
(Section \ref{subsec:back}).
The faint blue counts, on the other hand, are nearly constant with
cluster radius, as expected by their steep count slope
(Table \ref{tab:musample}).
A more quantitative magnification analysis is given
in Section \ref{subsec:wl1d}.

For validation purposes, we compare in Figure \ref{fig:CCsel}
our background samples with a dynamically-selected spectroscopic sample
of 
377
cluster galaxies (black) found within a projected distance of
$12\arcmin$ ($\sim \rhalo$) from the cluster center.
We find that our background selection procedure
successfully excludes all except 2 spectroscopically confirmed cluster
members \citep[see also][]{Coe+2012A2261,Umetsu+2012},
corresponding to a negligible contamination fraction of $\sim 0.5\%$.
We note that, in the blue background region, there are 4 cluster members,
of which two are excluded by the magnitude cuts used to reject bright
foreground/cluster galaxies.


We estimate the mean depths 
($\langle\beta \rangle,\langle \beta^2\rangle$)
of the background samples 
(Tables \ref{tab:gsample} and \ref{tab:musample}), 
which are necessary when converting the
observed lensing signal into physical mass units.
For this, we follow the prescription outlined in Section 4.4 of
\citet{Umetsu2014clash}.  
We utilize BPZ to measure photo-$z$s using
our PSF-corrected Subaru $BVR_\mathrm{C}i'z'$ photometry.  
Following \citet{Umetsu+2012},
we employ BPZ's ODDS parameter 
as the weight factor $w(z)$ in Equation (\ref{eq:depth}).
The resulting depth estimates are summarized in Tables \ref{tab:gsample}
and \ref{tab:musample}.

\section{Subaru Weak-lensing Analysis}
\label{sec:wlana}

We use our $z'$-band limited sample of red background galaxies (Table
\ref{tab:musample}) for magnifciation measurements and a full composite
sample of blue+red galaxies (Table \ref{tab:gsample}) for shear
measurements.  
In Section \ref{subsec:wl1d}, we perform a 1D weak-lensing analysis of
A1689 to derive azimuthally averaged lensing profiles from our new
Subaru data (Section \ref{sec:subaru}), and  examine the
consistency of complementary shear and magnification measurements.
In Section \ref{subsec:wl2d}, we apply the 2D inversion method developed 
in Section \ref{sec:method} and reconstruct the projected 2D mass
distribution from joint shear+magnification measurements.  

\subsection{Weak-lensing Profiles of A1689}
\label{subsec:wl1d}

A1689 exhibits a small offset
$d_\mathrm{off}\simeq 5$\,kpc\,$h^{-1}$ ($\simeq 2.3\arcsec$) between
the BCG and X-ray centroids (Table \ref{tab:cluster}), ensuring a
well-defined center. 
The X-ray and SZE centroids agree to within $1\arcsec$ (Table
\ref{tab:cluster}). 
Here we will adopt the BCG position as the cluster center
for a radial profile analysis. 


\begin{figure}[!htb] 
 \begin{center}
  \includegraphics[width=0.45\textwidth,angle=0,clip]{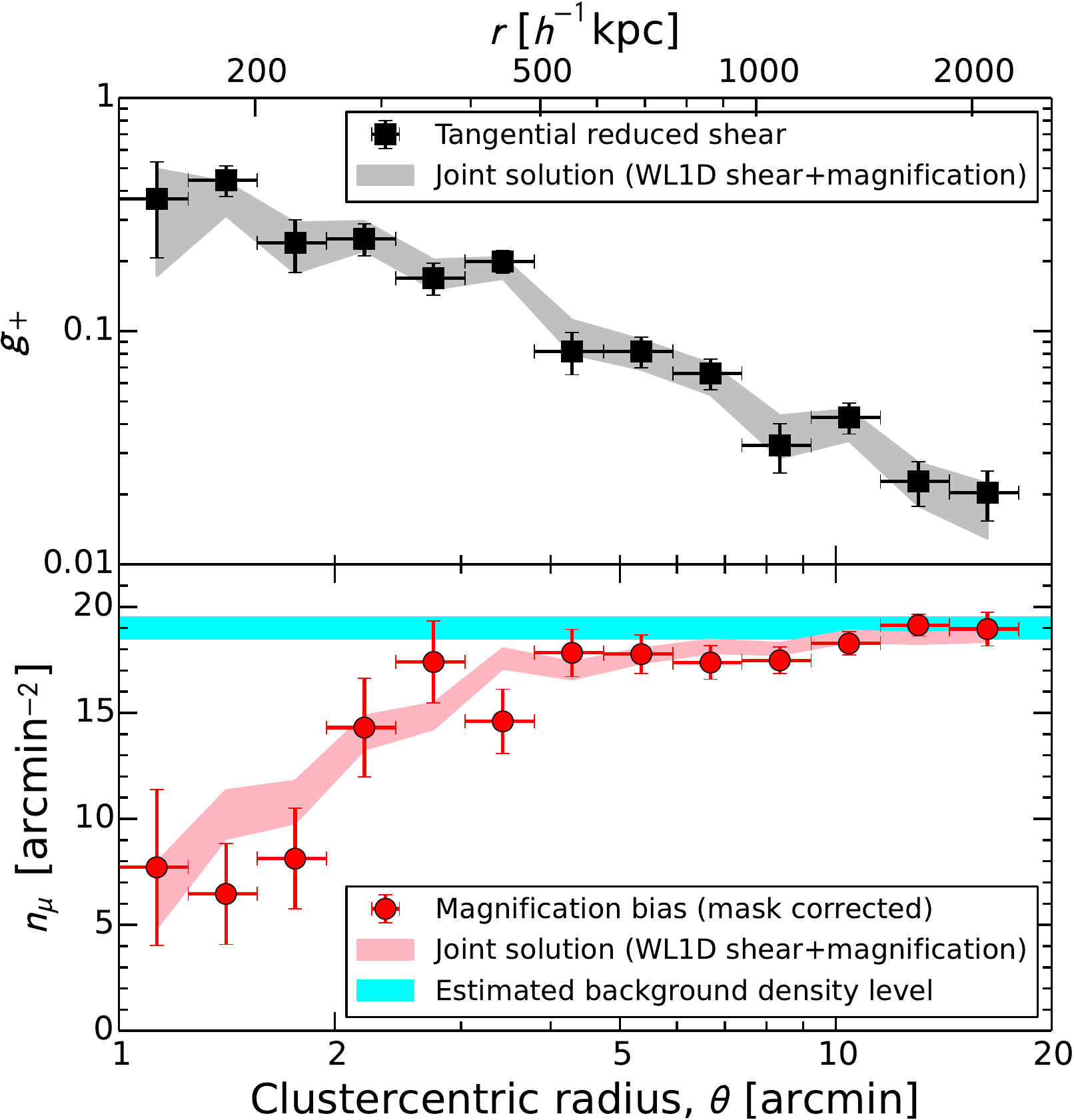}
 \end{center}
\caption{
\label{fig:wlplot}
Azimuthally averaged cluster weak-lensing profiles obtained from
Subaru multi-color observations of A1689. 
The upper panel shows the tangential reduced shear profile $g_+$ (black
squares) based on the full background sample.
The lower panel shows the magnification-bias profile $n_\mu$ (red circles)
of a $z'$-band limited sample of red background galaxies.
 For each observed profile, the shaded area represents 
 the joint reconstruction (68\% CL) from the
combined shear+magnification measurements. The
 horizontal bar (cyan shaded region) shows the constraints on the
 unlensed count normalization
estimated  from the source counts in cluster
 outskirts.    
} 
\end{figure} 
 

\begin{figure}[!htb] 
 \begin{center}
  \includegraphics[width=0.45\textwidth,angle=0,clip]{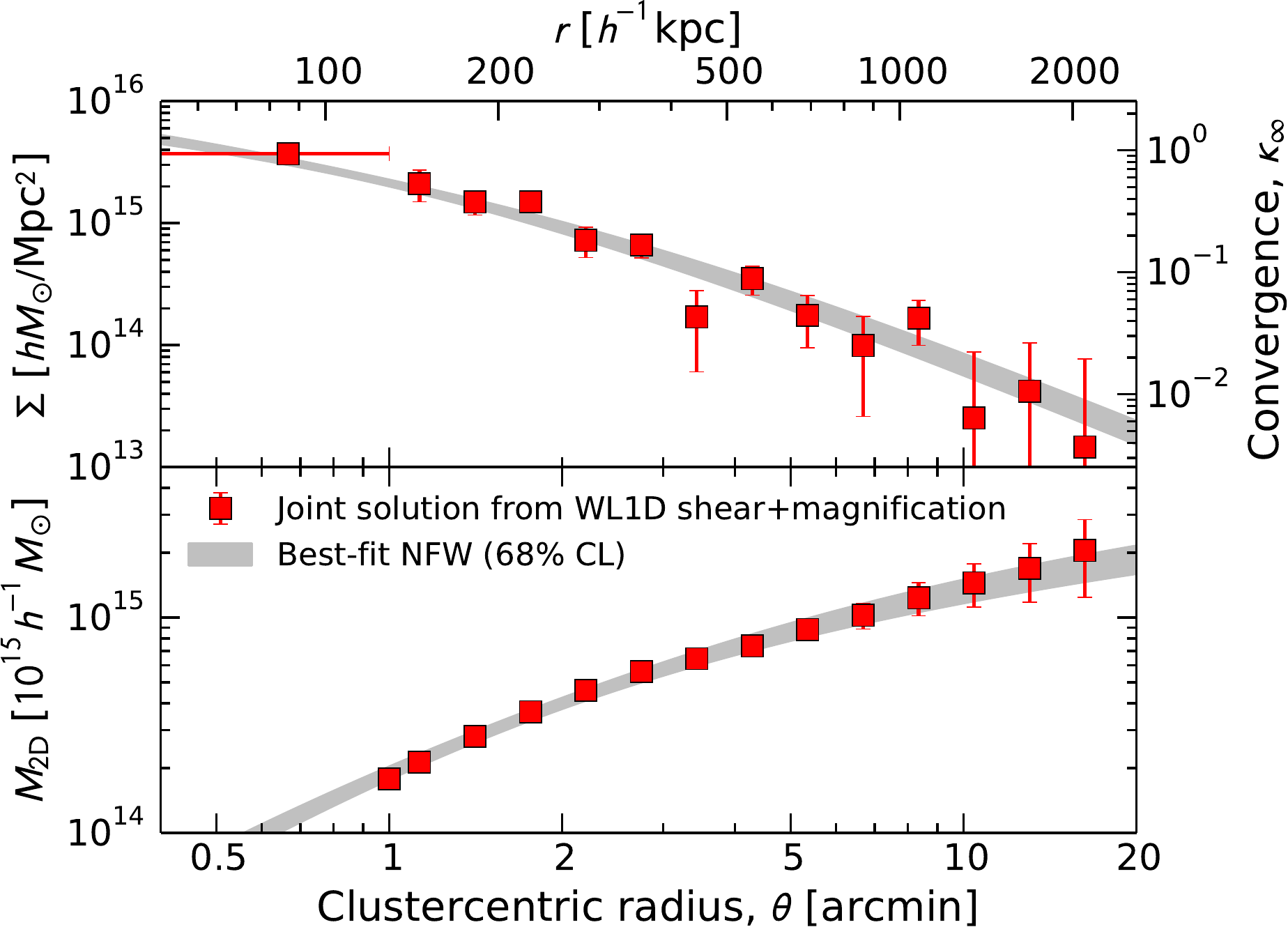}
 \end{center}
\caption{
\label{fig:kplot}
Surface mass density profile $\Sigma(\theta)$ (upper panel, red squares) 
derived from a Subaru 1D weak-lensing analysis of the combination of
 shear and magnification measurements shown in Figure \ref{fig:wlplot}.
The lower panel shows the corresponding 
cumulative mass profile $M_\mathrm{2D}(<\theta)$ (red squares).
 The gray area in each panel represents the
 best-fit projected Navarro--Frenk--White profile ($68\%$ CL) for the
 mass profile solution $\Sigma(\theta)$.  
} 
\end{figure} 

We derive azimuthally averaged radial profiles of tangential reduced
shear ($g_+$) and magnification bias ($n_\mu$) from Subaru data. 
We calculate the lensing profiles in $\Nbin = 13$ discrete
radial bins, spanning the range  
$[\theta_\mathrm{min},\theta_\mathrm{max}]=[1\arcmin,18\arcmin]$ with a constant
logarithmic spacing, 
$\Delta\ln\theta = \ln(\theta_\mathrm{max}/\theta_\mathrm{min})/\Nbin\simeq 0.22$.
The inner radial limit 
$r_\mathrm{min}\equiv D_\mathrm{l}\theta\mathrm{min}\simeq 129$\,\kpch
is sufficiently greater than the Einstein radius 
$\theta_\mathrm{Ein}=47.0\arcsec\pm 1.2\arcsec$ 
($z_\mathrm{s}=2$; Table \ref{tab:cluster}),
and it also satisfies
$r_\mathrm{min} > 2 d_\mathrm{off}\simeq 10$\,\kpch,  
so that the miscentering effects on mass profile reconstructions are
negligible \citep{Johnston+2007b,Umetsu+2011stack,Du+Fan2014}. 
The outer boundary $\theta_\mathrm{max}=18\arcmin$,
or $r_\mathrm{max}\equiv D_\mathrm{l}\theta_\mathrm{max}\simeq 2.3$\,\Mpch,
is large enough to encompass the entire virial region with
$\rvir \simeq 2$\,\Mpch \citep{UB2008}, 
but sufficiently small compared to the size of the Suprime-Cam field of
view so as to ensure accurate PSF anisotropy correction. 
The number of bins $\Nbin=13$ is chosen 
such that the detection signal-to-noise ratio (S/N) is of the
order of unity per bin, which is optimal for an inversion problem.  

In this work, we follow the prescription outlined in Section
3.2.2 of \citet{Umetsu2014clash} to perform magnification measurements
using the Subaru $BR_\mathrm{C}z'$-selected red galaxy 
sample (Table \ref{tab:musample}), which exhibits a clear depletion
signal (Figure \ref{fig:nplot}).
We have properly accounted and corrected for masking of background
galaxies due to cluster galaxies, foreground objects, and saturated
pixels (see also Section \ref{subsubsec:magbias}).
Unlike the nonlocal distortion signal,
the magnification signal falls off sharply with increasing cluster radius.
We thus estimate the count normalization and slope 
($\overline{n}_\mu,\alpha$) from the source counts in
cluster outskirts  
\citep{Umetsu+2011,Umetsu+2012,Umetsu2014clash,Medezinski+2013},
specifically at $12\arcmin$\,($\sim r_\mathrm{200c}) < \theta < \theta_\mathrm{max}$.

Figure \ref{fig:wlplot} shows the radial profiles of  ($g_+,n_\mu$).
A clear depletion of red galaxies is seen toward the 
center owing to geometric magnification of the sky area.
The statistical significance of the detection of the tangential
distortion is $22\sigma$.
The detection significance of the magnification signal is $9\sigma$,
which is $\sim 40\%$ of that of  distortion.

Here we construct the radial mass profile of A1689 from a joint
likelihood analysis of shear and magnification measurements (Figure
\ref{fig:wlplot}), using the method of \citet{Umetsu+2011}.
We have 26 constraints 
$\{g_{+,i}, n_{\mu,i}\}_{i=1}^{\Nbin}$
in 13 log-spaced clustercentric radial bins.
The model is described by $\Nbin+1=14$ parameters,
$\{\Sigma_\mathrm{min},\Sigma_i\}_{i=1}^{\Nbin}$,
where $\Sigma_\mathrm{min}\equiv \Sigma(<\theta_\mathrm{min})$ is the
average surface mass density 
interior to $\theta_\mathrm{min}$, and $\Sigma_i$ is the surface mass
density averaged in the $i$th radial bin.
To perform a reconstruction, we express the lensing observables 
($g_+, \mu^{-1}$)
in terms of $\Sigma$ using the relations given in
Appendix \ref{appendix:estimators}.
Additionally, we account for the calibration uncertainty in the
observational parameters 
$\bc=(\langle W\rangle_g, f_{W,g}, \langle W\rangle_\mu, \overline{n}_\mu, \alpha)$ as given in 
Tables \ref{tab:gsample} and \ref{tab:musample}.
Following \citet{Umetsu2014clash},
we fix $f_{W,g}$ to the observed value (Table \ref{tab:gsample}).

The results are shown in Figures \ref{fig:wlplot} and \ref{fig:kplot}.
The ML solution has a reduced $\chi^2$ of 11.5 for
12 degrees of freedom (dof),
indicating good consistency between the shear and magnification
measurements having different potential systematics.
This is demonstrated in Figure \ref{fig:wlplot}, which compares the
observed lensing profiles with the respective joint reconstructions (68\% CL). 
The resulting mass profile $\Sigma(\theta)$ is shown in the upper panel of 
Figure \ref{fig:kplot}.
The error bars represent the $1\sigma$ errors from the diagonal part
of the total covariance matrix $C$ \citep[][]{Umetsu2014clash}.
The corresponding cumulative mass profile is shown in the lower panel of
Figure \ref{fig:kplot}.

\subsection{Weak-lensing Mapmaking of A1689}
\label{subsec:wl2d}

We apply our 2D inversion method (Section \ref{sec:method}) to our new
Subaru observations (Sections \ref{sec:subaru}) for obtaining an
unbiased recovery of the projected matter distribution $\Sigma(\btheta)$
in A1689. 
In this approach, we combine the observed spatial shear pattern
$(g_1(\btheta),g_2(\btheta))$ 
with the azimuthally averaged magnification measurements 
$\{n_{\mu,i}\}_{i=1}^{\Nbin}$
(Section \ref{subsec:wl1d}), which impose a set of azimuthally integrated
constraints on the underlying
$\Sigma(\btheta)$ field, thus effectively breaking the mass-sheet
degeneracy. 
The algorithm takes into account
the nonlinear subcritical regime of the lensing properties.

For mapmaking, we pixelize the lensing fields into a $56\times 56$ grid 
with $\Delta\theta=0.5\arcmin$ spacing, 
covering the central $28\arcmin\times 28\arcmin$ field.
The model $\bm=(\bs,\bc)$ is specified by 
$\Npix=56^2$ parameters,
$\bs=\{\Sigma(\btheta_m)\}_{m=1}^{\Npix}$,
and a set of calibration parameters $\bc$ to marginalize over.
We utilize the FFTW implementation
of fast Fourier transforms (FFTs) to calculate 
$\gamma_\infty(\btheta)$  from $\kappa_\infty(\btheta)$ using Equation
(\ref{eq:shear2m}).
To minimize spurious aliasing effects from the periodic boundary
condition, the  maps are zero padded to twice the original length
in each spatial dimension
\cite[e.g.,][]{1998ApJ...506...64S,UB2008}.

\begin{figure}[!htb] 
 \begin{center}
  \includegraphics[width=0.45\textwidth,angle=0,clip]{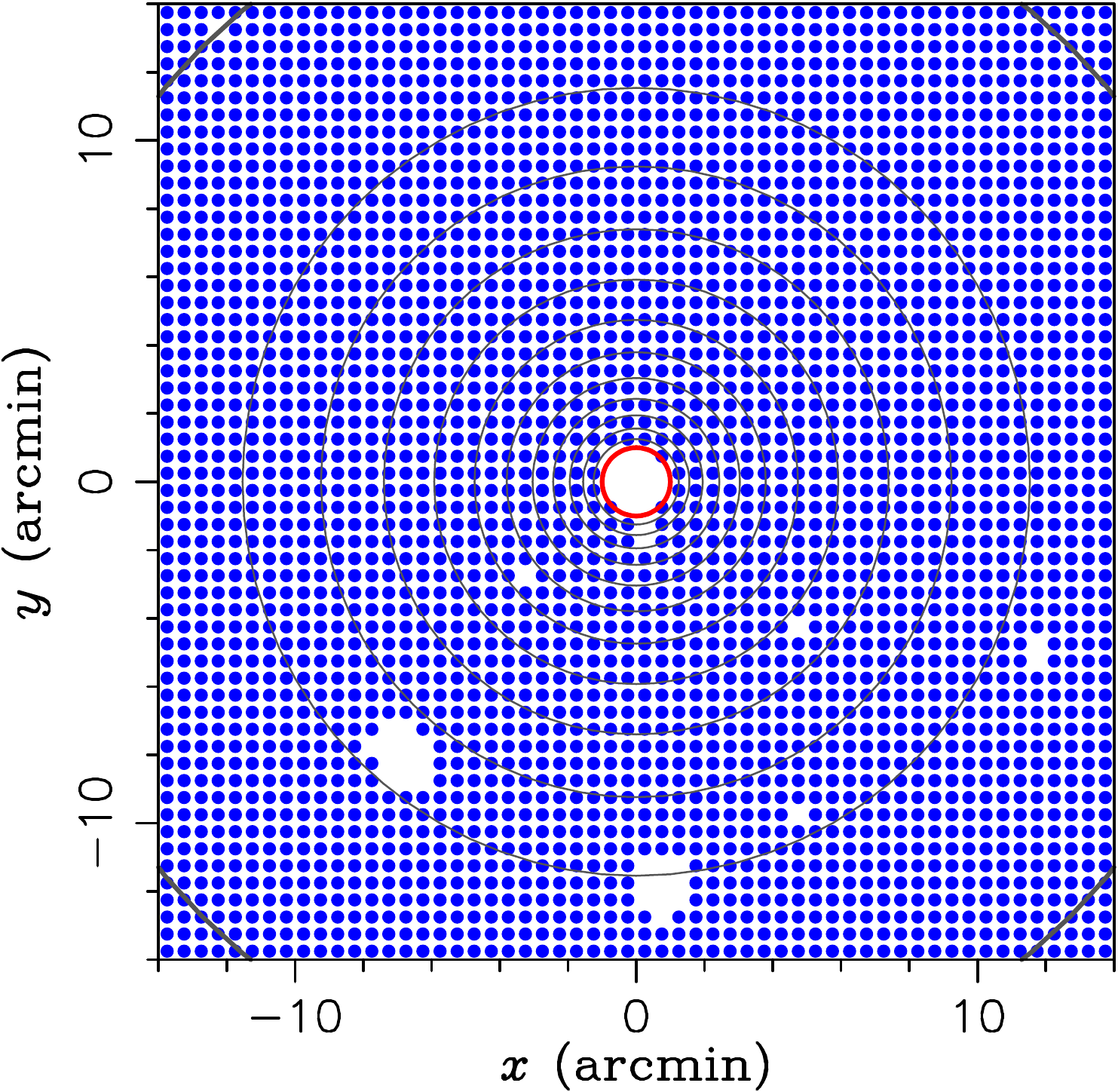}
 \end{center}
\caption{
Spatial distribution of weak-lensing constraints averaged onto a grid of
 $56\times 56$ pixels, covering a field of $28\arcmin \times 28\arcmin$
 centered on the BCG.
Each point represents a single pixel with two-component reduced
 shear constraints ($g_1,g_2$) averaged within a top-hat region with
 radius $\theta_\mathrm{f}=0.4\arcmin$.
We exclude from our analysis those pixels lying within the inner
$\theta_\mathrm{cut}=1\arcmin$ region (red circle) and those having no
 background galaxies with usable shape measurements (see Figure
 \ref{fig:subaru}). 
There are $3093$ pixels with reduced-shear constraints,
 yielding $6186$ constraints from 2D shear measurements.  
Azimuthally averaged magnification constraints are obtained in $13$
 logarithmically-spaced, clustercentric annuli spanning the range
 $[\theta_\mathrm{min},\theta_\mathrm{max}]=[1\arcmin,18\arcmin]$. 
\label{fig:grid}
} 
\end{figure}

We use a top-hat window of $\theta_\mathrm{f}=0.4\arcmin$ 
(Section \ref{subsubsec:shear}) to average over a local ensemble of
galaxy ellipticities  
\citep[$N=\pi\overline{n}_g\theta_\mathrm{f}^2\sim 10$;][]{Merten2014clash}
at each grid point, accounting for the intrinsic ellipticity
distribution of background sources. 
The filter size corresponds to an effective resolution of 
$2D_\mathrm{l}\theta_\mathrm{f}\simeq 100$\,\kpch 
at the cluster redshift. 
To avoid potential systematic errors,
we exclude from our analysis (Section \ref{subsubsec:lg}) those pixels
lying within central  
$\theta_{\rm cut}=1\arcmin$ where $\Sigma(\btheta)$
can be close to or greater than the critical value $\Sigma_{\rm c}$, 
as well as those containing no background galaxies with usable shape
measurements. 
For distortion measurements ($g_1(\btheta),g_2(\btheta)$) from the full
background sample (Table \ref{tab:gsample}),
this leaves us
with a total of $3093$ usable measurement pixels (blue points in Figure
\ref{fig:grid}), corresponding to $6186$ constraints.
For magnification measurements, we have 13 azimuthally averaged constraints 
$\{n_{\mu,i}\}_{i=1}^{N_{\rm bin}}$
in log-spaced clustercentric annuli (Figure \ref{fig:grid}).
The total number of constraints is thus 
$N_{\rm data}=6199$, yielding $N_{\rm data}-\Npix=3063$\,dof.

\begin{figure}[!htb] 
 \begin{center}
  \includegraphics[width=0.45\textwidth,angle=0,clip]{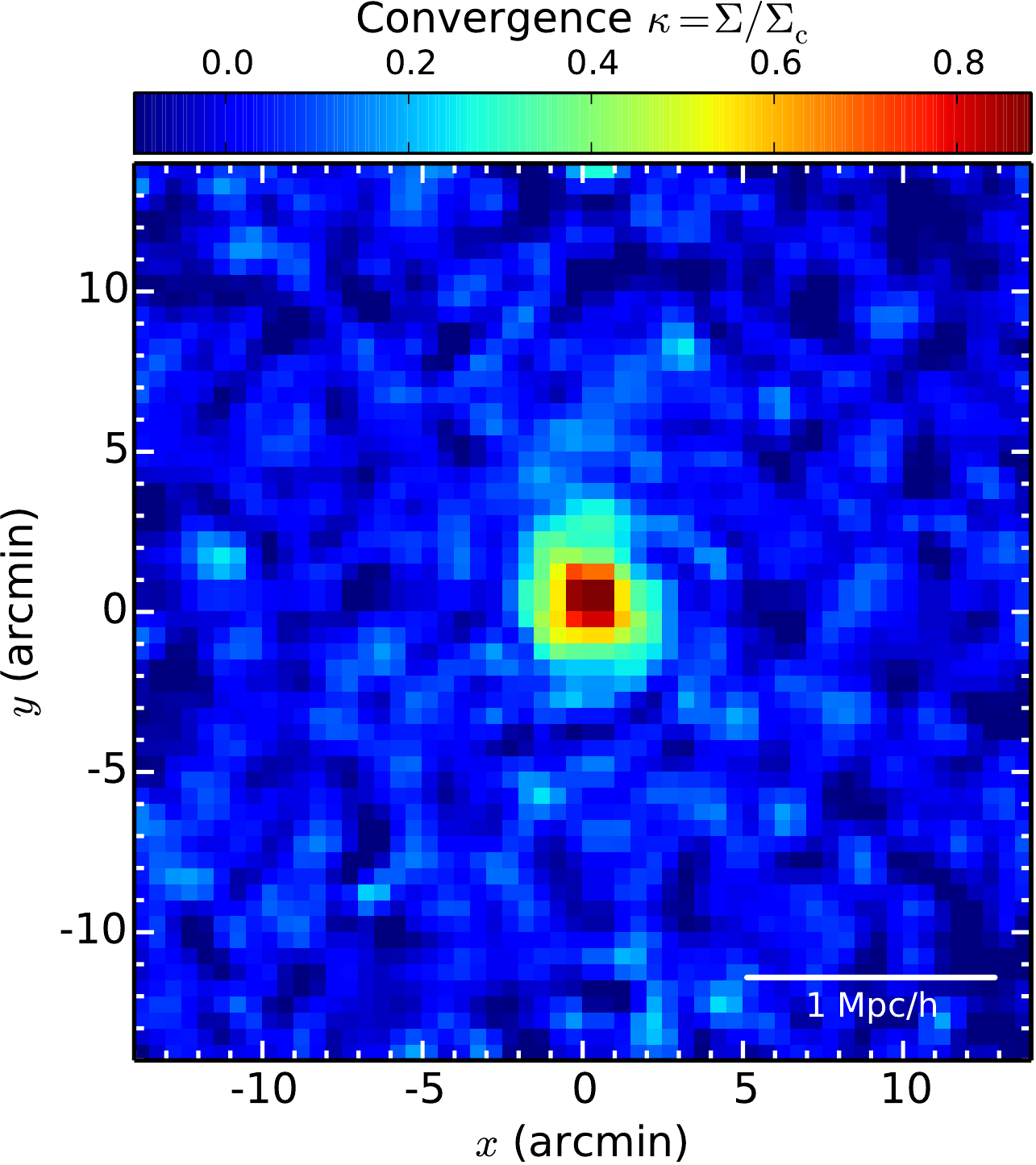}
 \end{center}
\caption{
\label{fig:kmap}
Projected mass distribution $\Sigma(\btheta)$ of A1689 
reconstructed from a Subaru weak-lensing analysis of 2D gravitational
 shear and azimuthally averaged magnification data.
The mass maps is $28\arcmin\times 28\arcmin$ in size 
($3.6$\,Mpc\,$h^{-1}$ on a side) 
and centered on the BCG.
The color bar indicates the lensing convergence
$\kappa=\langle \Sigma_\mathrm{c}^{-1}\rangle \Sigma$,
scaled to the mean depth of weak-lensing observations,
$1/\langle \Sigma_\mathrm{c}^{-1}\rangle=4.66\times 10^{15}hM_\odot$\,Mpc$^{-2}$.
For visualization purposes, the mass map is smoothed with a
 $1\arcmin$ FWHM Gaussian.
North is to the top, east to the left. The horizontal bar 
represents $1\,$Mpc\,$h^{-1}$ at the cluster redshift.
} 
\end{figure}

In Figure \ref{fig:kmap}, we show the resulting $\Sigma(\btheta)$ field
reconstructed from a joint analysis of the 2D shear and azimuthally
averaged magnification data. 
The $\chi^2$ value for the ML solution is
$\chi^2(\hat{\bm})=4046$ for 3063\,dof.
Here, for visualization purposes, the $\Sigma(\btheta)$ field is 
smoothed with a Gaussian of ${\rm FWHM}=1\arcmin$.
The main mass peak coincides well with the cluster center.
The projected mass distribution is elongated in the north-south
direction (Figure \ref{fig:subaru}; see also Section \ref{subsec:wl+sl})
and very similar to the distribution of cluster members  \citep{Kawaharada+2010}.


\begin{figure}[!htb] 
 \begin{center}
  \includegraphics[width=0.45\textwidth,angle=0,clip]{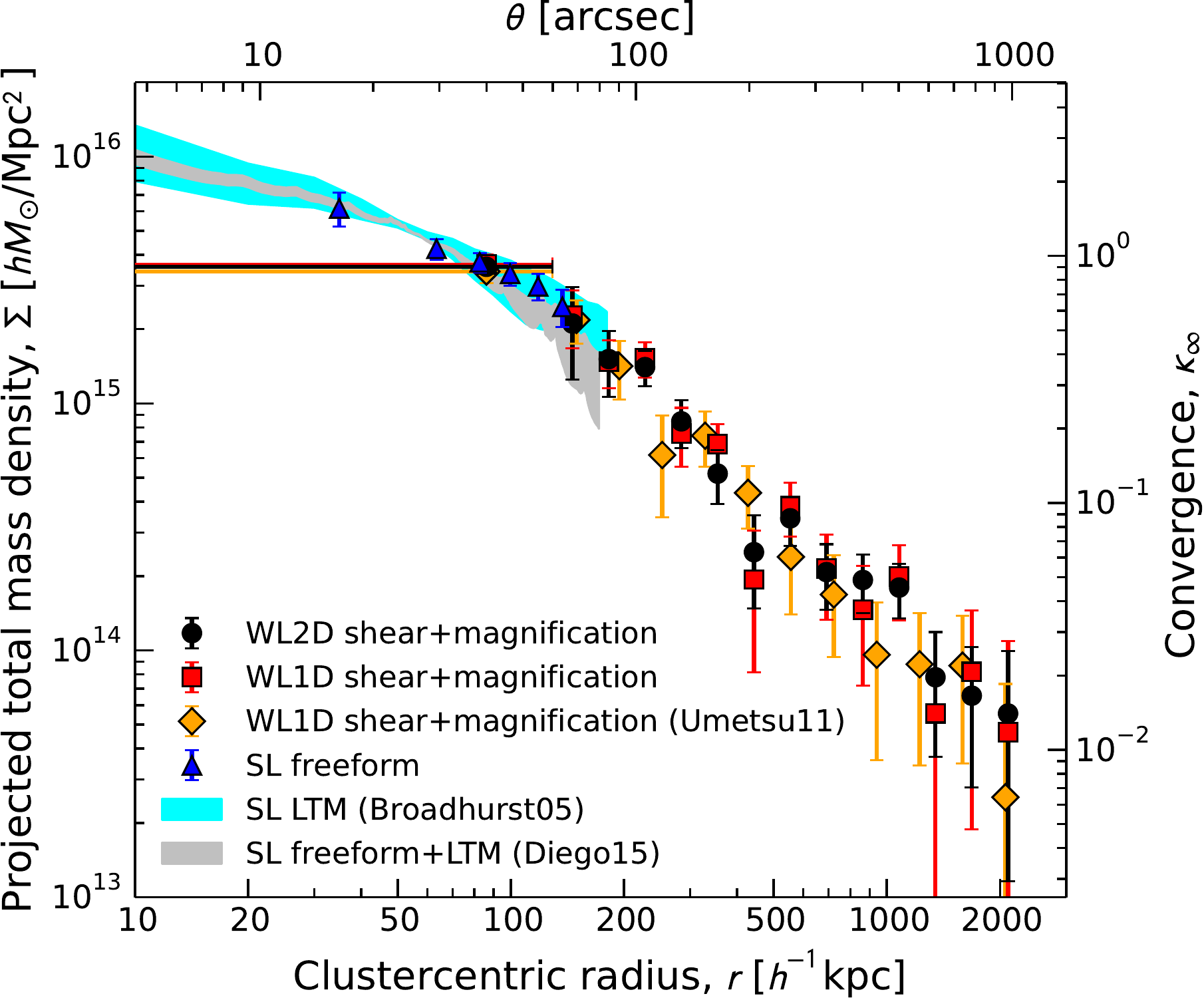}
 \end{center}
\caption{
\label{fig:kcomp}
Comparison of projected mass density profiles $\Sigma(r)$ derived
 from our 
Subaru 1D weak-lensing analysis (squares; Section \ref{subsec:wl1d}), 
Subaru 2D weak-lensing analysis (circles; Section \ref{subsec:wl2d}), and
free-form strong-lensing analysis of {\em HST} data
 (triangles; Section \ref{sec:slana}).
The cyan shaded area represents the mass profile with $1\sigma$ uncertainty 
from a strong-lensing analysis of \citet{2005ApJ...621...53B} based on
 the light-traces-mass (LTM) assumption. 
The gray shaded area shows the strong-lensing results ($68\%$ CL) from 
\citet{Diego2015a1689}
using a hybrid scheme combining both free-form grid and LTM substructure
 components.   
The diamonds with error bars show the results from our earlier
1D weak-lensing analysis \citep{Umetsu+2011} based on Subaru $Vi'$ data.
Good agreement between the strong and weak lensing results is seen in the
region of overlap. There is also good agreement between the different
 lensing methods and data sets.
} 
\end{figure} 


\begin{figure}[!htb] 
 \begin{center}
  \includegraphics[width=0.45\textwidth,angle=0,clip]{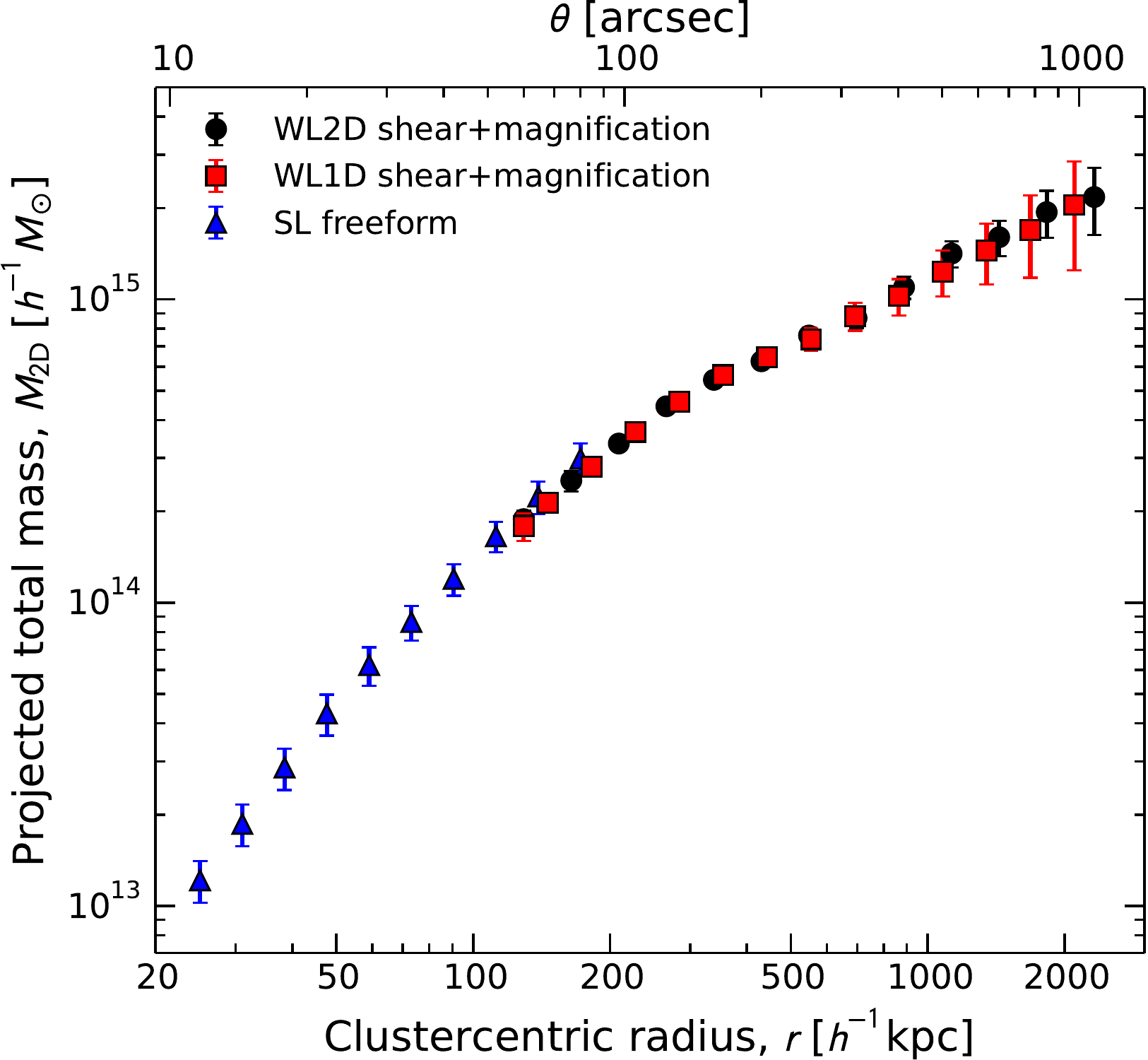}
 \end{center} 
\caption{
\label{fig:mplot}
Comparison of projected cumulative mass profiles $M_{\rm 2D}(<r)$ of
 A1689 derived from our Subaru 1D weak-lensing analsyis (squares;
 Section \ref{subsec:wl1d}), 
Subaru 2D weak-lensing analysis (circles; Section \ref{subsec:wl2d}), and
{\em HST} strong-lensing analysis (triangles; Section \ref{sec:slana}).
} 
\end{figure}

In Figure \ref{fig:kcomp}, we compare the projected mass profiles
$\Sigma(\theta)$ obtained from our 1D and 2D analyses of the
shear+magnification data.  Here we have used the method described in 
Appendix \ref{appendix:2dto1d} to construct an optimally weighted radial
projection of the $\bSigma$ map.
Our 1D- and 2D-based $\Sigma$ profiles are consistent within $1\sigma$ at
all cluster radii, and both are in good agreement with
the 1D results of \citet{Umetsu+2011} from the joint
shear+magnification analysis of the Subaru $Vi'$ data.
Similarly, our 1D and 2D weak-lensing results are in excellent agreement
with each other
in terms of the cumulative mass $M_{\rm 2D}(<\theta)$ as shown
in Figure \ref{fig:mplot}.

\section{{\em HST} Strong-lensing Analysis}
\label{sec:slana}

\subsection{Image Systems}
\label{subsec:slim}

A1689 has been a subject of detailed strong-lensing studies by
numerous authors 
\citep[e.g.,][]{2005ApJ...621...53B, 2006MNRAS.372.1425H,
2007ApJ...668..643L, Coe+2010, Diego2015a1689}. 
Thus far,
a total of 61 multiple-image candidate systems of 165 images
were identified from extremely deep optical and near-infrared data
from {\em HST} and Subaru \citep{Diego2015a1689}. 

To study global structural properties of the cluster,
we focus our strong-lensing analysis on the principal modes of the
cluster mass distribution, responsible for the massive, smooth halo
component (see Section \ref{subsubsec:sl}). 
For this aim,  
we conservatively select a subset of systems based on the following
criteria:  
{\em i}) We use only spectroscopically confirmed systems.  
{\em ii}) We consider only systems whose members were consistently
identified in different studies. 
{\em iii}) 
We limit our analysis to those lying within 80\arcsec
from the BCG, so that multiple images spread fairly evenly over
the analysis region.
{\em iv)} We discard systems of very close pairs.
They are primarily sensitive to substructures rather than the principal
modes of the mass distribution, which we are interested in. 

These criteria leave us with 12 systems (ID 1, 2, 4, 5, 6, 7, 11, 15,
18, 22, 24, 29, according to the original notation in
\citet{2005ApJ...621...53B}), for a total of 44 multiple images spanning
the range $1.4\arcsec$-$72.3\arcsec$ in cluster radius.

\subsection{{\sc PixeLens} Free-form Mass Reconstruction}
\label{subsec:sl2d}

Free-form models describe the lens on a grid of  pixels or a set of
basis functions, allowing for a wide range of
solutions \citep{Coles2008}. 
We have performed a free-form strong-lensing analysis of the central
region using the {\sc PixeLens} software \citep{Saha+Williams2004},
which produces pixelated maps of the surface mass density.
Each map is constrained to exactly reproduce the positions and parities 
of all given multiple images. 
{\sc PixeLens} generates a statistical ensemble
of models through which uncertainties and degeneracies in solutions can
be explored \citep{Coles2008}.  

Our {\sc PixeLens} analysis procedure largely follows 
\citet{Sereno+Zitrin2012} and \citet{Sereno2013glszx}.
To determine robust sampling strategies optimized to recover the smooth
cluster signal, we tested the {\sc PixeLens} algorithm using simulated
sets of multiple images in analytic lenses. 
The results suggest that the best strategy is to limit each analysis to
three image systems, for a total of a dozen of images, and to
reconstruct maps with $\sim 10$ pixels in the radial direction, avoiding
oversampling 
\citep{Lubini+Coles2012}.
We thus divide the strongly-lensed images in four groups of three
systems each and analyze each group separately.  
We end up with four triples consisting of systems 1, 5, and 11 (11
images), systems 2, 6, and 22 (11 images), systems 4, 15, and 29 (12
images), and systems 7, 18, and 24 (10 images). 
Image systems with similar configurations are divided into different
groups.


For each group, we compute 500 $\kappa$ maps within $80\arcsec$ from the 
BCG on a circular grid of 349 pixels ($10$ pixels along the radial
direction) with a pixel size of $8\arcsec$ ($\simeq 17.2$\,kpc\,$h^{-1}$).
These optimal settings allow us to avoid the known problem of too flat
density profiles recovered with {\sc PixeLens} modeling 
\citep[see][]{Grillo2010Pixelens,Umetsu+2012}, which otherwise could
bias cluster mass estimates.
As discussed by \citet[][see their Appendix]{Grillo2010Pixelens},
this bias can arise from  a combination of the mass-sheet degeneracy
\citep{Schneider+Seitz1995} and the assumed prior on the positive
definiteness of every pixel of the surface mass density map.

In the following, we restrict our analysis 
to the region where the cluster mass distribution is accurately
recovered by {\sc PixeLens}. 
We exclude the central 20\,kpc\,$h^{-1}$ region to
minimize the effects of miscentering and baryonic physics
\citep{Umetsu+2012,Umetsu2014clash}. 
For each group of reconstruction, we determine the outer cutoff radius
beyond which the logarithmic density slope is steeper than -2, the
asymptotic minimum slope for the projected Navarro--Frenk--White 
density profile \citep[NFW,][]{1997ApJ...490..493N}.
The maximum radius is $63.7\arcsec$ (188 mass pixels)
in three cases and $54.9\arcsec$ (140 mass pixels) for 
the group with the triple 4--15--29. 

\subsection{Comparison of Weak and Strong Lensing Results}
\label{subsec:wlslcomp}

We show in Figure \ref{fig:kcomp} the radial mass distribution of A1689
from our {\em HST} strong-lensing analysis.
The results are shown along with the previous strong-lensing results by
\citet{2005ApJ...621...53B} and \citet{Diego2015a1689}, as well as with
independent weak-lensing results from shear and magnification
information (Sections \ref{subsec:wl1d} and \ref{subsec:wl2d}).
The strong-lensing model of \citet{2005ApJ...621...53B} is based on the 
light-traces-mass (LTM) assumption, so that the {\em HST} photometry of 
cluster red-sequence galaxies was used as an initial guess for their
lens solution.
\citet{Diego2015a1689} used a hybrid (free-form + LTM) approach
 combining Gaussian pixel grid and cluster member components for
 describing large- and small-scale contributions to the deflection
 field, respectively. They constrained the range of solutions with
 sufficient accuracy to allow the detection of new counter images
 for further improving  the lensing solution of A1689.  
This comparison shows clear consistency among a wide variety of lensing
methods with different assumptions and potential systematics, 
demonstrating the robustness of our results
(see also Figure \ref{fig:mplot}).
Excellent agreement is also found between our strong-lensing mass profile
and that of \citet{2007ApJ...668..643L}.

\section{Triaxial Modeling of the Cluster Matter Distribution}
\label{sec:3dmodel}

Since we can only observe clusters in projection, determining the
intrinsic 3D shape and orientation of an aspherical cluster is an
intrinsically underconstrained problem \citep{2007MNRAS.380.1207S}. 
In this section, we describe the modeling of the 3D cluster matter
distribution as an ellipsoidal halo following \citet{Sereno2013glszx}.
In this approach, we exploit the combination of X-ray and SZE
observations to constrain the elongation of the ICM along the line of
sight.  
We use minimal geometric assumptions about the matter and gas
distributions to couple the constraints from lensing and SZE/X-ray
data. The parameter space is explored in a Bayesian inference framework.   
This multi-probe method allows us to improve constraints on the
intrinsic shape and orientation of the cluster mass distribution without
assuming HSE. 

\subsection{Matter Distribution}
\label{subsec:matter}

We model the cluster mass distribution with a triaxial
NFW density profile as motivated by cosmological
$N$-body simulations  \citep{2002ApJ...574..538J,Kasun+Evrard2005}.
The radial dependence of the spherical NFW density profile is given by
\citep{1996ApJ...462..563N,1997ApJ...490..493N} 
\begin{equation}
\rho(r) = \frac{\rho_\mathrm{s}}{(r/r_\mathrm{s})(1+r/r_\mathrm{s})^2}
\end{equation}
with $\rho_\mathrm{s}$ the characteristic density and $r_\mathrm{s}$ the 
inner characteristic radius at which the logarithmic slope of the
density profile is -2.
We generalize the spherical NFW model to obtain a triaxial density
 profile by replacing
$r$ and $r_\mathrm{s}$ with the respective ellipsoidal radii 
$R$ and $R_\mathrm{s}$, defined such that
\begin{equation}
R^2 = c^2\left( \frac{X^2}{a^2} + \frac{Y^2}{b^2} + \frac{Z^2}{c^2}\right)
= \frac{X^2}{q_a^2} + \frac{Y^2}{q_b^2} + Z^2,
\end{equation}
where $q_a = a/c$ and $q_b =  b/c$  ($a\le b\le c$) are the
minor--major and intermediate--major axis ratios,
respectively.\footnote{The intrinsic axis ratios $(q_a,q_b)$ here
correspond to $(\eta_{{\rm DM},a},\eta_{{\rm DM},b})$ of
\cite{Limousin2013} in their notation.}  
The corresponding eccentricities are $e_a=\sqrt{1-q_a^2}$ and
$e_b=\sqrt{1-q_b^2}$. 
The degree of triaxiality is defined as 
${\cal T}=e_b^2/e_a^2$ \citep{Sereno2013glszx}.

We define an ellipsoidal overdensity radius $R_{\Delta}$ 
\citep[e.g.,][]{Corless2009triaxial,Sereno+Umetsu2011,Buote+Humphrey2012p2}
such that the mean interior density contained within a ellipsoidal
volume of semimajor axis 
$R_{\Delta}$ is $\Delta\times \rho_\mathrm{c}$.
The total mass enclosed within $R_{\Delta}$ is 
$M_{\Delta}=(4\pi/3)\Delta q_a q_b \rho_\mathrm{c}
R_{\Delta}^3$.
We use $\Delta=200$ to define the halo mass, $\Mhalo$.
The triaxial concentration parameter is defined by
$\chalo = R_\mathrm{200c}/R_\mathrm{s}$.
The characteristic density  is then expressed as
$\rho_\mathrm{s}=M_\Delta/(4\pi q_a q_b R_\Delta^3)\times c_\Delta^3/[\ln(1+c_\Delta)-c_\Delta/(1+c_\Delta)]$
\citep{Buote+Humphrey2012p2}.

A triaxial halo is projected on to the sky plane as elliptical
isodensity contours \citep{Stark1977}, which can be expressed as a
function of the intrinsic halo axis ratios ($a/c,b/c$) and 
orientation angles ($\vartheta,\phi,\psi$)
with respect to the observer's line of sight.  
Here we adopt the $z$-$x$-$z$ convention of Euler angles to be
consistent with \citet{Stark1977}
\citep[see, e.g.,][]{Sereno+Ettori+Baldi2012}. 
The angle $\vartheta$ describes the inclination of the major ($Z$) axis with
respect to the line of sight.

For a given projection, the elliptical projected mass distribution can
be described as a function of the elliptical radius $\zeta$
defined in terms of the observer's coordinates $(X',Y')$ in the plane of the sky:
\begin{equation}
\label{eq:rellip}
\zeta^2 = 
\frac{1}{f}\left(
j X'^2 + 2k X'Y' + l Y'^2
\right)
\equiv
\frac{X''^2}{q_{\perp X}^2} + \frac{Y''^2}{q_{\perp Y}^2} 
\end{equation}
where 
$q_{\perp X}$ and $q_{\perp Y}$ ($q_{\perp X}\ge q_{\perp Y}$) are 
\begin{equation}
 \begin{aligned}
  q^2_{\perp X} &= \frac{2f}{j + l - \sqrt{(j-l)^2+4{k}^2}},\\ 
  q^2_{\perp Y} &= \frac{2f}{j + l + \sqrt{(j-l)^2+4{k}^2}}.
 \end{aligned}
\end{equation} 
with
\begin{equation}
 \label{eq:ABC}
 \begin{aligned}
j &=\cos^2\vartheta\left(\frac{c^2}{a^2}\cos^2\phi + \frac{c^2}{b^2}\sin^2\phi\right) + \frac{c^2}{a^2}\frac{c^2}{b^2}\sin^2\vartheta,\\
k &=  \sin\phi \cos\phi \cos\vartheta \left(\frac{c^2}{a^2}-\frac{c^2}{b^2}\right),\\
l &= \frac{c^2}{a^2}\sin^2\phi + \frac{c^2}{b^2}\cos^2\phi,\\
f &= \sin^2\vartheta\left(
\frac{c^2}{a^2} \sin^2\phi 
+ 
\frac{c^2}{b^2} \cos^2\phi
\right) + \cos^2\vartheta.
 \end{aligned}
\end{equation}
Here we have chosen the new coordinate system ($X'',Y''$) such that the
$X''$ axis is aligned with the major axis of the projected ellipse.
The minor--major axis ratio $q_\perp\equiv q_{\perp Y}/q_{\perp X}$ of the elliptical
density contours is given by\footnote{Note the projected axis ratio
$q_\perp$ is equivalent to $1/e_P$ of \citet{2007MNRAS.380.1207S}.}
\begin{equation}
q_\perp(a/c, b/c, \vartheta,\phi) = \left[
\frac{j+l-\sqrt{(j-l)^2+4k^2}}{j+l+\sqrt{(j-l)^2+4k^2}}
\right]^{1/2}.
\end{equation}
The principal axes of the isodensities are rotated by an angle $\psi$ with respect
to the projection on to the sky of the intrinsic major axis $Z$,
where $2\psi = {\rm arctan}[2k/(j-l)]$
\citep[][]{2007MNRAS.380.1207S}.
As observable parameters to describe the projected mass distribution,
we use the ellipticity
\begin{equation}
 \epsilon = 1-q_\perp
\end{equation}
and the position angle  $\psi_\epsilon$ of the projected major axis.

The projected surface mass density
$\Sigma(\zeta)$ as a function of the elliptical radius $\zeta$
 is related to the triaxial density profile
$\rho(R)$ by \citep{Stark1977}
\begin{equation}
  \Sigma(\zeta) =  \frac{2}{\sqrt{f}}
  \int_\eta^{\infty}\frac{\rho(R)RdR}{\sqrt{R^2-\zeta^2}} 
  = \frac{2R_{\rm s}}{\sqrt{f}}
  \int_{\zeta/R_{\rm s}}^{\infty}\frac{\rho(R_{\rm s}
  x)xdx}{\sqrt{x^2-(\xi/\xi_{\rm s})^2}},
\end{equation}
where 
$\xi \equiv q_{\perp X} \zeta = \sqrt{X''^2+Y''^2/q_\perp^2}$  
is the {\it observable} elliptical radius,
and
$\xi_{\rm s} = q_{\perp X} R_{\rm s}$ is the observable scale length
(semi-major axis) in
the sky plane \citep{2007MNRAS.380.1207S}.
The quantity $l_{\parallel} = R_{\rm s}/\sqrt{f}$
represents the line-of-sight half length of the ellipsoid of radius $R=R_{\rm s}$
\citep{2007MNRAS.380.1207S}.
It is useful to introduce the dimensionless scale factor
$e_\parallel$ that quantifies the extent of the cluster along the line of
sight \citep{2007MNRAS.380.1207S},
\begin{equation}
\label{eq:elong}
e_\parallel(a/c,b/c,\vartheta,\phi) =\frac{l_\parallel}{\xi_{\rm s}} =
 \left(
  \frac{q_\perp}{q_a q_b}
 \right)^{1/2}
 f^{-3/4}.
\end{equation}
The larger $e_\parallel$, the larger the elongation along the line
of sight. 
The quantity $e_\parallel$ corresponds to the
inverse of the elongation parameter $e_\Delta$ of
\citet{2007MNRAS.380.1207S}: $e_\parallel=1/e_\Delta$.

For a self-similar model
$\rho(R) = \rho_{\rm s} f_{\rm 3D}(R/R_{\rm s})$,
the projected mass density profile is expressed as 
\begin{equation}
\Sigma(\zeta) 
=\frac{2R_{\rm s}\rho_{\rm s}}{\sqrt{f}}
\int_{\zeta/R_{\rm s}}^{\infty}\frac{f_{\rm
3D}(x)xdx}{\sqrt{x^2-(\xi/\xi_{\rm s})^2}}
\equiv \Sigma_{\rm s} f_\mathrm{2D}(\xi/\xi_{\rm s}),
\end{equation}
where we have defined the scale surface mass density 
\begin{equation}
\Sigma_{\rm s}\equiv 2\rho_{\rm s} R_{\rm s} /\sqrt{f}
=2\rho_{\rm s} \xi_{\rm s} e_\parallel
= 2f_{\rm geo} \rho_{\rm s} \sqrt{q_\perp}\xi_{\rm s}
\end{equation} 
with $f_{\rm geo}\equiv e_\parallel/\sqrt{q_\perp}$
\citep[][]{Sereno2010AC114}.
Since $r_\mathrm{s,2D}\equiv \sqrt{q_\perp}\xi_\mathrm{s}$ is the
geometric-mean scale radius in projection, 
the geometrical factor $f_\mathrm{geo}$ represents the degree of
correction due to the line-of-sight elongation of the cluster.
The halo mass, $\Mhalo$, can then be expressed as 
$\Mhalo=(4\pi/3)200\rho_{\rm c} (\chalo r_\mathrm{s,2D})^3 f_\mathrm{geo}$.
In this work, we employ the radial dependence of the projected
NFW profile $f_\mathrm{2D}(x)$ as given by  \citet{2000ApJ...534...34W}.
For $f_\mathrm{geo}=1$, this reduces to a projected  (circular or
elliptical) mass model.
An elliptical mass density model
can be described by ($\Mhalo, \chalo, \epsilon, \psi_\epsilon$)
\citep{Oguri2010LoCuSS,Umetsu+2012}. 

\subsection{Intracluster Gas}
\label{subsec:ICM}

Both observations and theory indicate that the ICM density is nearly
constant on a family of concentric, coaxial ellipsoids 
\citep{Kawahara2010,Buote+Humphrey2012p1,Buote+Humphrey2012p2}. 
Although modeling both the gas and matter distributions as ellipsoids with
constant axis ratios is not strictly valid for halos in HSE \citep[][]{Sereno2013glszx}, an
ellipsoidal approximation for the ICM is suitable when
systems with modest eccentricities are considered \citep{Lee+Suto2003}.

Following \citet{Sereno2013glszx}, we make a few simplifying but
non-informative working hypotheses to relate the matter and gas distributions.
First, we assume that the matter and gas distributions in the cluster
are ellipsoidal with constant but different axis ratios and co-aligned
with each other. 
Second, the two distributions are assumed to have the same degree of
triaxiality, that is, 
${\cal T}(q_a,q_b)={\cal T}^\mathrm{ICM}(q_a^\mathrm{ICM},q_b^\mathrm{ICM})$ with 
${\cal T}^\mathrm{ICM} \equiv (e_b^\mathrm{ICM}/e_a^\mathrm{ICM})^2=[1-(q_b^{\rm ICM})^2]/[1-(q_a^{\rm ICM})^2]$
and 
$q_a^\mathrm{ICM} \le q_b^\mathrm{ICM}$.
If two ellipsoids have the same degree of triaxiality, then the
misalignment angle between their major axes in the plane of the sky is
zero \citep{Romanowsky+Kochanek1998},
which is consistent with what has been observed in A1689
\citep{Sereno+Umetsu2011,Sereno+Ettori+Baldi2012}. 
If ${\cal T}={\cal T}^\mathrm{ICM}$, we have the following relation
for the ratio of eccentricities between ICM and matter \citep{Sereno2013glszx}:
\begin{equation}
\label{eq:eccentricity} 
e_a^\mathrm{ICM}/e_a=e_b^\mathrm{ICM}/e_b\equiv e^\mathrm{ICM}/e.
\end{equation}
The intracluster gas in HSE is rounder
than the underlying matter distribution: $e^\mathrm{ICM}/e\simeq 0.7$
\citep{Lee+Suto2003}.

With these assumptions, the number of independent axis ratios 
is reduced to three.  Here we use $q_a$, $q_b$, 
and $q_a^{\rm ICM}$ as free parameters. Hence, the intermediate--major
axis ratio $q_b^{\rm ICM}$ of the ICM
is determined by ${\cal T}(q_a,q_b)$ and $q_a^\mathrm{ICM}$:
\begin{equation}
 q_b^\mathrm{ICM}= \sqrt{1-\frac{1-(q_a^\mathrm{ICM})^2}{{\cal T}^2}}.
\end{equation}
Finally, as supported by both theory and observations,
we assume that the gas distribution is rounder than the matter
distribution:
$q_a \le q_a^\mathrm{ICM}$.

Under these hypotheses, the projected matter and gas distributions of
the cluster have different ellipticities 
($\epsilon\ne \epsilon^\mathrm{ICM}$) and elongations 
($e_\parallel \ne e_\parallel^\mathrm{ICM}$) but share the same orientation
of the projected major axis, $\psi_\epsilon = \psi_\epsilon^\mathrm{ICM}$.
There are a total of six parameters 
($q_a,q_b,q_a^{\rm ICM},\vartheta,\phi,\psi$)
needed to describe the intrinsic shape and
orientation of the cluster system, compared to four
observable geometric constraints, ($\epsilon,\epsilon^\mathrm{ICM},
\psi_\epsilon=\psi_\epsilon^\mathrm{ICM}, e_\parallel^\mathrm{ICM}$).

\subsection{Bayesian 3D Inversion}
\label{subsec:3Dinversion}

In our analysis, the cluster model $\bp$ is defined by seven {\em
fundamental} parameters describing the total matter ellipsoid 
and one parameter determining the shape of the ICM halo:
\begin{equation}
 \bp=(\Mhalo, \chalo, q_a,q_b, \vartheta,\phi,\psi, q_a^{\rm ICM}).
\end{equation}
Hence, the overall ellipsoidal model
has eight free parameters.
On the other hand, 2D lensing constraints reduce to four parameters
\citep{Sereno+Umetsu2011}, 
($\kappa_{\rm s}, \xi_{\rm s}, \epsilon, \psi_\epsilon$).
A joint X-ray and SZE analysis of the ICM yields two additional
constraints \citep{Sereno2013glszx},
namely the ellipticity $\epsilon^{\rm ICM}$ of the ICM in projection
and the elongation $e_\parallel^{\rm ICM}$ of the ICM along the
line of sight.  Accordingly, combined lensing and X-ray/SZE data sets
effectively provide six observationally accessible parameters,
\begin{equation}
 \bo=(\kappa_{\rm s}, \xi_{\rm s}, \epsilon, \psi_\epsilon,
  \epsilon^{\rm ICM}, e_\parallel^{\rm ICM}).
\end{equation}
That is, the problem is underconstrained.

To make robust inference on the intrinsic properties of the cluster,
we use a forward modeling approach with Bayesian inference 
for this underconstrained inversion problem \citep{Sereno2013glszx}.
The observational parameters $\bo=\bo(\bp)$ can be
uniquely specified by the intrinsic parameters $\bp$.
The total likelihood function of combined lensing and X-ray/SZE
observations 
can be formally written as \citep{Sereno2013glszx}
\begin{equation}
\label{eq:L_tot}
 {\cal L}[\bo(\bp)] 
= {\cal L}_{\rm GL} \times {\cal L}_{\rm ICM}
\end{equation}
with ${\cal L}_{\rm GL}$
the likelihood function of lensing observables 
and 
${\cal L}_{\rm ICM}$ that of X-ray/SZE observables. 

\subsection{Priors}
\label{subsec:priors}

For our base model,
we use uninformative priors for the intrinsic parameters $\bp$.
%
We adopt flat priors of $q_{\rm min}\le q_a\le 1$ and
$q_a\le q_b\le 1$ for the intrinsic axis ratios of the matter
distribution, 
where $q_{\rm min}$ is
introduced to exclude models with extremely small axis ratios because
such configurations would be dynamically unstable and not expected for
cluster halos. 
The probability functions can then be expressed as 
$P(q_a)=1/(1-q_{\rm min})$ for $q_{\rm min}\le q_a\le 1$ 
and
$P(q_b|q_a)=1/(1-q_a)^{-1}$
for $q_b\ge q_a$. In what follows, we fix $q_{\rm min}=0.1$
\citep{2005ApJ...632..841O,Sereno2013glszx}.
Alternatively, we may consider the axis-ratio priors that follow
distributions obtained from $\Lambda$CDM $N$-body simulations
\citep{2002ApJ...574..538J}. 

For the minor--major axis ratio of the ICM, we use a uniform
distribution in the interval $q_a\le q_a^{\rm ICM}\le 1$ (see Section
\ref{subsec:ICM}). The prior of $q_a^{\rm ICM}$, $P(q_a^{\rm ICM}|q_a)$
can then be defined in a similar way to that of $q_b$.
For the orientation angles, we consider a population of randomly
oriented halos with $P(\cos\vartheta)=1$ for $0\le \cos\vartheta\le 1$
and $P(\phi)=1/\pi$ for $-\pi/2\le \phi\le \phi/2$.
Finally, we employ uniform priors for the remaining parameters.

\section{Multi-probe Analysis of A1689}
\label{sec:results}

Here we apply the Bayesian inversion method outlined in Section
\ref{sec:3dmodel} to our multiwavelength observations of A1689. The
results are discussed in Section \ref{sec:discussion}.   

\subsection{Weak and Strong Lensing}
\label{subsec:wl+sl}

A full 2D lensing analysis is crucial for comparison with predictions of the properties of aspherical clusters
\citep{2005ApJ...632..841O}. 
In this work, we have employed free-form methods for both weak- and
strong-lensing mass reconstructions (Sections \ref{sec:wlana} and
\ref{sec:slana}), which provide a pixelated $\Sigma$ map and its
covariance matrix in each regime.

In this subsection, we derive constraints on the projected
halo properties (Section \ref{subsec:matter}) from lensing data.
We model the observed $\Sigma$ field with a projected ellipsoidal NFW 
profile (Section \ref{subsec:matter}), 
specified by 
$(\kappa_\mathrm{s}, \xi_\mathrm{s}, \epsilon, \psi_\epsilon)$. 
Additionally, we include the halo centroid $\btheta_\mathrm{c}$ as
parameters to conservatively account for the degree of miscentering.


\begin{figure}[!htb] 
\begin{tabular}{c}
 \includegraphics[width=0.45\textwidth,angle=0,clip]{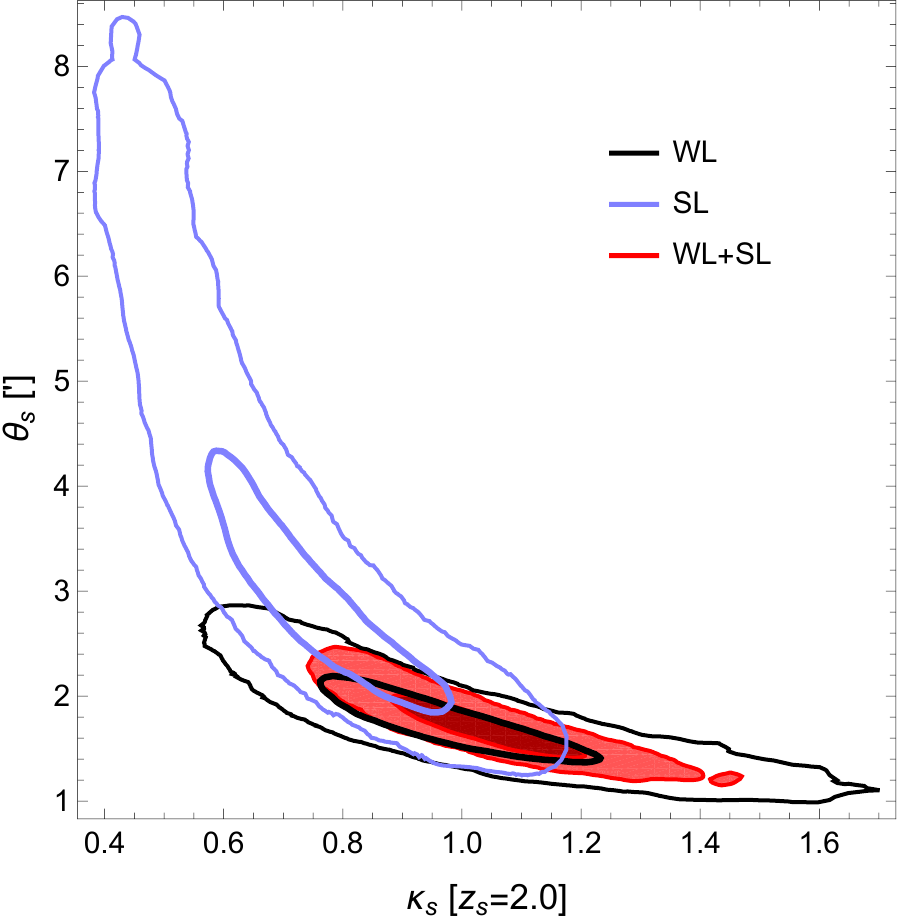}
\end{tabular}
\caption{
\label{fig:2DNFW}
Marginalized posterior distribution for the projected NFW parameters
($\kappa_\mathrm{s},\theta_\mathrm{s}$) 
obtained from three different lensing data sets
(see Table \ref{tab:2DNFW}) , namely
 weak-lensing-only (black; WL), strong-lensing-only (blue; SL), and combined weak and
 strong lensing (red shaded; WL+SL).
For each case, the contour levels are at $\exp(-2.3/2)$ and
 $\exp(-11.8/2)$ of the maximum, corresponding to the $1\sigma$ and
 $3\sigma$ confidence levels, respectively, for a Gaussian distribution.
The scale convergence
 $\kappa_\mathrm{s}=\Sigma_\mathrm{s}/\Sigma_\mathrm{c}$ is normalized
 to a fiducial source redshift of $z_\mathrm{s}=2$.
} 
\end{figure} 


\begin{deluxetable*}{ccccccc}
\tabletypesize{\footnotesize}
\tablecolumns{7} 
\tablecaption{
 \label{tab:2DNFW}
Parameters of the projected NFW model constrained from lensing observations
}  
\tablewidth{0pt}  
\tablehead{ 
 \multicolumn{1}{c}{Data\tablenotemark{a}} &
 \multicolumn{1}{c}{$\kappa_\mathrm{s}$\tablenotemark{b}} &
 \multicolumn{1}{c}{$\xi_\mathrm{s}$\tablenotemark{c}} &
 \multicolumn{1}{c}{$\epsilon$\tablenotemark{d}} & 
 \multicolumn{1}{c}{$\psi_\epsilon$\tablenotemark{e}} & 
 \multicolumn{1}{c}{$\btheta_\mathrm{c}$\tablenotemark{f}}
\\
 \colhead{} & 
 \colhead{} &
 \multicolumn{1}{c}{($\arcmin$)} &
 \colhead{} &
 \multicolumn{1}{c}{(deg)} &
 \multicolumn{1}{c}{($\arcsec$)} 
} 
\startdata
 WL    & $0.97\pm 0.16$ & $1.74\pm 0.27$ & $0.29\pm 0.07$ & $14.2\pm 8.4$ & $-1.2\pm 3.0,  4.9\pm 4.1$\\
 SL    & $0.73\pm 0.14$ & $3.00\pm 0.90$ & $0.27\pm 0.09$ & $13.0\pm 9.8$ & $-0.8\pm 1.9, -4.8\pm 2.3$\\
 GL    & $1.03\pm 0.11$ & $1.70\pm 0.20$ & $0.29\pm 0.05$ & $11.4\pm 4.9$ & $ 0.0\pm 1.3, -1.9\pm 1.4$
\enddata 
\tablenotetext{a}{WL: weak lensing shear and magnification; SL: strong lensing; GL: combined strong lensing, weak-lensing shear and magnification.}
\tablenotetext{b}{Scale convergence,
 $\kappa_\mathrm{s}=\Sigma_\mathrm{s}/\Sigma_\mathrm{c}$, normalized to
 a reference source redshift of $z_\mathrm{s}=2$.} 
\tablenotetext{c}{Projected scale radius of the elliptical NFW model
 measured along the major axis.} 
\tablenotetext{d}{Projected mass ellipticity, $\epsilon=1-q_\perp$, with
 $q_\perp$ the projected minor--major axis ratio.}
\tablenotetext{e}{Position angle of the major axis measured east of
 north.}
\tablenotetext{f}{Halo centroid position relative to the BCG position.}
\end{deluxetable*}

\subsubsection{Weak-lensing Data}
\label{subsubsec:wl}

The $\chi^2$ function for the Subaru weak-lensing observations
is expressed as \citep{2005ApJ...632..841O}
\begin{equation}
\label{eq:chi2wl}
 \chi^2_{\rm WL}=\sum_{m,n=1}^{\Npix}
\left[\Sigma(\btheta_m)-\hat{\Sigma}(\btheta_m)\right]
\left( C^{-1} \right)_{mn}
\left[\Sigma(\btheta_n)-\hat{\Sigma}(\btheta_n)\right],
\end{equation}
where $\bSigma=\{\Sigma(\btheta_m)\}_{m=1}^{\Npix}$ is 
the mass map from the 2D weak-lensing analysis (Section \ref{subsec:wl2d}),
$C^{-1}$ is the inverse of the error covariance matrix,
and the hat symbol denotes a modeled quantity. 
The corresponding likelihood is 
${\cal L}_\mathrm{WL}(\kappa_\mathrm{s}, \xi_\mathrm{s}, \epsilon,
\psi_\epsilon,\btheta_\mathrm{c}) 
\propto \exp(-\chi^2_\mathrm{WL}/2)$.

Figure \ref{fig:2DNFW} shows the results in terms of the marginalized
posterior distribution 
for the scale convergence,
$\kappa_\mathrm{s}=\Sigma_\mathrm{s}/\Sigma_{\rm c}$,
and the scale radius, $\theta_\mathrm{s}=\xi_s/D_\mathrm{l}$.
Table \ref{tab:2DNFW} summarizes marginalized constraints on the
individual parameters. 
In the present study, we employ the robust
biweight estimators of \citet{1990AJ....100...32B} for the central
location (mean) and scale (standard deviation) of the marginalized
posterior distributions
\citep[e.g.,][]{Sereno+Umetsu2011,Umetsu2014clash}.

\subsubsection{Strong-lensing Data}
\label{subsubsec:sl}

Mass maps derived from strong lensing exhibit a high degree
of correlation between adjacent regions.
The problem is exacerbated for parametric methods, 
which model the total mass distribution by a superposition of lens
components assuming parametric density profiles. 
This also persists in free-form modeling \citep{Lubini2014}, albeit to a 
lesser degree.

\begin{figure}[!htb] 
\begin{tabular}{c}
 \includegraphics[width=0.45\textwidth,angle=0,clip]{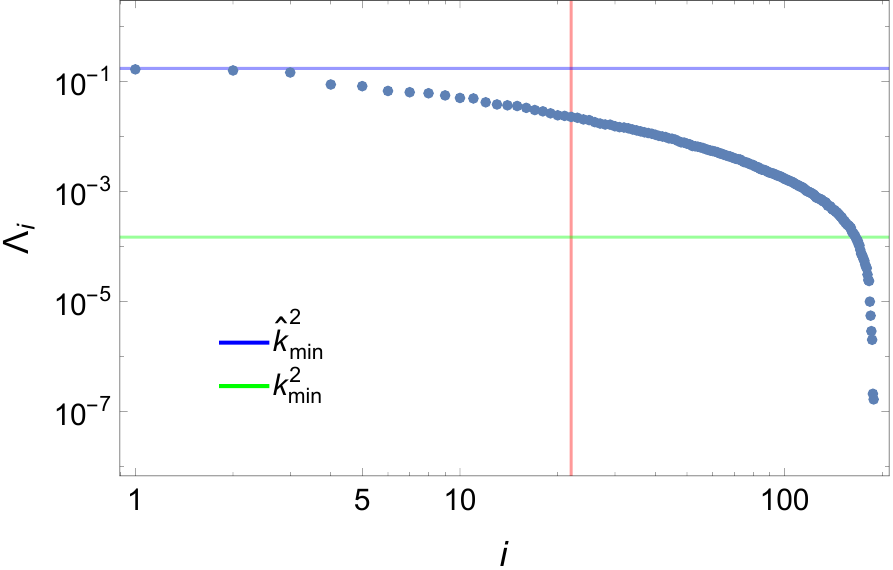}
\end{tabular}
\caption{
\label{fig:eigen}
Ordered eigenvalues $\mathsf{\Lambda}$ of the covariance matrix  for the
 {\sc PixeLens} mass reconstruction. The vertical red line indicates the
 maximum rank considered for our analysis, $N_\mathrm{max}=2N_\mathrm{im}$,
i.e., the 
 number of observational constraints on the image position. The blue
 horizontal line shows the minimum $\kappa^2$ value found in the   
 ensemble-averaged pixelated model.
The green horizontal line shows the minimum $\kappa^2$ value from the
 entire statistical ensemble of models generated by {\sc PixeLens}.
The results are shown for the covariance matrix as constrained by the
 systems 1, 5, and 11.
} 
\end{figure}

The degree of correlation can be examined by an eigenvalue analysis. 
Let us decompose the $C$ matrix as $C= U \Lambda U^{-1}$, 
with $\Lambda$ the diagonal matrix of eigenvalues
and $U$ the unitary matrix of eigenvectors. 
The first few eigenvalues describe the principal modes of variation of
the mass model \citep{Lubini2014,Mohammed2014}. Large eigenvalues
correspond to massive pixels, namely, those composing the inner part of
the mass distribution that is best constrained by strong lensing. The
ordered list of eigenvalues progressively decreases with increasing rank
and drops abruptly near the maximum rank, indicating a high degree
of correlation (Figure \ref{fig:eigen}).  

Here, we employ a regularization approach to conservatively account for 
the high degree of correlation of the covariance matrix.
This was first proposed by \citet{Umetsu+2012}
for the 1D analysis of strong-lensing mass profiles.
If the covariance matrix $C$ is not degenerate, we can construct a
$\chi^2$ function for each group of multiple images as  
\begin{equation}
 \begin{aligned}
\chi_{\mathrm{SL}, \alpha}^2 &=  
  \sum_{m,n}
  \left[
   \Sigma_{m} - \hat{\Sigma}_{m}
  \right]
 \left(
  C^{-1}
  \right)_{mn}
  \left[
  \hat{\Sigma}_{n} - \hat{\Sigma}_{n}
  \right], \\
 &=  \sum_{m} 
  \frac{
   \left[
      (\Sigma_U)_m - 
      (\hat{\Sigma}_U)_m
  \right]^2}{\Lambda_m},
  \end{aligned}
\end{equation}
where
$\Sigma_m$ is the observed $\Sigma$ value of the $m$th pixel,
$(\Sigma_U)_m=\sum_l U_{ml}\Sigma_{l}$ is the
projection onto the eigenbasis,
$\alpha$ runs over the four groups of images 
(Section \ref{sec:slana}), 
and the hat symbol is used to denote a modeled quantity.
Each group has its own $\bSigma$, $C$, $U$, and $\Lambda$.
Here we drop the index $\alpha$ on the right hand side
to simplify the notation.

In this approach, we limit ourselves to the principal modes 
and truncate the summation at $N_\mathrm{max}$ largest eigenvalues as
\begin{equation}
\chi_{\mathrm{SL},\alpha}^2 
 \approx  \sum_{m=1}^{N_{\mathrm{max}}} 
  \frac{
   \left[
      (\Sigma_{\mathsf{U}})_m - 
      (\hat{\Sigma}_{\mathsf{U}})_m
  \right]^2}{\Lambda_m}.
\end{equation}
A natural choice for $N_\mathrm{max}$ is the number of observational
constraints.
We thus set $N_\mathrm{max}=2N_\mathrm{im}$ with $N_\mathrm{im}$ the
number of multiple images used.
The total $\chi^2$ is given by
\begin{equation}
\label{eq:chi2sl}
\chi^2_\mathrm{SL}=\sum_\alpha \chi^2_{\mathrm{SL},\alpha}.
\end{equation}

We find that the eigenvalues before the drop range
approximately between the minimum $\kappa^2$ value in the
ensemble-averaged pixelated model and 
that found from the whole ensemble of models generated by {\sc
PixeLens} (Section \ref{subsec:sl2d}). 
This is demonstrated in Figure \ref{fig:eigen}. 
The $2N_\mathrm{im}$-th eigenvalue lies
approximately in the middle of this range and sets a conservative scale.
We checked the reliability and performance of this regularization method
using analytical models.   

Some multiple image systems share very similar configurations
(e.g., systems 1 and 2).
Such a redundancy is valuable for determining cosmological parameters
\citep{Lubini2014}, or for improving the sensitivity to local
substructures.
Assigning a full weight to systems having similar configurations would
inflate the relative contribution of strong lensing with respect to weak
lensing. 
To avoid this, we multiply $\chi^2_\mathrm{SL}$ by a weighting factor
$w_\mathrm{SL}$,
defined as the inverse of the geometrical average 
of the number of such redundant image systems.
We find $w_\mathrm{SL}=2/3$ for our analysis.
The likelihood is then defined as 
${\cal L}_\mathrm{SL}(\kappa_\mathrm{s}, \xi_\mathrm{s}, \epsilon,
\psi_\epsilon,\btheta_\mathrm{c}) 
\propto \exp(-w_\mathrm{SL} \chi^2_\mathrm{SL}/2)$. 

The results are summarized in Table \ref{tab:2DNFW} and Figure \ref{fig:2DNFW}.

\subsubsection{Combining Weak and Strong Lensing}
\label{subsubsec:wlsl}

We now combine the weak- and strong-lensing likelihoods
constructed in Sections \ref{subsubsec:wl} and \ref{subsubsec:sl},
respectively, to jointly constrain the projected NFW parameters. 
The likelihood function ${\cal L}_\mathrm{GL}$
for the combined weak plus strong lensing data
can be written as \citep{Sereno+Umetsu2011}
\begin{equation}
\label{eq:L_GL}
{\cal L}_\mathrm{GL} = {\cal L}_\mathrm{WL} \times {\cal L}_\mathrm{SL}
 \propto
\exp[-(\chi^2_\mathrm{WL}+ w_\mathrm{SL}\chi^2_\mathrm{SL})/2],
\end{equation}
where $\chi^2_\mathrm{WL}$ and $\chi^2_\mathrm{SL}$ are defined by
Equations (\ref{eq:chi2wl}) and (\ref{eq:chi2sl}), respectively.

Figure \ref{fig:2DNFW} shows that the scale radius ($\theta_\mathrm{s}$)
and the scale convergence ($\kappa_\mathrm{s}$) are highly degenerate
and anti-correlated.  In particular, the scale radius is poorly
constrained by strong lensing alone because of the limited coverage of
multiple images, $\theta\simlt 1.1\arcmin$ (Section \ref{subsec:sl2d}). 
The allowed range of $\theta_\mathrm{s}$ lies well outside the region
where the multiple images are observed. Thus, the inference of
parameters by strong lensing requires an extrapolation well beyond the
observed region.  
For this reason, in the present study, we do not consider
strong-lensing-only triaxial modeling (see Table \ref{tab:3DNFW}).
On the other hand, since the posterior distributions
from the independent weak-lensing and strong-lensing analyses are
compatible, combining weak lensing with strong lensing 
provides improved parameter constraints (Table \ref{tab:2DNFW}).

\subsection{Combined X-ray plus SZE Analysis}
\label{subsec:szx}


\begin{deluxetable}{lcc}
\tabletypesize{\footnotesize}
\tablecolumns{3} 
\tablecaption{
 \label{tab:Ycomp}
 Integrated Comptonization $Y$ parameter measured interior to a cylinder of radius $r$.
}  
\tablewidth{0pt}
\tablehead{ 
 \multicolumn{1}{c}{Instrument} &
 \multicolumn{1}{c}{$r$} &
 \multicolumn{1}{c}{$Y(<r)$} 
\\
 \colhead{} & 
 \multicolumn{1}{c}{($\arcmin$)} &
 \multicolumn{1}{c}{($10^{-10}$\,sr)} 
} 
\startdata
	BIMA/OVRO&	1.5&	$1.00 \pm 0.28$\\
	BIMA/OVRO&	3.0&    $2.64 \pm 0.97$\\
	SZA&		1.5&    $1.11 \pm 0.10$\\
	SZA&		3.0&    $2.83 \pm 0.42$\\
	SZA&		4.5&    $4.33 \pm 0.81$\\
	SZA&	 	6.0&	$5.50 \pm 1.18$
\enddata 
\end{deluxetable}

With a known halo geometry (e.g., sphericity) and under the ideal gas
assumption, the thermodynamic quantities of the ICM are overconstrained
by X-ray and SZE data. This is because the thermal pressure can be
independently determined from thermal SZE data and  X-ray
spectroscopy/imaging data. We can therefore relax the assumption of
spherical symmetry to solve for the elongation of the ICM
distribution \citep{Sereno+Ettori+Baldi2012}.
Combining gravitational lensing and X-ray/SZE observations 
with minimal geometric assumptions (Section \ref{subsec:ICM}) allows
us to break the degeneracy between mass and elongation for the total
matter distribution \citep{Sereno2013glszx}. 
Such a multi-probe approach based on lensing and X-ray/SZE data
is free from the assumption of HSE,
compared to the lensing plus X-ray analysis, which relies on equilibrium
assumptions between the gravitational potential and pressure components  
\citep[see][]{Limousin2013}.

In our multi-probe approach, 
the ICM distribution is modelled with an ellipsoidal
parametric profile which can fit X-ray surface-brightness ($S_X$) and
temperature ($T_X$) distributions.
Comparison with the SZE amplitude then determines
the elongation $e_\parallel^\mathrm{ICM}$
For example, for an isothermal plasma \citep{DeFilippis2005}, we have
\begin{equation}
 1/e_\parallel^\mathrm{ICM}
\propto
D_\mathrm{l}
\frac{S_X}{\Delta T_\mathrm{SZE}^2}\frac{T_X^2}{\Lambda_X}
\end{equation}
with $\Delta T_\mathrm{SZE}$ the SZE temperature decrement and
$\Lambda_X$ the X-ray cooling function of the ICM.
In this work, we rely on the X-ray data to constrain the ICM morphology
in projection space; we use aperture-integrated constraints on the SZE
signal (Table \ref{tab:Ycomp}) to determine the line-of-sight elongation
$e_\parallel^\mathrm{ICM}$. 

Our X-ray data are taken from \citet{Sereno+Ettori+Baldi2012}, who
performed an X-ray analysis on {\em Chandra} and {\em XMM-Newton}
observations. 
Here we briefly summarize essential results needed for this study. For
details, we refer to \citet{Sereno+Ettori+Baldi2012}. 
\citet{Sereno+Ettori+Baldi2012} showed that
exposure corrected and point-source removed
{\em Chandra} X-ray images in the 0.7--2.0\,keV band 
are well described by concentric ellipses with ellipticity
$\epsilon^X=0.15\pm 0.03$ and orientation angle
$\psi_\epsilon^X=(12\pm 3)$\,degrees measured east of north.
Following \citet{Sereno+Ettori+Baldi2012} and \citet{Sereno2013glszx},
we model the 3D electron density in the intrinsic coordinate system with
the following parametric form
\citep{2006ApJ...640..691V,Ettori2009fgas}:
\begin{equation}
\label{eq:ne3d}
 n_e= n_0\left[
1+\left(\frac{R}{r_{\rm c}}\right)^2
\right]^{-3\beta/2}
\left[
1+\left(\frac{R}{r_{\rm t}}\right)^2
\right]^{-\gamma/3},
\end{equation}
where $n_0$ is the central
electron density, $r_{\rm c}$ is the ellipsoidal core radius, 
$r_{\rm t} (>r_{\rm c})$ is the ellipsoidal truncation radius,
$\beta$ is the slope in the intermediate density regions,
and
$\gamma$ is the outer slope.
The 3D gas density is parametrized as
\citep{Sereno2013glszx}
\begin{equation}
\label{eq:te3d}
 T = \frac{T_0}
{\left[ 1+\left(R/r_T\right)^2 \right]^{0.45}},
\end{equation}
where $T_0$ is the central gas temperature, and $r_T$ describes a
temperature decline at large cluster radii.
The parametrizations of Equations (\ref{eq:ne3d}) and (\ref{eq:te3d})
were motivated by the absence of cool-core features in our data. For
further justification, see Section 5 of \citet{Sereno+Ettori+Baldi2012}.

\begin{figure}[!htb] 
 \begin{center}
  \includegraphics[width=0.45\textwidth,angle=0,clip]{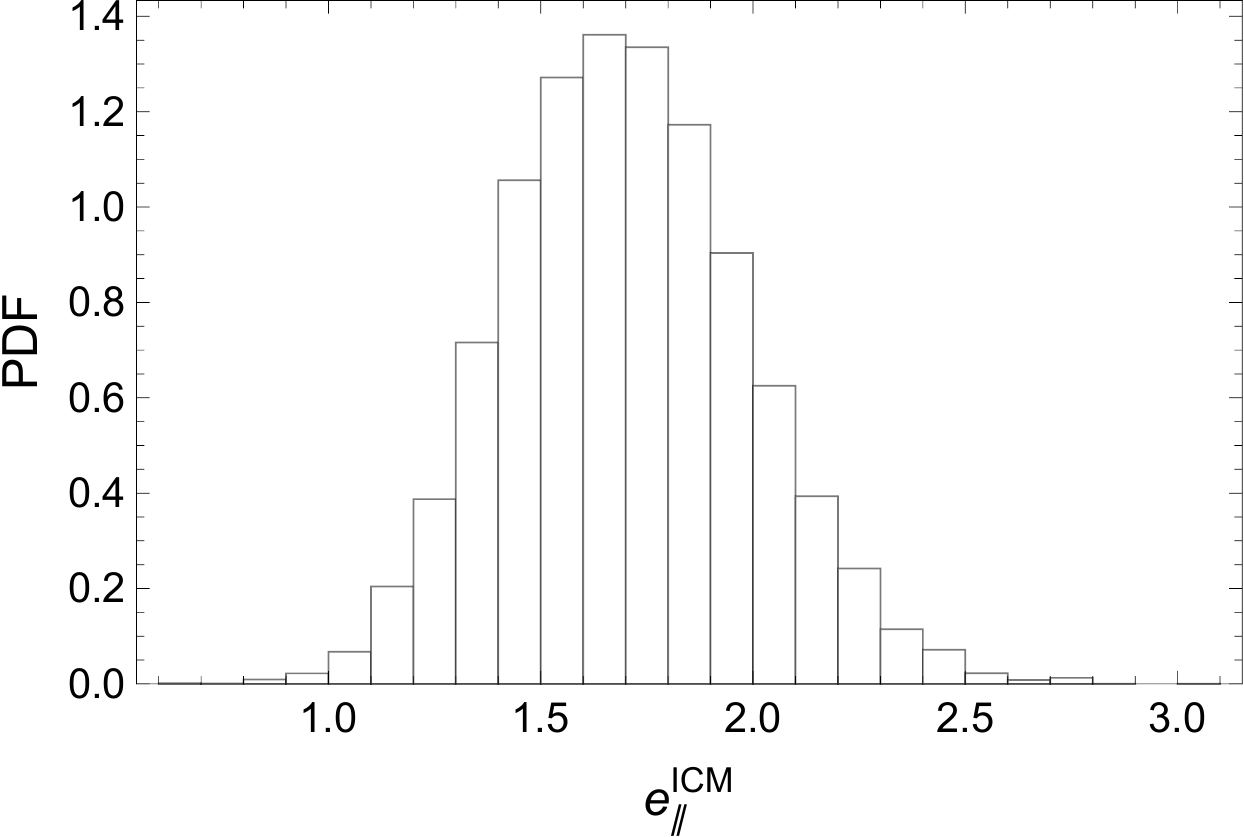}
 \end{center}
\caption{
\label{fig:elong}
Marginalized posterior probability distribution of the elongation
 $e^\mathrm{ICM}_\parallel$ as derived from the combined X-ray plus SZE 
 analysis (Section \ref{subsec:szx}). 
} 
\end{figure} 

The thermal SZE provides a complementary measure of the thermal energy
content in a cluster. 
In this study, we perform a self-consistent multi-scale analysis of 
high-significance 30\,GHz interferometric SZE observations of A1689 obtained with 
the Berkeley-Illinois-Maryland Array (BIMA), the Owens Valley Radio
Observatory (OVRO), and the Sunyaev-Zel'dovich Array (SZA).  
The BIMA and OVRO observations of A1689 are presented in 
\cite{LaRoque2006}, while the SZA observations of A1689 are presented in \cite{Gralla2011}.
Owing to the different scales probed by the instruments, we fit the
OVRO/BIMA and SZA data separately using the spherical \cite{Arnaud2010}
pressure profile. This profile is an adaptation  
of the generalized NFW pressure profile first proposed by
\cite{Nagai2007}, and first  
fitted to SZE observations in \cite{Mroczkowski2009}.
A joint fit to the OVRO, BIMA, and SZA data was also performed to
determine the best-fit SZE centroid reported in Table
\ref{tab:cluster}. 

As in \cite{Mroczkowski2009}, a model for the cluster and contaminating radio
sources is computed in the image plane, then Fourier transformed for comparison
to the interferometric data.  The best-fit model and $1\sigma$ confidence intervals
are determined using a Markov chain Monte Carlo (MCMC) procedure.
The OVRO and BIMA data measure radial scales from $0.5\arcmin$--$4\arcmin$, while
the SZA data probe radial scales from $1\arcmin$--$6\arcmin$. 
\cite{Bonamente2012} showed that the adoption of the \cite{Arnaud2010}
profile versus other non-isothermal pressure profiles accurate out to
$r_\mathrm{500c}$ does not significantly impact the parameters derived
from the fits when the radii for which the results are computed are at
scales accessible to the instruments.

A summary of the SZE data used is given in Table \ref{tab:Ycomp}.
The integrated Comptonization parameter $Y(<r)$ 
interior to a cylinder of radius $r$ is written in terms of the
electron density and temperature profiles (Equations (\ref{eq:ne3d}) and
(\ref{eq:te3d})) as
\begin{equation}
\label{eq_sze1}
Y = \frac{\sigma_{\rm T} k_{\rm B} }{m_{e} c^2}
\int_{\Omega_r}\!d\Omega \int\!dl\,n_e T
\end{equation}
with $\sigma_\mathrm{T}$ the Thomson cross section,
$k_\mathrm{B}$ the Boltzmann constant,
$m_e$ the electron mass,
and
$c$ the speed of light in vacuum;
$\Omega_r$ is the solid angle of the integration aperture.

The model profiles given by Equations (\ref{eq:ne3d}), (\ref{eq:te3d}), and
(\ref{eq_sze1}) are then compared with combined X-ray surface brightness
($S_X$), X-ray spectroscopic temperature ($T_X$), and thermal SZE
decrement ($Y$) observations.  
Briefly summarizing, the X-ray surface brightness profile 
$\{S_{X,i}\}_{i=1}^{N_S}$
observed by {\em Chandra} was extracted from $N_S=68$
elliptical annuli out to an elliptical radius of
$\xi=900$\,kpc\,$h_{70}^{-1}$ ($\sim 5\arcmin$),
and the {\em XMM-Newton} temperature profile $\{T_{X,i}\}_{i=1}^{N_T}$
was measured in $N_T=5$
elliptical annual bins out to $\xi=900$\,kpc\,$h_{70}^{-1}$ 
\citep[][]{Sereno+Ettori+Baldi2012}. 
Thanks to the improved SZE analysis, the $Y$ parameter is measured at
several apertures from BIMA/OVRO and SZA data as summarized in Table
\ref{tab:Ycomp}. 
We find good consistency between the BIMA/OVRO and SZA
results at $r=1.5\arcmin$ and $3\arcmin$ where these independent data
overlap. At an integration radius of $r=3\arcmin$, our results are also
in excellent agreement with $Y(<3\arcmin)=(2.5\pm 0.6)\times
10^{-10}$\,sr from 94\,GHz interferometric observations with the
7-element AMiBA \citep[][their Table 5]{Umetsu+2009}. 

The X-ray part of the $\chi^2$ function
can be written as \citep{Sereno+Ettori+Baldi2012}
\begin{equation}
 \label{eq:chi2_X}
  \chi^2_X 
   = \sum_{i=1}^{N_{S}}
  \left(\frac{S_{X,i}-\hat{S}_{X,i}}{\sigma_{S,i}}\right)^2 
  +
 \sum_{i=1}^{N_T} \left(\frac{T_{X,i}-\hat{T}_{X,i}}{\sigma_{T,i}}\right)^2
\end{equation}
with ($\hat{S}_X, \hat{T}_X$) model predictions for the corresponding
X-ray observables and ($\sigma_S,\sigma_T$) their corresponding errors.

The $\chi^2$ function for the SZE observations is written as
\begin{equation}
\label{eq:chi2sze}
\chi_\mathrm{SZE}^2 = \sum_j \sum_i
\left( 
 \frac{\Delta Y_{ji} -\hat{\Delta Y}_i}
      {\sigma_{\Delta,ji}} 
\right)^2,
\end{equation}
where $\Delta Y_{ji}$ is the differential $Y$ parameter
for the $j$th instrument (BIMA/OVRO or SZA)
in the $i$th annular ring,
$\Delta Y_{ji}\equiv Y_j(<r_{i+1})-Y_j(<r_{i})$,
and $\sigma_{\Delta,ji}$ is its $1\sigma$ uncertainty.
The $Y$ values are sampled at every $1.5\arcmin$ (Table
\ref{tab:Ycomp}), which is sufficiently larger than the synthesized
beam. 
Hence, differential $\Delta Y$measurements in adjacent annuli are
approximately uncorrelated given the annulus size considered.

A combined analysis of the X-ray and SZE data is performed 
using the combined function $\chi^2 = \chi_X^2+\chi_\mathrm{SZE}^2$.
The parameter space is explored using an MCMC approach as described in
\citet{Sereno2013glszx}. 
Since parameter constraints on the $n_e$ and $T$ models are dominated
by the {\em Chandra} surface brightness and {\em XMM-Newton} temperature
data, respectively, we find our results are fully
consistent with those of \citet{Sereno+Ettori+Baldi2012} based on the
same X-ray data.
The best-fit central temperature 
\citep[$T_0=9.8\pm 0.2$\,keV,][]{Sereno+Ettori+Baldi2012} is in good
agreement with the {\em Suzaku} X-ray results of
\citet{Kawaharada+2010}.
On the other hand, using the improved SZE data, we obtain tighter
constraints on the elongation $e^\mathrm{ICM}_\parallel$. The resulting
posterior distribution of  
$e_\parallel^\mathrm{ICM}$ is shown in Figure \ref{fig:elong}. 
The posterior mean and standard deviation are
$e_\parallel^\mathrm{ICM}=1.70\pm 0.29$. 


\begin{deluxetable*}{lcccccc}
\tabletypesize{\footnotesize}
\tablecolumns{7} 
\tablecaption{
 \label{tab:3DNFW}
Intrinsic parameters of the total matter distribution
obtained using different data sets and different priors
}  
\tablewidth{0pt}  
\tablehead{ 
 \multicolumn{1}{c}{Data\tablenotemark{a}} &
 \multicolumn{1}{c}{Prior} &
 \multicolumn{1}{c}{$\Mhalo$} &
 \multicolumn{1}{c}{$\chalo$} &
 \multicolumn{1}{c}{$q_a$} & 
 \multicolumn{1}{c}{$q_b$} & 
 \multicolumn{1}{c}{$\cos\vartheta$\tablenotemark{b}}
\\
 \colhead{} & 
 \colhead{} &
 \multicolumn{1}{c}{($10^{15}M_\odot h^{-1}$)} &
 \colhead{} &
 \colhead{} &
 \colhead{} &
 \colhead{} 
} 
\startdata 
 WL            & Spherical & $1.31\pm 0.11$ & $8.87\pm 1.11$  & $1$            & $1$            & --- \\
 WL            & Flat      & $1.28\pm 0.26$ & $10.70\pm 2.85$ & $0.39\pm 0.18$ & $0.77\pm 0.15$ & $0.54\pm 0.29$ \\
 WL            & $N$-body  & $1.22\pm 0.23$ & $9.15\pm 1.77$  & $0.47\pm 0.08$ & $0.66\pm 0.12$ & $0.60\pm 0.30$ \\
 SL            & Spherical & $1.79\pm 0.31$ & $8.69\pm 1.26$  & $1$            & $1$            & --- \\
 GL         & Spherical & $1.32\pm 0.09$ & $10.10\pm 0.82$ & $1$            & $1$            & --- \\
 GL         & Flat      & $1.49\pm 0.25$ & $10.30\pm 2.52$ & $0.45\pm0.20$  & $0.77\pm 0.14$ & $0.47\pm 0.29$ \\
 GL         & $N$-body  & $1.41\pm 0.19$ & $ 9.65\pm 1.54$ & $0.47\pm0.08$  & $0.66\pm 0.12$ & $0.60\pm 0.29$ \\
 WL + X/SZ     & Flat      & $1.21\pm 0.19$ & $7.91\pm 1.41$  & $0.39\pm 0.16$ & $0.56\pm 0.20$ & $0.93\pm 0.06$ \\
 WL + X/SZ     & $N$-body  & $1.16\pm 0.17$ & $7.42\pm 1.21$  & $0.40\pm 0.08$ & $0.52\pm 0.12$ & $0.94\pm 0.05$ \\
 GL + X/SZ  & Flat      & $1.24\pm 0.16$ & $8.36\pm 1.27$  & $0.39\pm 0.15$ & $0.57\pm 0.19$ & $0.93\pm 0.06$ \\
 GL + X/SZ  & $N$-body  & $1.20\pm 0.13$ & $7.89\pm 0.96$  & $0.40\pm 0.08$ & $0.52\pm 0.12$ & $0.94\pm 0.05$ 
\enddata 
\tablecomments{Intrinsic parameters of the total matter distribution of A1689
 derived from a triaxial analysis of multiwavelength data sets, using spherical, flat, and $N$-body priors on the distribution of axis ratios ($q_a,q_b$).
}
\tablenotetext{a}{WL: weak-lensing shear and magnification; SL: strong
 lensing; GL: combined strong lensing, weak-lensing shear and magnification;  X/SZ: combined X-ray and SZE measurements.
}
\tablenotetext{b}{Cosine of the angle between the major axis and the line of sight.}
\end{deluxetable*}


\begin{deluxetable}{lccc}
\tabletypesize{\footnotesize}
\tablecolumns{4} 
\tablecaption{
 \label{tab:ICM}
Intrinsic shapes of the ICM distribution
}  
\tablewidth{0pt}  
\tablehead{ 
 \multicolumn{1}{c}{Priors} &
 \multicolumn{1}{c}{$q_a^\mathrm{ICM}$} &
 \multicolumn{1}{c}{$q_b^\mathrm{ICM}$} &
 \multicolumn{1}{c}{$e^\mathrm{ICM}/e$}  
} 
\startdata 
 Flat      & $0.60\pm 0.14$ & $0.70\pm 0.16$ & $0.87\pm 0.07$\\
 $N$-body  & $0.58\pm 0.10$ & $0.65\pm 0.11$ & $0.89\pm 0.06$ 
\enddata 
\tablecomments{
Constraints on the intrinsic axis ratios
 ($q_a^\mathrm{ICM}, q_b^\mathrm{ICM}$) of the ICM distribution
 and the relation with the total matter distribution ($e^\mathrm{ICM}/e$),
 obtained from the full triaxial analysis of combined weak/strong-lensing and X-ray/SZE data sets (Section \ref{sec:results}).
 $q_b^\mathrm{ICM}$ and $e^\mathrm{ICM}/e$ are derived parameters.
 }
\end{deluxetable}

\subsection{Multi-probe Deprojection}
\label{subsec:full}

Here we perform joint likelihood analyses of combined lensing and
X-ray/SZE data, using different combinations of lensing data sets
(Section \ref{subsec:wl+sl}). 

The likelihood ${\cal L}_\mathrm{ICM}$ of the X-ray/SZE data is written
in terms of two observable ICM parameters (Section
\ref{subsec:3Dinversion}), namely 
the ellipticity $\epsilon^\mathrm{ICM}$ and 
line-of-sight elongation $e_\parallel^\mathrm{ICM}$ of the ICM.
Following \citet{Sereno+Ettori+Baldi2012,Sereno2013glszx},
we include a nuisance parameter $\Delta
e_\parallel^{\rm sys}$ that quantifies the additional 
uncertainty on $e_\parallel^{\rm ICM}$, 
accounting for potential
calibration systematics in the X-ray/SZE measurements.
It is assumed to follow a normal distribution 
with zero mean and standard deviation
$\sigma_\parallel^\mathrm{sys}=0.07$.
Since the systematic uncertainty is quite small compared to the width of
the marginalized posterior distribution $P(e_\parallel^\mathrm{ICM})$
(Figure \ref{fig:elong}),
the impact on 
the final results is minor.
The X-ray/SZE part of the likelihood ${\cal
L}_\mathrm{ICM}(\epsilon^\mathrm{ICM}, e_\parallel^\mathrm{ICM}; \Delta
e_\parallel^\mathrm{sys})$ is written as \citep{Sereno2013glszx}
\begin{equation}
\label{eq:L_ICM}
 \begin{aligned}
  {\cal L}_\mathrm{ICM}
&=
  \frac{1}{\sqrt{2\pi}\sigma_{\epsilon,X}}
  \exp\left[-\frac{(\epsilon^X-\epsilon^\mathrm{ICM})^2}{2\sigma^2_{\epsilon,X}}\right]\\
&\times P(e_\parallel^\mathrm{ICM}-\Delta e_\parallel^\mathrm{sys})\\
&\times   \frac{1}{\sqrt{2\pi}\sigma_\parallel^\mathrm{sys}}\exp\left[-\frac{1}{2}
  \left(
    \frac{\Delta
  e_\parallel^\mathrm{sys}}{\sigma_\parallel^\mathrm{sys}}
  \right)^2
 \right],
 \end{aligned}
\end{equation}
where $\epsilon^X$ and $\sigma_{\epsilon,X}$ are the measured value of
the ICM ellipticity and its uncertainty, respectively (Section
\ref{subsec:szx}). 

To perform a joint analysis with the X-ray/SZE data,
we consider three different likelihood functions for the lensing part,
namely, 
${\cal L}_\mathrm{WL}$,
${\cal L}_\mathrm{SL}$,
and
${\cal L}_\mathrm{GL}={\cal L}_\mathrm{WL}{\cal L}_\mathrm{SL}$,
which are all functions of the projected NFW parameters
$\kappa_\mathrm{s},\xi_\mathrm{s},\epsilon,\psi_\epsilon$, and
$\btheta_\mathrm{c}$. 
Following \citet[][]{Sereno2013glszx}, we exploit
constraints from the X-ray analysis about the gas centroid $\btheta_\mathrm{c}^X$
and position angle $\psi_\epsilon^X$
(Section \ref{subsec:szx}), which are used as priors for the
centroid $\btheta_\mathrm{c}$
and position angle $\psi_\epsilon$
of the underlying halo
\citep[see Section 4 of][]{Sereno2013glszx}.
These priors are consistent with the 
geometric assumptions we have made in Section \ref{subsec:ICM}.

For our base model, we use
flat priors for the intrinsic axis ratios of the underlying
halo (Section \ref{subsec:priors}). 
We also consider an alternative prior distribution predicted
by cosmological $N$-body simulations of \citet{2002ApJ...574..538J}.
For details, we refer to \citet{Sereno+Umetsu2011} and
\citet{Sereno2013glszx}. 

\begin{figure*}[!htb] 
\begin{tabular}{c}
 \includegraphics[width=0.9\textwidth,angle=0,clip]{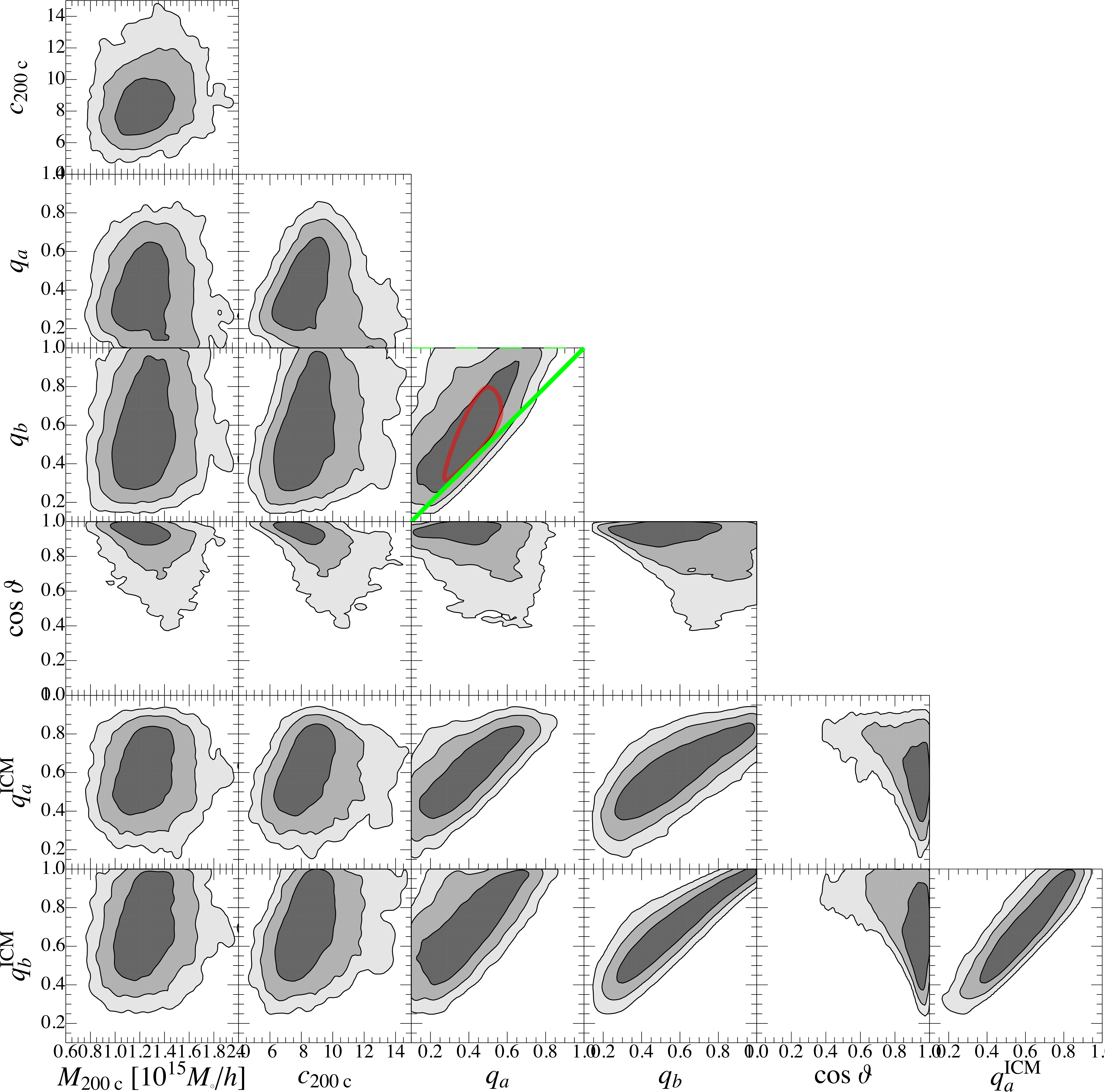}
\end{tabular}
\caption{
\label{fig:fullPDF}
Marginalized posterior distributions for the intrinsic parameters of the
 triaxial cluster model obtained from a joint analysis of the
 weak/strong-lensing and X-ray/SZE data.
In each panel, the contours levels are shown at $\exp(-2.3/2)$,
 $\exp(-6.17/2)$, and $\exp(-11.8/2)$ of the maximum, corresponding to
 the $1\sigma$, $2\sigma$, and $3\sigma$ confidence levels,
 respectively, for a Gaussian distribution.
In the $q_b$ versus $q_a$ plane, the green solid (diagonal) and dashed
 (horizontal) lines represent prolate ($q_a=q_b$) and oblate ($q_b=1$)
 configurations, 
 respectively, and the thick red line shows the
 $1\sigma$ contour for the axis-ratio distribution in $\Lambda$CDM
 $N$-body simulations of \citet{2002ApJ...574..538J}.  
} 
\end{figure*}

\section{Results and Discussions}
\label{sec:discussion}
 
\begin{figure}[!htb] 
 \begin{center}
  \includegraphics[width=0.45\textwidth,angle=0,clip]{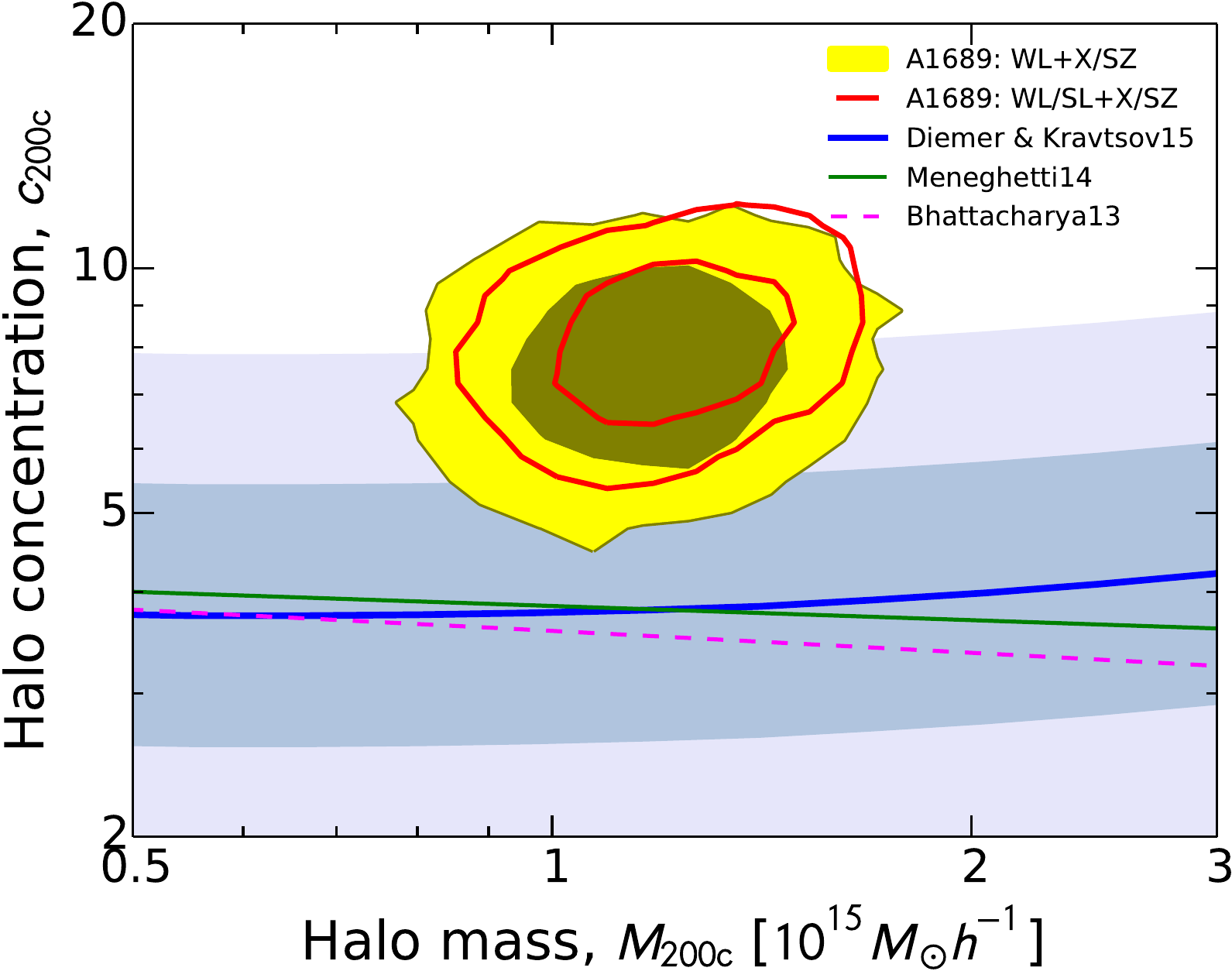}
 \end{center}
\caption{
\label{fig:cM}
Marginalized constraints on the ellipsoidal NFW model parameters
 ($\Mhalo,\chalo$) for A1689 compared to the $c$--$M$ relations 
 predicted for the full population of halos in $\Lambda$CDM cosmological
 simulations 
 \citep{Bhatt+2013,Meneghetti2014clash,Diemer+Kravtsov2015}. 
 The yellow shaded regions show the results from weak lensing combined with
 X-ray/SZE data.
The red contours are from the full analysis of weak/strong-lensing and
 X-ray/SZE data. 
For each case, the contours show the 68.3\% and 95.4\% confidence levels
 in the $c$--$M$ plane.
The light blue areas show the $1\sigma$ and $2\sigma$ ranges
 of intrinsic halo concentrations  (with a 68\% scatter of 0.16 dex), respectively, as
 obtained by \citet{Diemer+Kravtsov2015}. 
All model predictions are evaluated at the cluster redshift
 $z_\mathrm{l}=0.183$. 
Overall, the inferred range of $\chalo$ is high but
 overlaps with the $\sim 2\sigma$ tail of the predicted distribution for
 high-mass cluster halos.
} 
\end{figure} 

The resulting constraints on the intrinsic parameters for the underlying 
halo ($\Mhalo,\chalo,q_a,q_b,\cos\vartheta$) are given in Table
\ref{tab:3DNFW}, for different combinations of data sets and three
different priors on the axis-ratio distribution:  
(1) spherical prior ($q_a=q_b=1$);
(2) flat distribution of axis ratios and random distribution of halo
orientations (Section \ref{subsec:priors}); 
(3) $N$-body $\Lambda$CDM predictions \citep{2002ApJ...574..538J}.
The baseline results for the combined weak/strong-lensing and X-ray/SZE
analysis obtained with flat priors are shown in Figure \ref{fig:fullPDF}.
Table \ref{tab:ICM} gives a summary of our baseline constraints on the
intrinsic axis ratios of the ICM halo,
($q_a^\mathrm{ICM},q_b^\mathrm{ICM}$), and on the ICM-to-matter ratio of
halo eccentricities, $e^\mathrm{ICM}/e$.
Table \ref{tab:cMtable} lists the published ($\Mhalo,\chalo$)
measurements for A1689 based on the combination of both weak and strong
lensing.  
For previous compilations, see
\citet[][their Table A1]{2007MNRAS.379..190C}
\citet[][their Table 4]{2007ApJ...668..643L},
\citet[][their Table 5]{UB2008},
\citet[][their Table 4]{Corless2009triaxial}, and
\citet[][their Table 2]{Coe+2010}.

\subsection{Mass and Concentration}

\begin{deluxetable*}{ccccc}
\tabletypesize{\footnotesize}
\tablecolumns{6} 
\tablecaption{
 \label{tab:cMtable}
Published mass and concentration measurements of A1689 from combined
 weak and strong lensing
}  
\tablewidth{0pt}  
\tablehead{ 
 \multicolumn{1}{c}{Author} &
 \multicolumn{1}{c}{$\Mhalo$} &
 \multicolumn{1}{c}{$\chalo$} &
 \multicolumn{1}{c}{Prior\tablenotemark{a}} &
 \multicolumn{1}{c}{External data\tablenotemark{b}}
\\
 \colhead{} &
 \multicolumn{1}{c}{($10^{15}M_\odot h^{-1}$)} &
 \colhead{} &
 \colhead{} &
 \colhead{} 
} 
\startdata 
Spherical modeling\\
\citet{BTU+05}              & $1.20\pm 0.13$ & $10.9^{+1.1}_{-0.9}$ & Spherical & ----\\ 
\citet{2006MNRAS.372.1425H} & $1.58\pm 0.14$ & $7.6\pm 0.5$         & Spherical & ----\\ 
\citet{UB2008}\tablenotemark{c}              & $1.30\pm 0.11$ & $10.1^{+0.8}_{-0.7}\pm 2.2$ &  Spherical & ----\\ 
\citet{Coe+2010}            & $1.3^{+0.3}_{-0.2}$ & $9.2\pm 1.2$    & Spherical & ----\\ 
This work                   & $1.32\pm 0.09$ & $10.10\pm 0.82$      & Spherical & ----\\
\hline
Triaxial modeling\\
\citet{2005ApJ...632..841O}\tablenotemark{d} & $1.14^{+0.26}_{-0.51}$ & $13.6^{+1,8}_{-10.5}$ & Flat & ----\\ 
\citet{Sereno+Umetsu2011}                    & $1.07\pm 0.23$         & $9.3\pm 2.0$          & Flat & ----\\ 
This work                                    & $1.49\pm 0.25$         & $10.30\pm 2.52$       & Flat & ----\\ 
\hline
With line-of-sight information\\
\citet{Corless2009triaxial} & $0.83\pm 0.16$ & $12.2\pm 6.7$  & Flat + $\cos\vartheta$ & ----\\ 
\citet{Sereno+Umetsu2011}   & $0.99\pm 0.17$ & $7.7\pm 1.1$   & Flat + $\cos\vartheta$ & ----\\
\citet{Morandi2011A1689}  & $1.81\pm 0.06$ & $5.71\pm 0.47$ & Flat                   & X-ray\\ 
\citet{Sereno2013glszx}     & $0.93\pm 0.12$ & $7.8\pm 0.7$   & Flat                   & X-ray/SZE\\
This work                   & $1.24\pm 0.16$ & $8.36\pm 1.27$ & Flat                   & X-ray/SZE
\enddata 
\tablecomments{The results based on the combination of both weak and
 strong lensing are summarized (converted from quoted values assuming an NFW density profile if necessary).
}
\tablenotetext{a}{Spherical: spherical prior on the intrinsic
 axis-ratios;
 Flat: flat prior on the intrinsic axis ratios; 
 $\cos\vartheta$: $\Lambda$CDM-like prior on the biased orientation
 of strong-lensing cluster halos \citep{Corless2009triaxial}.
}
\tablenotetext{b}{External data sets used in combination with lensing for constraining the line-of-sight elongation.}
\tablenotetext{c}{The weak-lensing mass map of \citet{UB2008} was used
 in the triaxial analyses by \citet{2005ApJ...632..841O}, \citet{Sereno+Umetsu2011}, \citet{Morandi2011A1689}, and \citet{Sereno2013glszx}.}
\tablenotetext{d}{NFW-equivalent of triaxial model parameters from \citet{2005ApJ...632..841O}.}
\end{deluxetable*}

\subsubsection{Spherical Modeling}
\label{subsubsec:sph}

The degree of concentration of A1689 has been a subject of controversy.  
Here we first compare the results obtained assuming a spherical NFW halo
(Table \ref{tab:3DNFW}) to those of previous work.
Our full 2D weak-lensing analysis based on Subaru $BVR_\mathrm{C}i'z'$
data yields
a projected concentration of 
$\chalo=8.9\pm 1.1$ 
($\cvir=11.2\pm 1.4$) at
$\Mhalo=(1.31\pm0.11)\times \Munit$.
This is in excellent agreement with, and improved from, our earlier  
weak-lensing work:
$\chalo=10.7^{+4.5}_{-2.7}$ \citep{UB2008}
and
$\chalo=10.2^{+2.5}_{-2.0}$ \citep{Umetsu+2011},
both of which are based on the joint analysis of shear and magnification
data from Subaru $Vi'$ imaging.\footnote{\citet{UB2008} derived a
$\kappa(\btheta)$ map for the cluster using an entropy-regularized
maximum-likelihood combination of 2D shear and magnification
maps. \citet{Umetsu+2011} derived a $\kappa(\theta)$ profile from a joint 
likelihood analysis of azimuthally-averaged shear and magnification
measurements.}  
This accurate agreement comes in spite of using
different data reduction procedures 
and mass reconstruction methods (Sections \ref{sec:method},
\ref{sec:subaru}, and \ref{sec:wlana}).  

Combining weak and strong lensing reduces the uncertainties on the
concentration.
The {\em HST} strong-lensing data alone also favor a high degree of
projected concentration, $\chalo=8.69\pm 1.26$, but with a somewhat
higher halo mass, $\Mhalo=(1.79\pm 0.31)\times \Munit$.
The combined weak and strong lensing data yield 
$\chalo=10.10\pm 0.82$ at $\Mhalo=(1.32\pm 0.09)\times \Munit$,
corresponding to the Einstein radius of
$\theta_\mathrm{Ein}=52^{+6}_{-7}\arcsec$ at $z_\mathrm{s}=2$. 
Our analysis thus reproduces the correct size of the
observed Einstein radius (Table \ref{tab:cluster}).  
These results are
in good agreement with those of \citet{UB2008} and
\citet{Coe+2010} (Table \ref{tab:cMtable}), in spite of using completely
independent approaches to strong lens modeling (Section \ref{sec:slana}).
Most recent weak-and-strong lensing studies of A1689
appear to converge toward $\chalo\sim 9$--$10$ with a typical
measurement uncertainty of $10\%$ (Table \ref{tab:cMtable};
with the spherical prior), thanks to the advanced analysis methods and
greatly improved quality of data.

\subsubsection{Triaxial Modeling}

Including triaxiality weakens parameter constraints from
lensing data \citep{2005ApJ...632..841O,Corless2009triaxial}, compared to those derived
assuming spherical symmetry.
The parameter constraints become more degenerate and less restrictive
 because of the
lack of information of the halo elongation along the line of sight
(Table \ref{tab:3DNFW}).
These trends are also found
in the posterior distributions from our data (Tables
\ref{tab:3DNFW} and \ref{tab:cMtable}).

Now we consider the results from full triaxial analyses combining
lensing with X-ray/SZE data.
Table \ref{tab:3DNFW} shows that our posterior inference of the intrinsic
parameters is insensitive to the assumed choice of priors (``Flat'' or
``$N$-body'') when the line-of-sight information from X-ray/SZE data is
combined with lensing, 
suggesting that 
the posterior constraints are dominated by the likelihood (i.e.,
information from data) rather than the prior \citep{Sereno2013glszx}.
Whatever the assumptions regarding the axis ratios,
we find the posteriors
(Table \ref{tab:3DNFW}) to be statistically compatible with the
predicted  distribution $c(M)$ for the full population of halos in
$\Lambda$CDM cosmological simulations
\citep{Bhatt+2013,Meneghetti2014clash,Diemer+Kravtsov2015}.\footnote{The
theoretical predictions from \citet{Bhatt+2013} and
\citet{Diemer+Kravtsov2015} are based on DM-only simulations, and those
from \citet{Meneghetti2014clash} are based on nonradiative simulations of DM and baryons.}
This is demonstrated in Figure \ref{fig:cM} for the weak-lensing plus
X-ray/SZE analysis and for the weak/strong-lensing plus X-ray/SZE
analysis, both based on the uninformative priors.
Here we adopt the median $c$--$M$ relation obtained by
\citet{Diemer+Kravtsov2015} as a reference model for comparison.

A1689 appears to be a high mass cluster of $\Mhalo\sim \Munit$ in the
high-concentration tail of the predicted $c(M)$ distribution (Figure
\ref{fig:cM}).  
The posterior tail at lower concentrations of A1689 is only 
$\simgt 1\sigma$ away from the predicted median concentration
($\log_{10}{\overline{c}_\mathrm{200c}}\simeq 0.58\pm 0.16$; Figure \ref{fig:cM}).
Our results are also in agreement with those obtained by a multi-probe
analysis of \citet{Sereno2013glszx} (see Table \ref{tab:cMtable}), who
developed the triaxial inversion algorithm used in this work.

The halo concentration and orientation are strongly
correlated \citep{Sereno+Umetsu2011,Sereno2013glszx}. 
For the posterior range $0^\circ\le \vartheta\le 5^\circ$
assuming a nearly perfect alignment between the halo major axis and the
line of sight,
we find $\chalo=7.4\pm 1.0$ ($6.7\pm 1.1$) from weak/strong lensing
(weak lensing) combined with the X-ray/SZE data.

\subsection{Intrinsic Shape and Orientation of A1689}
\label{subsec:intshape}

We have obtained evidence for a triaxial mass distribution of A1689. 
The projected mass distribution derived from weak-lensing shear and
magnification reveals a north--south elongation 
($\psi_\epsilon=14.2^\circ\pm 8.4^\circ$ east of north, see Table
\ref{tab:2DNFW} and Figure \ref{fig:subaru}). 
We have determined the ellipticity of the projected
mass distribution to be $\epsilon=0.29\pm 0.07$ (Table \ref{tab:2DNFW}),
which is typical for the population of collisionless CDM halos
\citep{2002ApJ...574..538J} but slightly rounder than
the standard CDM prediction for the mean halo ellipticity, 
$\langle\epsilon\rangle\sim 0.4$ \citep{Oguri2010LoCuSS}.
The matter ellipticity is detected at the $4\sigma$ level from weak
lensing alone, thanks to the greatly improved quality of Subaru
data. Our free-form reconstruction from {\em HST} strong lensing gives a
consistent estimate of $\epsilon=0.27\pm 0.09$. 
The ICM and matter distributions are co-aligned in projection
($\psi_\epsilon^X=12^\circ\pm 3^\circ$) but with different ellipticities
($\epsilon_X=0.15\pm 0.03$), which is consistent with the geometric
assumptions made (Section \ref{subsec:ICM}).

When combined with X-ray/SZE observations, our lensing data favor a
triaxial geometry of the matter distribution with minor--major axis ratio
$q_a\sim 0.4$ and major axis closely aligned with the line of
sight ($\vartheta=22^\circ\pm 10^\circ$, Table \ref{tab:3DNFW}).   
These results are robust against the choice of priors 
and combinations of lensing data sets.
Despite that the intermediate--major axis ratio $q_b$ is less
constrained, the data prefer prolate ($q_a=q_b$) over oblate ($q_b=1$)
configurations. A spherical configuration for A1689 is strongly ruled out.
Overall, triaxial configurations fit the combined
lensing and X-ray/SZE data much better than axially symmetric halos do
\citep{Sereno2013glszx}. 

Our analysis shows that A1689 is elongated along the line of sight, as
found by previous studies
\citep{Sereno+Ettori+Baldi2012,Sereno2013glszx,Limousin2013}. 
From the posterior samples, we find
$e_\parallel=1.19\pm 0.37$ 
($1.20\pm 0.34$)
and
$e_\parallel^\mathrm{ICM}=1.22\pm 0.24$ 
($1.24\pm 0.25$),
as constrained by the combined 
weak/strong-lensing (weak lensing) and X-ray/SZE data sets.  
Such biased orientations are favored, although the intrinsic
orientations are a priori assumed to be random. 
The a priori probability of a randomly oriented halo to have
$\vartheta<45^\circ$ is $\sim 29\%$ \citep{Sereno2013glszx}.
The a posteriori probability of such a configuration is found to be 96\%
(99\%) assuming a flat ($N$-body-like) distribution of axis ratios.
We emphasize that the use of X-ray plus SZE data is essential for
obtaining data-driven constraints on the line-of-sight elongation.
To break parameter degeneracies in a lensing-only triaxial analysis, one
would have to assume informative priors on the halo shape and
orientation \citep{Corless2009triaxial,Sereno+Umetsu2011}.


We find that the ICM is mildly triaxial with
$q_a^\mathrm{ICM}\sim 0.6$ and 
$q_b^\mathrm{ICM}\sim 0.7$ (Table \ref{tab:ICM}).
The ratio of ICM to matter eccentricities is 
$e^\mathrm{ICM}/e=0.87\pm 0.07$ (Table \ref{tab:ICM}),
supporting the theoretical assumption we have made that
the shape of the gas distribution is rounder than the
underlying matter (Section \ref{subsec:ICM}). 
On the other hand, we find that  the gas distribution is more elongated
than the gravitational potential \citep[$e^\mathrm{ICM}/e\simgt 0.7
$,][]{Lee+Suto2003}, suggesting a deviation from HSE.
These results are again insensitive to the choice of the
priors.
The inferred values of $q_a^\mathrm{ICM}$ and $q_b^\mathrm{ICM}$
are somewhat lower (more elongated) than, but consistent within errors
with, the results of \citet{Sereno+Ettori+Baldi2012,Sereno2013glszx} 
based on the same X-ray data.
The difference is mainly due to the improved, self-consistent SZE
analysis.



\begin{figure}[!htb] 
 \begin{center}
  \includegraphics[width=0.45\textwidth,angle=0,clip]{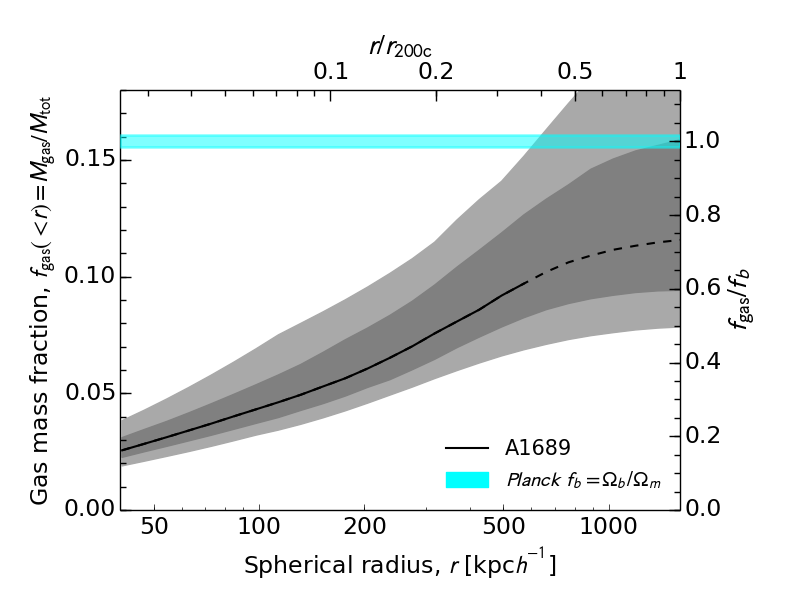}
 \end{center}
\caption{
\label{fig:fgas}
Ratio of spherically-enclosed gas mass ($M_\mathrm{gas}$) to total mass
 ($M_\mathrm{tot}$) as a function of spherical radius $r$, derived from
 the full triaxial analysis of weak/strong-lensing and X-ray/SZE data.
The middle line tracks the median. The gray shaded regions represent
 the 68.3\% and 95.4\% quantiles of the distribution.
Portions of these lines are dashed to indicate extrapolations to larger
 cluster radii. The horizontal bar shows the cosmic baryon fraction
 $f_b=\Omega_b/\Omega_m$ 
 determined by \citet{Planck2015XIII}.
} 
\end{figure} 

\subsection{Gas Mass Fraction}

\begin{deluxetable}{ccccc}
\tabletypesize{\footnotesize}
\tablecolumns{5} 
\tablecaption{
 \label{tab:enclosed}
Ellipsoidal and spherically-enclosed mass estimates for A1689
}  
\tablewidth{0pt}  
\tablehead{ 
 \multicolumn{1}{c}{Overdensity\tablenotemark{a}} &
 \multicolumn{2}{c}{Ellipsoidal\tablenotemark{b}} &
 \multicolumn{2}{c}{Spherically enclosed\tablenotemark{c}} 
\\
 \multicolumn{1}{c}{$\Delta$} &
 \multicolumn{1}{c}{$R_\Delta$} &
 \multicolumn{1}{c}{$M(<R_\Delta)$} &
 \multicolumn{1}{c}{$r_\Delta$} &
 \multicolumn{1}{c}{$M_\mathrm{sph}(<r_\Delta)$} 
} 
\startdata 
 $500$  & $1.89\pm 0.46$ & $0.97\pm 0.13$ & $1.08\pm 0.06$ & $0.88\pm 0.13$\\
 $200$  & $2.79\pm 0.69$ & $1.24\pm 0.16$ & $1.60\pm 0.16$ & $1.15\pm0.16$
\enddata 
\tablecomments{The overdensity radii are given in units of Mpc\,$h^{-1}$. The enclosed masses are in units of $\Munit$.}
\tablenotetext{a}{Mean interior overdensity with respect to the critical density $\rho_\mathrm{c}$ for closure of the universe at $z=0.183$.}
\tablenotetext{b}{Ellipsoidal overdensity radius $R_\Delta$ and total mass enclosed within $R_\Delta$.}
\tablenotetext{c}{Spherical overdensity radius $r_\Delta$ and spherically-enclosed total mass within $r_\Delta$.}
\end{deluxetable}

We compute the ratio 
of spherically-enclosed gas mass $M_\mathrm{sph,gas}(<r)$ to total
mass $M_\mathrm{sph,tot}(<r)$ using the posterior samples of the
ellipsoidal cluster model:
\begin{equation}
 \fgas(<r) \equiv \frac{M_\mathrm{sph,gas}(<r)}{M_\mathrm{sph,tot}(<r)},
\end{equation}
where $M_\mathrm{sph}(<r)$ denotes the total mass enclosed within a
sphere of radius $r$, $M_\mathrm{sph}(<r)=\int_{4\pi}\!d\Omega\int_0^r dr'r'^2\rho(\br')$
with $d\Omega$ the solid angle. 
In Table \ref{tab:enclosed}, we list the values of ellipsoidal and spherical
overdensity mass of the cluster evaluated at $\Delta=200$ and $500$.


The resulting $\fgas$ profile is shown in Figure
\ref{fig:fgas} as a function of integration radius $r$.
The gas mass fraction within 
$0.9\mathrm{Mpc}\sim 1.2r_\mathrm{2500c}$ is estimated as 
$\fgas(<0.9\mathrm{Mpc})= 0.100^{+0.031}_{-0.016}$.
When the gas mass measurements are extrapolated to $r_\mathrm{500c}$
(Table \ref{tab:enclosed}), we find 
$\fgas(<r_\mathrm{500c})= 0.112^{+0.039}_{-0.020}$.
When compared to the cosmic baryon fraction $f_b$ inferred from
\citet{Planck2015XIII},
$f_\mathrm{gas}(<r_\mathrm{500c})/f_b=0.71^{+0.25}_{-0.12}$.
These are consistent with typical values observed for high-mass clusters
\citep{Allen2008fgas,Umetsu+2012,Okabe2014suzaku}. 

Previous studies based on X-ray and lensing data found relatively low
$\fgas$ values for A1689 using lensing total mass estimates, but 
assuming spherical symmetry:
$\fgas(<0.25 \rhalo)=(0.0557\pm 0.0039)h_{70}^{-3/2}$\footnote{\citet{2008MNRAS.386.1092L} found
$\rhalo=1.71$\,\Mpch from their analysis.} 
\citep{2008MNRAS.386.1092L};
 $\fgas(<r_\mathrm{2500c})=0.0552^{+0.0056}_{-0.0062}$,
 $\fgas(<r_\mathrm{500c}) =0.0812^{+0.0145}_{-0.0157}$,
and
 $\fgas(<r_\mathrm{200c}) =0.1053^{+0.0227}_{-0.0246}$
\citep[][see also \citet{Kawaharada+2010}]{Okabe2014suzaku}.

\citet{Umetsu+2009} measured gas fractions for a sample of four
high-mass clusters including A1689 from a joint analysis of AMiBA SZE
and Subaru weak-lensing observations, combined with published X-ray
temperature measurements. Assuming spherical symmetry, they found for A1689
$\fgas(<r_\mathrm{2500c})=0.098^{+0.025}_{-0.026}$
and
$\fgas(<r_\mathrm{500c})=0.115\pm 0.029$, 
in excellent agreement with our results.
Their gas fraction measurements are expected to be less sensitive to
triaxiality because their $\fgas$ estimator depends on the ratio of the
SZE and lensing signals, which are subject to similar projection effects 
albeit with somewhat different degrees of impact.
 
\begin{figure}[!htb] 
 \begin{center}
  \includegraphics[width=0.45\textwidth,angle=0,clip]{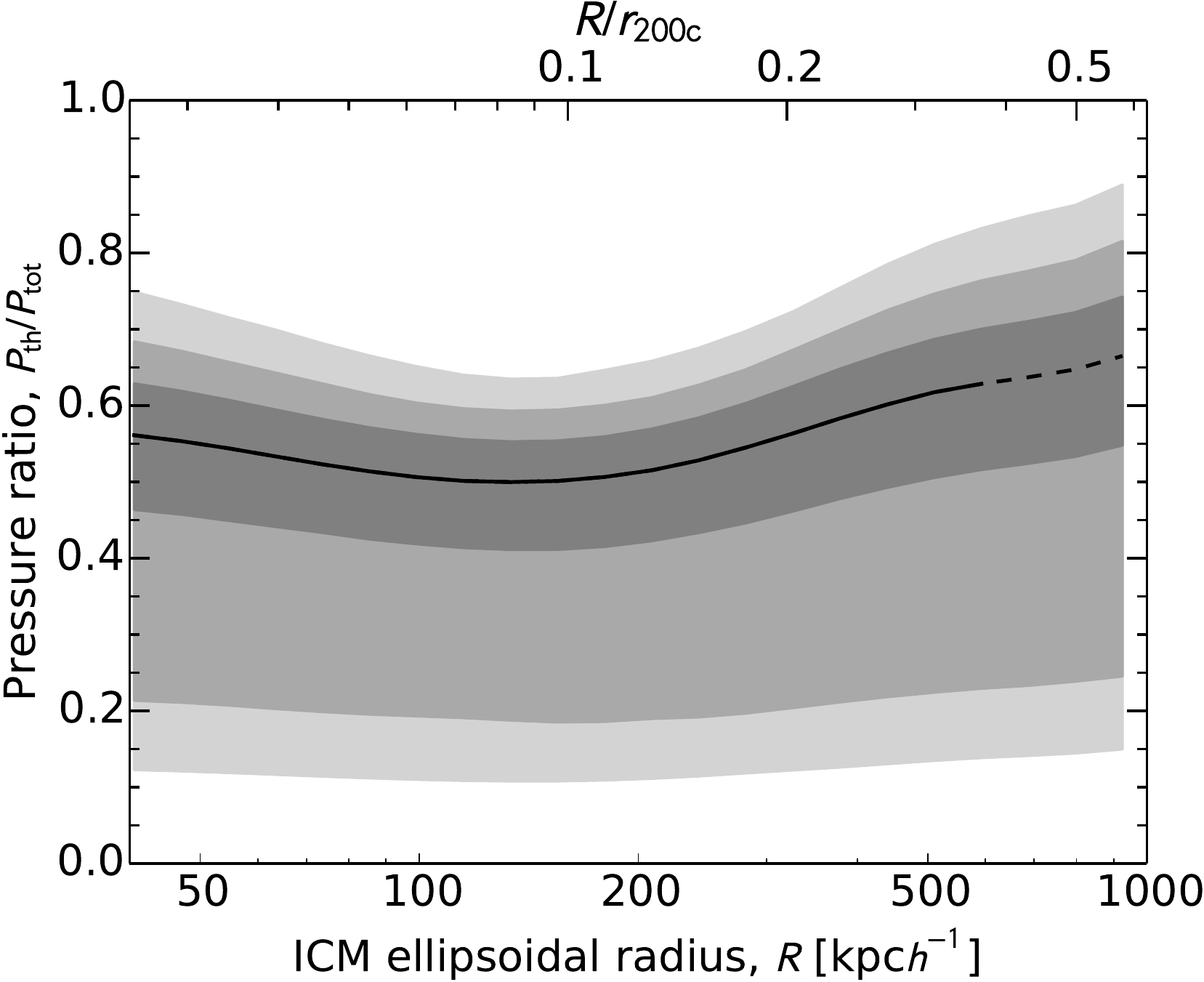}
 \end{center}
\caption{
\label{fig:pratio}
Ratio of the thermal gas pressure ($P_\mathrm{th}$) to the total
 equilibrium pressure ($P_\mathrm{tot}$) in
 A1689 as a function of the ellipsoidal radius $R$ measured
 along the major axis of the ICM halo.
The middle line tracks the median. The gray shaded regions show
 the 68.3\%, 99.4\%, and 99.7\% quantiles of the distribution,
 respectively. Portions of these lines are dashed to indicate
 extrapolations to larger  cluster radii.
} 
\end{figure} 

\subsection{Degree of Hydrostatic Equilibrium}
\label{subsec:HSE}

A quantitative assessment of the degree of equilibrium in the ICM is a
critical issue for cluster cosmology based on hydrostatic mass estimates 
\citep[e.g.,][]{Planck2014XX,CoMaLit1}.    
A significant advantage of our method is the ability to
determine the intrinsic structure, shape, and orientation of the cluster
system without a priori assuming HSE \citep{Sereno2013glszx}. 
This allows us to compare the ICM properties
directly to the gravitating mass corrected for projection effects, and
thus to quantify the contribution of the thermal gas 
pressure $P_\mathrm{th}$ to the total equilibrium pressure $P_\mathrm{tot}$
\citep[][]{Molnar2010HSE,Kawaharada+2010}.
Here $P_\mathrm{tot}$ is determined by the gravitational potential $\Phi$ through
$\bnabla P_\mathrm{tot}=-\rho_\mathrm{gas} \bnabla \Phi$ with
$\rho_\mathrm{gas}$ the gas mass density.
A consequence of the pressure equilibrium is the {\em X-ray shape
theorem} \citep{Buote+Canizares1994}, namely, that the gas in strict HSE
is expected to follow iso-potential surfaces of the underlying matter
distribution. For A1689, we find that the gas is more elongated than the
gravitational potential (see Section \ref{subsec:intshape}), which
points to a deviation from equilibrium.

In Figure \ref{fig:pratio}, we show the ratio of thermal to equilibrium
gas pressure, $P_\mathrm{th}/P_\mathrm{tot}$, as a function of
ellipsoidal radius $R$ of the ICM distribution. For this aim, we
have recomputed the posterior probability distributions for the cluster
parameters, by imposing a sharp prior of $e^\mathrm{ICM}/e=0.7$ 
\citep[see][]{Sereno2013glszx}, corresponding to the
assumption that the gas shape follows the gravitational potential.
We find $P_\mathrm{th}/P_\mathrm{tot}\sim 0.6$ 
out to $\sim 0.9$\,Mpc ($\sim 0.4\rhalo$),
indicating a significant level ($\sim 40\%$) of non-thermal pressure
support. The results here are consistent with \citet{Sereno2013glszx},
although our analysis favors a slightly higher level of non-thermal
pressure support. 
We find no significant radial trend in the
$P_\mathrm{th}/P_\mathrm{tot}$ ratio profile.

Our results are in agreement with \citet{Molnar2010HSE}, who
 analyzed a simulated sample of 
massive regular clusters of $(1-2)\times \Munit$ having a smooth density
profile, drawn from high-resolution cosmological simulations.
Their simulations are therefore highly relevant to interpreting the
 observations of A1689.
They found a significant non-thermal contribution due to
 subsonic  gas motions in the
 core region  (20\%--45\%), a minimum contribution (5\%--30\%) at about
 $0.1\rvir$ \citep{Lau+2009}, growing outward to about 30\%--45\% at
 the virial  radius $\rvir$ \citep{Nelson2014}. 

\citet{Molnar2010HSE} also tested the validity of HSE in A1689 
using gravitational lensing \citep[see][]{UB2008,Kawaharada+2010} and
{\em Chandra} X-ray observations under the assumption of spherical
geometry, finding a non-thermal contribution of $\simlt 40\%$.
As discussed by 
\citet{Sereno2013glszx}, this however indicates that this test is highly
sensitive to biases in the X-ray temperature
measurements \citep{Donahue2014clash}. For the cluster, we find the {\em
Chandra} temperatures are about $10\%$ higher than the {\em XMM-Newton}
results used here \citep{Sereno+Ettori+Baldi2012}, so that 
the thermal contribution $P_\mathrm{th}/P_\mathrm{tot}\simgt 0.6$
obtained by \citet{Molnar2010HSE} could be correspondingly
overestimated relative to our results based on the {\em XMM-Newton}
temperatures.

By combining {\em Suzaku} X-ray observations with the
same lensing data as used in \citet{Molnar2010HSE},
\citet{Kawaharada+2010} showed, assuming spherical symmetry, that the
thermal gas pressure within $r_\mathrm{500c}$ is at most 
$40$\%--60\% of the equilibrium pressure and $30$\%--40\% around the
virial radius. Intriguingly, their {\em Suzaku} observations reveal
anisotropic distributions of gas temperature and entropy in cluster
outskirts at $\simgt r_\mathrm{500c}$, correlated with large-scale
structure of galaxies surrounding the cluster. The outskirt
regions in contact with low-density void environments have low gas
temperatures and entropies, indicating that the outskirts of A1689 are
in the process of being thermalized \citep{Kawaharada+2010}.
Their {\em Suzaku} temperature measurements are in agreement with the
{\em XMM-Newton} results \citep{Sereno+Ettori+Baldi2012}.

\citet[][see also \citet{Limousin2013}]{Morandi2011A1689} obtained  
$\Mhalo=(1.81\pm 0.06)\times \Munit$,
$\chalo=5.71\pm 0.47$, and
$q_a\sim 0.5$ for A1689 from a joint analysis of {\em Chandra}
X-ray, weak-lensing, and strong-lensing data (see Table
\ref{tab:cMtable}). The inferred level of triaxiality is 
similar to what we have found (Table \ref{tab:3DNFW}), whereas the
concentration is somewhat smaller and the mass is significantly higher
than our results. 
They found that about 20\,percent
of the total ICM pressure is in non-thermal form, by assuming that
$P_\mathrm{th}/P_\mathrm{tot}$ is constant with radius and the gas shape
follows the form expected for HSE. 
We note again that the $P_\mathrm{th}/P_\mathrm{tot}$ results are
also sensitive to calibration biases in the X-ray temperature
measurements. 

The mass discrepancy between the present results and those by
\citet[][]{Morandi2011A1689} 
can be explained by the difference in their relative weights 
assigned to the weak- and strong-lensing data sets.
As we have seen in Section \ref{subsubsec:sph},
the {\em HST} strong-lensing data favor higher values of
 $M(<r_\mathrm{200c})$
(Table \ref{tab:3DNFW}), although this represents a significant
extrapolation beyond the radial range covered by the multiple images. 
Hence, if the parameter constraints are highly
dominated by strong lensing, this could lead to an overestimate of
$\Mhalo$.


\subsection{Comparison with {\em Planck} data}

We compare the SZE measurements from the interferometric data presented 
in Section \ref{subsec:szx} with a total power estimate based on the
recent {\em Planck} data \citep{Planck2015I}.  
A1689 is detected by {\em Planck} with high significance
\citep[$\mathrm{S/N}>15$,][]{Planck2015XXVII}.  
We construct {\em Planck} SZE maps in two different ways with different
assumptions, using the data in the 143\,GHz, 217\,GHz, and 353\,GHz
channels. The 217\,GHz and 353\,GHz bands are used primarily to remove
the CMB and Galactic foregrounds. 
The difference between the two maps accounts for different assumptions
about the Galactic components: one is based on local estimates of the dust
properties, and the other is on global properties. The resulting SZE
maps are obtained at an effective resolution of $8\arcmin$\,FWHM.
The SZE signal is integrated as a function of clustercentric radius.
We obtain a direct estimate for the total Compton $Y$ parameter of 
 $Y_\mathrm{\it Planck}=(3.8\pm 0.8)\times 10^{-10}$\,sr
integrated out to a sufficiently large radius $13\arcmin$ 
($\sim \rhalo$), beyond
which the integrated SZE signal converges. Here the error is estimated 
 from aperture photometry in the background regions

This direct {\em Planck} measurement of the total SZE signal can be
compared to the results inferred from the interferometric SZA
observations (Section \ref{subsec:szx}). Taking
$Y_\mathrm{SZA}(<6\arcmin)$ (Table \ref{tab:Ycomp}) as a lower limit on
the total SZE flux, we find 
$Y_\mathrm{SZA}(<6\arcmin)/Y_\mathrm{\it Planck}=1.45\pm 0.44$.
Hence, the results from two independent SZE instruments operating at
different angular scales are compatible with each other at $1\sigma$.
The relatively low $Y$ value derived from the {\em Planck} data could be 
understood in light of the low gas temperature and entropy  
at $\simgt r_\mathrm{500c}$ observed by the {\em Suzaku} X-ray satellite
(Section \ref{subsec:HSE}).
The {\em Suzaku} X-ray observations are in agreement with the thermal
pressure profile of A1689 obtained from {\em Planck} data out to $\sim
2r_\mathrm{500c}$ (Y. Mochizuki et al. 2014, submitted to ApJ).  


When compared to {\it Planck}'s hydrostatic mass estimate, 
$M_\mathrm{500c}=(8.77\pm 0.34)\times 10^{14}M_\odot h_{70}^{-1}$, 
our lensing mass measurements (Table \ref{tab:enclosed}) give a
spherical mass ratio of  
$M_\mathrm{\it Planck}/M_\mathrm{GL}=0.70\pm 0.15$ 
and 
$0.58 \pm 0.10$
with and without corrections for lensing projection effects,
respectively.

\section{Summary}
\label{sec:summary}

We have carried out a 3D multi-probe analysis of the rich cluster A1689, 
 one of the most powerful known lenses on the sky
 ($\theta_\mathrm{Ein}=47.0\arcsec\pm 1.2\arcsec$ at $z_\mathrm{s}=2$,
 Table \ref{tab:cluster}), 
 by combining improved weak-lensing data from new wide-field 
$BVR_\mathrm{C}i'z'$ Subaru/Suprime-Cam observations (Sections
 \ref{sec:subaru} and \ref{sec:wlana}) with complementary
 strong-lensing (Section \ref{sec:slana}),  X-ray and SZE (Section 
 \ref{subsec:szx}) data sets. 

We have generalized the 1D weak-lensing inversion method of
\citet{Umetsu+2011} to a 2D description of the mass distribution without
assuming particular functional forms 
(Section \ref{sec:method}).
This free-form method combines the spatial shear pattern with azimuthally
averaged magnification information, the combination of
 which breaks the mass-sheet degeneracy.

We have reconstructed the projected matter distribution from a 
 joint weak-lensing analysis of 2D shear and
 azimuthally integrated magnification constraints (Section \ref{sec:wlana}).
The resulting mass distribution reveals elongation with an axis ratio of
 $q_\perp \sim 0.7$ in projection (Figures \ref{fig:subaru} and
 \ref{fig:kmap}), aligned well with the distributions of cluster
 galaxies and ICM \citep[see][]{Kawaharada+2010}. 
When assuming a spherical NFW halo, our full weak-lensing analysis 
yields a projected halo concentration of $\chalo^\mathrm{2D}=8.9\pm 1.1$ 
($\cvir^\mathrm{2D} \sim 11$),
which is consistent with and improved from earlier
 weak-lensing work based on Subaru $Vi'$ imaging \citep{UB2008,Umetsu+2011}.

We obtain excellent consistency between weak and strong
 lensing in the region where these independent data overlap, 
$\simlt 200$\,kpc (Figures \ref{fig:kplot} and \ref{fig:mplot}).
We also find an improved agreement between weak and strong lensing in terms
 of constraints on projected NFW parameters (Figure \ref{fig:2DNFW}) 
 relative to previous work \citep{Sereno+Umetsu2011}. This is largely
 due to improved techniques for strong-lensing reconstruction
 and to careful regularization of the covariance matrix (Section
 \ref{subsubsec:sl}). 

In a parametric triaxial framework, we have determined the intrinsic
 structure, shape, and orientation of the matter and gas distributions
 of the cluster, by combining weak/strong lensing with X-ray/SZE data
 under minimal geometric assumptions (Section \ref{sec:results}).
We have shown that the data favor a triaxial
geometry with minor--major axis ratio $q_a=0.39\pm 0.15$
 and major axis closely aligned with the line of sight
 ($\vartheta = 22^\circ\pm 10^\circ$).
A spherical
configuration for A1689 has been strongly ruled out.
We obtain
 a halo mass  $\Mhalo=(1.24\pm 0.16)\times \Munit$
and
 a halo concentration $\chalo=8.36\pm 1.27$,
which is higher than typical concentrations 
found for high-mass clusters \citep[$3\simlt \chalo \simlt 6$;
e.g.,][]{Okabe+2013,Umetsu2014clash,Merten2014clash},
but overlaps well with the $\simgt 1\sigma$ tail of the
predicted distribution
\citep[Figure \ref{fig:cM};][]{Bhatt+2013,Meneghetti2014clash,Diemer+Kravtsov2015}.

We find that the ICM is mildly triaxial with
$q_a^\mathrm{ICM}=0.60\pm 0.14$
and
$q_b^\mathrm{ICM}=0.70\pm 0.16$ (Table \ref{tab:ICM}).
The gas distribution is rounder than the
underlying matter, 
$e^\mathrm{ICM}/e=0.87\pm 0.07$,
but more elongated than the gravitational potential 
($e^\mathrm{ICM}/e\simgt 0.7$),
suggesting a deviation from equilibrium.
The gas mass fraction enclosed within a sphere of radius 
$r=0.9\mathrm{Mpc}\sim 1.2r_\mathrm{2500c}$ is found to be
$f_\mathrm{gas}=10.0^{+3.1}_{-1.6}\%$.
When the gas mass measurements are extrapolated to $r_\mathrm{500c}$,
$\fgas(<r_\mathrm{500c})= 11.2^{+3.9}_{-2.0}\%$.
When compared to the cosmic baryon fraction $f_b$ 
\citep{Planck2015XIII}, we find 
$f_\mathrm{gas}(<r_\mathrm{500c})/f_b=0.71^{+0.25}_{-0.12}$ 
(Figure \ref{fig:fgas}). 
These are consistent with typical values observed for high-mass clusters.
The thermal gas pressure contributes to $\sim 60\%$ of the total
 pressure out to $\sim 0.9$\,Mpc (Figure \ref{fig:pratio}), indicating a
 significant level of non-thermal pressure support. 
The results are, however, sensitive to calibration biases in the X-ray
temperature measurements \citep{Donahue2014clash}.  
When compared to {\em Planck}'s hydrostatic mass
 estimate, our lensing mass measurements  yield a spherical mass ratio of 
$M_\mathrm{\it Planck}/M_\mathrm{GL}=0.70\pm 0.15$ and $0.58\pm 0.10$
with and without corrections for lensing projection effects,
 respectively.

Extending this work to larger samples of clusters will enable us to
recover intrinsic distributions of cluster structural properties
(e.g., $\Mhalo, \chalo$) and axis ratios ($q_a,q_b$), for a direct
statistical comparison with the standard $\Lambda$CDM paradigm
and for a wider examination of alternative DM scenarios
\citep[e.g.,][]{Schive2014psiDM}. 
The CLASH survey
\citep{Postman+2012CLASH} provides such ideal multiwavelength data sets
of high quality
\citep{Donahue2014clash,Umetsu2014clash,Zitrin2015clash,Czakon2015,Rosati2014VLT},  
for a sizable sample of 25 high-mass clusters.


\acknowledgments
We thank the anonymous referee for constructive comments and suggestions.
We are grateful for discussions with Sherry Suyu, Adi Zitrin, and Radek
Wojtak. 
We acknowledge the Subaru Support Astronomers plus Kai-Yang Lin and
Hiroaki Nishioka, for assistance with our Subaru observations. 
We thank Nick Kaiser for making the {\sc IMCAT} package publicly
available.
We thank Oliver Czoske for providing the redshift survey information for
A1689.
The work is partially supported by the Ministry of Science and
Technology of Taiwan 
under the grant MOST\,103-2112-M-001-030-MY3.
M.S. acknowledges financial contributions from contracts ASI/INAF
I/023/12/0, by the PRIN MIUR 2010-2011 `The dark Universe and the cosmic
evolution of baryons: from current surveys to Euclid' and by the PRIN
INAF 2012 `The Universe in the box: multiscale simulations of cosmic
structure'.  
M.N. acknowledges financial support from PRIN INAF 2014.
J.M.D acknowledges support of the consolider project 
CSD2010-00064
and
AYA2012-39475-C02-01 funded by the Ministerio de Economia y
Competitividad.  
N.O. is supported by a Grant-in-Aid from the Ministry of Education,
Culture, Sports, Science, and Technology of Japan (26800097).
This work was partially supported by ``World Premier International
Research Center Initiative (WPI Initiative)`` and the Funds for the
Development of Human 
Resources in Science and Technology under MEXT, Japan.
This research was performed while T.M. held a National Research Council
Research Associateship Award at the Naval Research Laboratory (NRL).   
We thank John Carlstrom, Megan Gralla, Marshall Joy, Dan Marrone, and
the entire SZA and OVRO/BIMA teams for providing the SZA and OVRO/BIMA
data used in this study.  
Support for the SZA observations presented in this work was provided by
NSF through award AST-0838187 and PHY-0114422 at the University of
Chicago. 
The OVRO and BIMA observations presented here were supported by National
Science Foundation grants AST 99-81546 and 02-28963.

\appendix

\section{Nonlinear Effect on the Source-averaged Lensing Fields}
\label{appendix:nonlin}

\subsection{Reduced Gravitational Shear}
\label{appendix:nonlin_g}

The reduced shear, $g=\gamma/(1-\kappa)$, is nonlinear with $\kappa$, so
that the averaging operator with respect to the source 
redshift acts nonlinearly on $\kappa$.
In general, a spread of the source redshift distribution, in combination
with the single source-plane approximation, may  
lead to an overestimation of the gravitational shear in the nonlinear
regime.

Let us expand the reduced shear $g=g(z)$ with respect
to $\kappa(z)=W(z)\kappa_\infty$  and $\gamma(z)=W(z)\gamma_\infty$ as
\begin{equation}
g=\gamma/(1-\kappa)=W\gamma_\infty(1-W\kappa_\infty)^{-1}
=W\gamma_\infty\sum_{k=0}^\infty\left(W\kappa_\infty\right)^k.
\end{equation}
The reduced shear averaged over the source redshift distribution
is expressed as
\begin{equation}
\langle g\rangle 
=\gamma_\infty\sum_{k=0}^\infty \langle W^{k+1}\rangle_g \kappa_\infty^k,
\end{equation}
where the angular brackets
represent an ensemble average over the redshift
distribution of background sources.
In the weak-lensing limit where $\kappa_\infty\ll 1$,
$\langle g\rangle\approx \langle W\rangle_g \gamma_\infty
\equiv \langle\gamma\rangle$.   The next order of approximation is
\begin{equation}
\label{eq:s97}
\langle g\rangle_g \approx \gamma_\infty\left(
\langle W\rangle_g  + \langle W^2\rangle_g
\kappa_\infty^2 \right)
\approx \frac{\langle W\rangle_g \gamma_\infty}{1-\kappa_\infty\langle
W^2\rangle_g/\langle W\rangle_g}.
\end{equation}
\citet{Seitz+Schneider1997} showed 
that Equation (\ref{eq:s97}) yields an excellent
approximation in the mildly-nonlinear regime with $\kappa_\infty\simlt 0.6$. 
Defining $f_{W,g}\equiv \langle W^2\rangle_g/\langle W\rangle^2_g$, 
we have the
following expression for the source-averaged reduced shear valid in the
mildly-nonlinear regime:
\begin{equation}
\langle g\rangle
\approx \frac{\langle\gamma\rangle}{1-f_{W,g}\langle\kappa\rangle},
\end{equation}
with 
$\langle\kappa\rangle=\langle W\rangle_g \kappa_\infty$.
For a lens  at relatively low redshift, 
$\langle W^2\rangle_g \approx \langle W\rangle^2_g$
and 
$f_{W,g}\approx 1$, leading to the single source-plane approximation:
$\langle g\rangle\approx \langle \gamma\rangle/(1-\langle \kappa\rangle)$. 
The level of bias introduced by this approximation is 
$\Delta g/g\approx (f_{W,g}-1)\langle \kappa\rangle$.
In typical ground-based deep observations of $z_\mathrm{l}\simlt 0.5$
clusters,
$\Delta f_W\equiv f_W-1$ is found to be of the order of several percent
\citep{Umetsu2014clash}, so that the relative error 
is negligibly small in the mildly-nonlinear regime.

\subsection{Magnification Bias}
\label{appendix:nonlin_mu}

Let us consider a maximally-depleted sample of background sources
with $\alpha=-d\log \overline{N}_\mu(>F)/dF=0$, for which the effect of magnification bias is purely
geometric, $b_\mu =\mu^{-1}$, and insensitive to the intrinsic source
luminosity function.
In the nonlinear subcritical regime, the source-averaged
magnification bias is expressed as \citep{Umetsu2013,Umetsu2014clash}
\begin{equation}
\langle \mu^{-1}\rangle
=
(1-\langle \kappa\rangle)^2-|\langle\gamma\rangle|^2
+(f_{W,\mu} -1)\left(
\langle\kappa\rangle^2-\langle\gamma\rangle^2
\right)
\approx
(1-\langle \kappa\rangle)^2-|\langle\gamma\rangle|^2,
\end{equation}
where $f_{W,\mu}\equiv \langle W^2\rangle_{\mu}/\langle W\rangle_\mu^2$
is of the order of unity, 
$\langle \kappa\rangle = \langle W\rangle_\mu \kappa_\infty$, and 
$\langle \gamma\rangle = \langle W\rangle_\mu \gamma_\infty$.
Hence, the error associated with  the single source-plane approximation is
$\langle \Delta \mu^{-1} \rangle = (f_{W,\mu}-1)(\langle
\kappa\rangle^2-\langle\gamma\rangle^2) \equiv \Delta f_{W,\mu}
(\langle \kappa\rangle^2-\langle\gamma\rangle^2)$, 
which is much smaller than unity for background populations of
our concern ($\Delta f_{W,\mu}\sim O(10^{-2})$) in the mildly-nonlinear subcritical regime
where $\langle\kappa\rangle\sim |\langle\gamma\rangle|\sim O(10^{-1})$.
It is therefore reasonable to use the single source-plane approximation
for calculating the magnification bias of depleted source populations
with $\alpha \ll 1$.

\section{Discretized Expressions for Cluster Lensing Profiles}
\label{appendix:estimators}

First, we derive a discrete expression 
for the mean interior convergence $\kappa_\infty(<\theta)$
as a function of clustercentric radius $\theta$
using the azimuthally averaged convergence $\kappa_\infty(\theta)$.
In the continuous limit,
the mean convergence $\kappa_\infty(<\theta)$ interior to radius $\theta$
can be expressed in terms of $\kappa_\infty(\theta)$ 
as
\begin{equation}
\kappa_\infty(<\theta)=\frac{2}{\theta^2}\int_0^{\theta}
\!d\ln\theta'\theta'^2\kappa_\infty(\theta').
\end{equation}
For a given set of $(\Nbin+1)$ concentric radii $\theta_i$
$(i=1,...,\Nbin+1)$,
defining $\Nbin$ radial bands in the range 
$\theta_\mathrm{min}\equiv\theta_1\le\theta\le \theta_{N_\mathrm{bin}+1}\equiv \theta_\mathrm{max}$,
a discretized estimator for $\kappa_\infty(<\theta)$
can be written in the following way:
\begin{equation}
\label{eq:avkappa_d}
\kappa_\infty(<\theta_i)=
\left(\frac{\theta_\mathrm{min}}{\theta_i}\right)^2
%
\kappa_\infty(<\theta_\mathrm{min})+
\frac{2}{\theta_i^2}\sum_{j=1}^{i-1}
\Delta\ln\theta_{j}
\overline\theta_{j}^2
\kappa_\infty(\overline\theta_{j}),
\end{equation}
with
$\Delta\ln\theta_i \equiv (\theta_{i+1}-\theta_i)/\overline\theta_i$
and $\overline\theta_i$
the area-weighted center of the $i$th
annular bin defined by $[\theta_i,\theta_{i+1}]$.
In the continuous limit, we have
\begin{equation}
\label{eq:medianr}
\overline{\theta}_i
=
2\int_{\theta_i}^{\theta_{i+1}}\!d\theta'\theta'^2/
(\theta_{i+1}^2-\theta_{i}^2)\nonumber\\ 
=
\frac{2}{3}
\frac{\theta_{i}^2+\theta_{i+1}^2+\theta_{i}\theta_{i+1}}
{ \theta_{i}+\theta_{i+1} }. 
\end{equation}

Next, we derive discretized expressions for the tangential
reduced shear $g_{+}(\theta)$ and 
the inverse magnification $\mu^{-1}(\theta)$ in terms of the binned
convergence $\kappa_\infty(\overline{\theta}_i)$, using the following relations:
\begin{eqnarray}
g_+(\overline\theta_i) &=&
\frac{
\langle W\rangle_g \left[
 \kappa_\infty(<\overline\theta_i)-\kappa_\infty(\overline\theta_i)
\right]
}{1- f_{W,g}\langle W\rangle_g\kappa_\infty(\overline\theta_i)},\\
\mu^{-1}(\overline\theta_i)&=&
\left[1-\langle W\rangle_\mu\kappa_\infty(\overline\theta_i)\right]^2
-
\langle W\rangle_\mu^2\left[
 \kappa_\infty(<\overline\theta_i)-\kappa_\infty(\overline\theta_i)
\right]^2,
\end{eqnarray}
where both the quantities depend on the mean convergence
interior to the radius $\overline\theta_i$, 
$\kappa_\infty(<\overline{\theta}_i)$.
By assuming a constant density in each radial band,
we find the following expression for 
$\kappa_\infty(<\overline\theta_i)$:
\begin{equation} 
\kappa_\infty(<\overline\theta_i) = 
\frac{1}{2}\Big[
\left(\theta_{i} / \overline{\theta}_i\right)^2
\kappa_\infty(<\theta_i) 
+
\left(\theta_{i+1} / \overline{\theta}_i\right)^2
\kappa_\infty(<\theta_{i+1})
\Big],
\end{equation}
where $\kappa_\infty(<\theta_i)$ and $\kappa_\infty(<\theta_{i+1})$
can be computed using Equation (\ref{eq:avkappa_d}).

Accordingly, all relevant cluster lensing observables,
$g_+(\theta)$ and $n_\mu(\theta)$,
can be uniquely specified by the binned convergence profile
$\{\kappa_{\infty,\mathrm{min}},\kappa_{\infty,i}\}_{i=1}^{\Nbin}$
with $\kappa_{\infty,\mathrm{min}}\equiv \kappa_\infty(<\theta_\mathrm{min})$ and 
$\kappa_{\infty,i}\equiv \kappa_\infty(\overline{\theta}_i)$.

\section{Two-dimensional to One-dimensional Projection}
\label{appendix:2dto1d}

To make a direct comparison between the results from 1D and 2D
weak-lensing analyses, we construct a projected mass profile $\Sigma(\theta)$ 
from an optimally weighted radial projection of the $\Sigma(\btheta)$ field
as \citep[][]{Morandi2011A1689} 
\begin{equation}
 \bSigma_{(1)}=
\left[
 A^t C_{(2)}^{-1}A
\right]^{-1} A^t C_{(2)}^{-1} \bSigma_{(2)}
\end{equation}
where 
$\bSigma_{(2)}=\{\Sigma(\btheta_m)\}_{m=1}^{\Npix}$ is a pixelated mass
map, $C_{(2)}$ is the pixel--pixel covariance matrix of $\bSigma_{(2)}$,
$\bSigma_{(1)}$ is a vector of radially binned $\Sigma$ values, and
$A$ is a mapping matrix whose elements $A_{mi}$
represent the fraction of the area of the $m$th pixel lying within the
$i$th clustercentric radial bin (Section \ref{subsubsec:magbias}).
The covariance matrix for $\bSigma_{(1)}$ is given by
\begin{equation}
C_{(1)}=\left[A^tC_{(2)}^{-1}A\right]^{-1}.
\end{equation}

\end{document}